\documentclass[a4paper,fleqn,usenatbib]{mnras}
\usepackage{txfonts}

\usepackage[T1]{fontenc}
\usepackage{ae,aecompl}
\usepackage{graphicx}
\usepackage{ctable}
\usepackage{longtable}
\newcommand{\HI}{\textsc{H~i}}
\newcommand{\OI}{\textsc{O~i}}
\newcommand{\SII}{\textsc{S~ii}}
\newcommand{\OVI}{\textsc{O~vi}}
\newcommand{\CIV}{\textsc{C~iv}}
\newcommand{\MgII}{Mg~\textsc{ii}}
\newcommand{\AlII}{Al~\textsc{ii}}
\newcommand{\SiII}{Si~\textsc{ii}}
\newcommand{\FeII}{Fe~\textsc{ii}}
\newcommand{\SiIV}{Si~\textsc{iv}}

\newcommand{\kms}{km~s$^{-1}$}
\newcommand{\nav}{$N_{\rm a}({\emph v})$}

\title[Precipitating high-velocity clouds]{The high-velocity clouds above the disk of the outer Milky Way: misty precipitating gas in a region roiled by stellar streams}

\author[T. M. Tripp]{%
Todd M. Tripp$^{1}$\thanks{E-mail: tripp@astro.umass.edu}
\\
$^{1}$Department of Astronomy, University of Massachusetts, Amherst, MA 01003\\
}

\date{Accepted 2022 January 3. Received 2022 January 3; in original form 2021 December 14}

\pubyear{2022}

\begin{document}
\label{firstpage}
\pagerange{\pageref{firstpage}--\pageref{lastpage}}
\maketitle

\begin{abstract}
The high-velocity clouds (HVCs) in the outer Milky Way at $20^{\circ} < l < 190^{\circ}$ have similar spatial locations, metallicities, and kinematics. Moreover, their locations and kinematics are coincident with several extraplanar stellar streams. The HVC origins may be connected to the stellar streams, either stripped directly from them or precipitated by the aggregate dynamical roiling of the region by the stream progenitors. This paper suggests that these HVCs are ``misty'' precipitation in the stream wakes based on the following observations. New high-resolution (2.6 \kms) ultraviolet spectroscopy of the QSO H1821+643 resolves what appears to be a single HVC absorption cloud (at 7 \kms\ resolution) into five components with $T \lesssim 3\times 10^{4}$ K. Photoionization models can explain the low-ionization components but require some depletion of refractory elements by dust, and model degeneracies allow a large range of metallicity.  High-ionization absorption lines (\SiIV, \CIV, and \OVI) are kinematically aligned with the lower-ionization lines and cannot be easily explained with photoionization or equilibrium collisional ionization; these lines are best matched by non-equilibrium rapidly cooling models, i.e., condensing/precipitating gas, with high metallicity and a significant amount of \HI. Both the low- and high-ionization phases have low ratios of cooling time to freefall time and cooling time to sound-crossing time, which enables fragmentation and precipitation. The H1821+643 results are corroborated by spectroscopy of six other nearby targets that likewise show kinematically correlated low- and high-ionization absorption lines with evidence of dust depletion and rapid cooling. 
\end{abstract}


%\clearpage
\section{Introduction}
\label{sec:intro}

Circumgalactic media (CGM), the large gaseous envelopes of galaxies, were initially revealed by ground-based detections of Ca~\textsc{ii} and \MgII\ absorption in the spectra of quasars projected behind low redshift galaxies at impact parameters $\rho \gtrsim$ 20 kpc \citep{boksenberg78,bergeron86,bergeron91,bowen91}.  After the launch of the \textsl{Hubble Space Telescope (HST)} in 1990, ultraviolet absorption spectra likewise showed that \HI\ is also ubiquitously detected in galaxy halos out to $\rho \approx$ 200 kpc with a high-covering fraction \citep{morris93,lanzetta95,stocke95,chen98,tripp98}. Contemporaneously, measurements of $dN/dz$ (number per unit redshift) of \MgII\ \citep{churchill99} and \OVI\ absorbers \citep{tripp00a,tripp00b,savage02} in high-resolution quasar spectra obtained with \textsl{HST}, the \textsl{Far Ultraviolet Spectroscopic Explorer (FUSE)}, and ground-based telescopes also indirectly indicated that large ($\rho \lesssim 200$ kpc) galaxy halos are pervasively filled with low- and high-ionization metals \citep[e.g.,][]{rigby02,tumlinson05}; this was subsequently confirmed in surveys of QSOs behind low$-z$ galaxies \citep[e.g.][]{prochaska11,tumlinson11,nielsen13,stocke13,bordoloi14,liang14,johnson15,burchett16,heckman17}.

% Goofy latex problem here.  Need extra line here to prevent fatal compiling (latex -> pdf) problem.
 \ 
 
\subsection{Circumgalactic Medium Physics}
\label{sec:cgm_physics}

The discovery of the CGM opened a new window for understanding galaxy evolution \citep[see reviews by, e.g.,][]{bregman07,putman12,naab17,tumlinson17,peroux20}. The CGM plays a critical role in galaxy evolution by regulating the galactic ``baryon cycle'', and studies of this topic are now increasingly focusing on relevant gas microphysics.  

Consider inflows, for example. New star formation occurs throughout a significant portion of a galaxy's history \citep[e.g.,][]{larson72,fenner03,weisz11}, and this requires continuing accretion of gas from the intergalactic medium, recycling of stellar wind and supernova ejecta, acquisition of gas from satellite galaxies, and/or major mergers.  The CGM gas physics could profoundly affect the outcomes of these processes.  For example, the \citet{white78} idea that gas within a ``cooling radius'' falls into a galaxy and forms stars leads to too much star formation if it is simply assumed that all of the gas in this region eventually forms stars \citep[see Fig.12 in][]{maller04}.  Instead, the physics of the cooling gas, and/or feedback, must prevent some of it from turning into stars.  \citet{maller04} suggest that thermal instability causes CGM gas to fragment so that while a smaller portion cools rapidly, falls in to the galaxy, and fuels star formation, much of the CGM gas ends up in a stable regime that cools very slowly and is effectively removed from the baryon cycle.

Outflows present similarly confounding puzzles, e.g., the existence and survival of cool and low-ionization gas moving at high speeds in observed outflows \citep[e.g.,][]{veilleux05,tremonti07,cashman21,zhang17} and high-velocity clouds \citep[HVCs, e.g.,][]{putman12,richter17}. These clouds should be shredded and destroyed before they can attain the observed velocities if they are accelerated by the ram pressure of a hot wind \citep[e.g.,][]{zhang17}.  Possibly the material is accelerated by radiation pressure \citep[e.g.,][]{murray05,murray11,hopkins12} or cosmic rays \citep[e.g.,][]{everett08,bruggen20,quataert21}, but alternatively the high-speed cool gas may be due rapid radiative cooling in some situations.  If a cloud is sufficiently large, then mixing of the cold cloud with the hotter wind can lead to a region of mixed material with a cooling time less than the cloud-destruction time, and this cooling region can cause a cloud's mass to grow (\citealp{armillotta16,gronke18,gronke20,schneider18,kanjilal21}, but see caveats in S5.4 of \citealp{schneider20}). This multiphase cloud growth could explain the presence of cool gas at high speeds and the pervasive low- and high-ionization metals found in the CGM \citep[e.g.,][]{tumlinson11,werk13,burchett19}. Similar conclusions have been reached about inflowing gas: galactic accretion is now recognized to sometimes occur in cold, filamentary streams \citep{keres05,dekel09}, and \citet{mandelker20} have shown that if a cold stream has a large enough radius, it can grow in mass due to cooling in a mixing layer. Models of HVCs moving through a galactic halo also find that the HVCs may be disrupted (or grow very little) if they are small \citep{heitsch09,marinacci10}, but they can grow substantially if they are large enough \citep{fraternali15,gritton17}. \citet{bruggen16} have shown that thermal conduction, while causing evaporation, can actually \textsl{extend} the lifetime of cold clouds (see also \citealp{vieser07} in a somewhat different context).

Thus how gases interact and mix in the multiphase CGM is a potentially crucial question, but the details are a matter of debate. Mixing and cooling gas could reside in turbulent mixing layers (TMLs) that arise in the shear flows between hot and cool regions with some relative motion between them \citep{begelman90,slavin93,fielding20}, it could arise from classical thermal instability \citep{field65}, or it could be driven by pressure differences between the phases \citep{gronke20}. \citet{mccourt18} argue that cooling will cause an optically thin plasma at $T \leq 10^{6}$ K to naturally fragment into a mist-like arrangement of small and denser ``cloudlets'' that comprise an astrophysical ``cloud''; because the fragmentation occurs rapidly, they refer to this as cloud ``shattering'' \citep[see also][]{liang20,gronke21}. In a galactic atmosphere with gravity, it has been argued that circumgalactic ``precipitation'' \citep{sharma12,mccourt12,voit15,voit21} regulates galaxy evolution.  In brief, in this model the ratio of the cooling time, 
\begin{equation}
t_{\rm cool} = \frac{3}{2} \frac{nkT}{{n_{\rm e}n_{\rm i}\Lambda}}, \label{cooling_eqn}
%t_{\rm cool} = 3/2 (nkT/{n_{\rm e}n_{\rm i}\Lambda}),
\end{equation} 
where $T$ is the temperature, $n_{\rm e}$ and $n_{\rm i}$ are the electron and ion number densities, and $\Lambda$ is the cooling function, to the free-fall time, $t_{\rm ff} = \sqrt{2r/g}$, where $g$ is the local gravitational acceleration, indicates how the circumgalactic gas will behave: if $t_{\rm cool}/t_{\rm ff} \gtrsim 10$, then the region is expected to be thermally stable, but if $t_{\rm cool}/t_{\rm ff} \lesssim 1$, cool gas is expected to condense out of the ambient CGM and rain down on the galaxy, thereby fueling star and black-hole formation.  Subsequent feedback from the stars and/or black-hole activity then increases $t_{\rm cool}/t_{\rm ff}$ until the CGM returns to the thermally stable regime.  This concept provides a neat solution to some galaxy-evolution problems: the stability of the CGM in the high-$t_{\rm cool}/t_{\rm ff}$ regime explains the persistence of the CGM and the small fraction of the baryons that turn into stars, and the periodic rain of (low-$t_{\rm cool}/t_{\rm ff}$) cool gas enables prolonged star formation. \citet{voit21} argues that convection will sort a circumgalactic atmosphere so that it has an entropy gradient with a constant median $t_{\rm cool}/t_{\rm ff} \approx 10$; in this situation, ``buoyancy damping'' \citep[see][]{voit21} often saturates the growth of a thermal instability before condensation occurs and multiphase gas develops.  In this case, \citet{voit21} suggests that dynamical disturbances from satellites, bulk flows, and turbulence may be important because they boost the likelihood of precipitation and growth of multiphase gas.  However, other studies \citep[e.g.,][]{sharma12} favor a constant-entropy CGM, which allows cooling to occur more extensively. To prevent overcooling and overproduction of stars, these isentropic precipitation models appeal to cloud shredding.  Precipitation could also depend on the mass of a galaxy and its halo \citep{fielding17,stern21}. Cosmological simulations also find condensation due to thermal instability \citep[e.g.,][]{nelson20,esmerian21}, although these simulations show that the history and destiny of a parcel of gas may be more complex than indicated by $t_{\rm cool}/t_{\rm ff}$.

Finally, cosmic rays (CRs), magnetic fields, and plasma-physics processes may further complicate what happens in the CGM \citep[e.g.,][]{zhang17,mccourt15,ji20,bustard21,butsky20,butsky21,vandevoort21}.

\subsection{Complex C and the Outer Galaxy High-Velocity Clouds: A Laboratory for Testing CGM Physics}
\label{sec:hvc_summary}
It is important to seek observational opportunities to test CGM gas physics theory.  In many ways, the HVCs of the Milky Way provide a valuable ``laboratory'' for studying gas in the CGM.  This study focuses in particular on the Galactic HVCs known as Complex C and the ``Outer Arm'' (OA) because these clouds offer several advantages as probes of CGM physics.  To aid the discussion in this section, it may be useful for the reader to examine Figure~1 in \citet{tripp12} and Figure~11 in \citet{sembach03}; these all-sky maps show the large-scale properties and kinematical similarities of the HVCs, and the \citet{sembach03} map shows how the highly-ionized gas generally moves at similar velocities as the 21cm-emitting gas. The Heliocentric distances of Complex C and the OA are reasonably well constrained (see below), which allows angular dimensions to be converted to physical scales, and their proximity has enabled detailed high-resolution 21cm emission mapping \citep[e.g.,][]{marchal21}.  This also places the clouds in a Galactic context, and spatial sizes provide important constraints on their physical origins.  For example, models that produce highly ionized gas (e.g., \OVI ) by photoionization often require very large clouds \citep[e.g.,][]{tripp08}; such models are unlikely if the absorption arises in a nearby cloud that is known to be much smaller from its angular dimensions and location. This is also the region of a galaxy where effects from cosmic rays and magnetic fields may be more important, so these clouds could test plasma-physics predictions.  The known location also constrains the ionizing flux impinging on the HVCs.  Photoionization models can provide insight on gas densities and radial cloud sizes based on ion ratios, but since the ratios homologously depend on the ionization parameter $U$ ($\equiv n_{\gamma}/n_{\rm H} = $ ionizing photon density/total H number density), if the shape and intensity of the ionizing radiation field are significantly uncertain, which is often the case, then the inferred $n_{\rm H}$ will be uncertain by the same amount.  In turn, the cloud size = $N_{\rm H}/n_{\rm H}$, where $N_{\rm H}$ is the total H column density, is likewise uncertain. For Complex C and the OA, the \citet{fox05} models of the ionizing radiation field emerging from the Galaxy can be applied using the constrained location of the HVCs, and this reduces the uncertainty in $n_{\gamma}$ and hence in the $n_{\rm H}$ inferred from photoionization models.

% TABLE: Outer-Galaxy clouds and stellar streams
\ctable[
caption={Properties of Outer-Galaxy High-Velocity Gas Clouds and Stellar Streams},
label={tab:clouds},
doinside=\footnotesize,
star
]{lccccccc}{
\tnote[a]{Range of Heliocentric distance ($d_{\odot}$) constraints from the literature and Galactocentric radius ($R_{\rm G}$) of the HVCs and stellar streams, calculated assuming the distance from the Sun to the Galactic center is 8.5 kpc.} 
\tnote[b]{Source of the Heliocentric distance constraint: (1) \citet{lehner10}, (2) \citet{tripp12}, (3) \citet{wakker07}, (4) \citet{thom08}, (5) \citet{morganson16}, (6) \citet{ibata21}, (7) \citet{malhan19}.}
\tnote[c]{Logarithmic metallicity in the usual notation, [M/H] = log (M/H) - log (M/H)$_{\odot}$.  The metallicities in the literature are based on various elements (see cited sources); here M generically represents various metals. For the HVCs, the listed range of metallicities reflects results obtained from different analyses and different sightlines. In the cases of the Monoceros North and Anticenter Streams, the listed numbers are the median (spread) of metallicities of stars from the \textsc{segue} spectroscopic survey. \citet{laporte20} obtain similar results using stars from the \textsc{lamost} survey.}
\tnote[d]{Source of the metallicity measurement: (7)  \citet{tripp12}, (8) \citet{tripp03}, (9) \citet{collins07}, (10) \citet{laporte20}, (11) \citet{ibata21}, (12) \citet{malhan19}.}
\tnote[e]{Metallicity favored by the stellar-stream identification algorithm of \citet{malhan18}, which agrees with the spectroscopic metallicities of three stars from \textsc{segue} and \textsc{lamost}.}
\tnote[f]{Metallicity adopted by the \citet{malhan18} algorithm.  However, unlike the Hr\'{i}d stream, the Gaia-9 metallicity was not corroborated with spectroscopic stellar metallicities.}
}
{\FL
   \       &   \multicolumn{2}{c}{\underline{Galactic Coordinates}} &  \multicolumn{3}{c}{\underline{\ \ \ \ \ \ Distance\tmark[a] \ \ \ \ \ \ }} & \multicolumn{2}{c}{\underline{\ \ \ \ Metallicity \ \ \ \ }} \NN
Object & Long. & Lat. & $d_{\odot}$ & $R_{\rm G}$ & Ref.\tmark[b]        & [M/H]\tmark[c] & Ref.\tmark[d] \NN
   \      & (deg.) & (deg.) & (kpc) & (kpc) & \ & \ & \ML
\underline{High-Velocity Cloud} & \ & \ & \  & \ & \NN
Outer Arm & $50 - 190$ & $0 - 30$ & $2-15$ & $10-18$ & 1,2 & $-0.7$ to $-0.3$ & 7 \NN
Complex C & $15 - 150$ & $15 - 60$ & $7-13$ & $11-15$ & 3,4 & $-1.0$ to $-0.5$ & 8,9 \NN
\underline{Stellar Stream} & \ & \ & \  & \ & \NN
Monoceros (North) & $80 - 220$ & $15 - 35$ & $8-9$ & $11-16$ & 5 & $-0.59(0.32)$ & 10 \NN
Anticenter Stream & $75 - 240$ & $15 - 40$ & $9-10$ & $11-17$ & 5 & $ -0.73(0.26)$ & 10 \NN
Hr\'{i}d & $45 - 105$ & $8 - 25$ & $\approx 3$ & $7-10$ & 6 & $-1.1$\tmark[e] & 11 \NN
Gaia-9 & $60 - 115$ & $40 - 50$ & $3-7$ & $8-10$ & 6 & $-1.9$?\tmark[f] & 11 \NN
GD-1   & $100 - 210$ & $35-65$ & $7-12$& $5-13$ & 7 & $-2.24\pm0.21$ & 12 \LL
}

\begin{figure*}
\includegraphics[width=15.5cm]{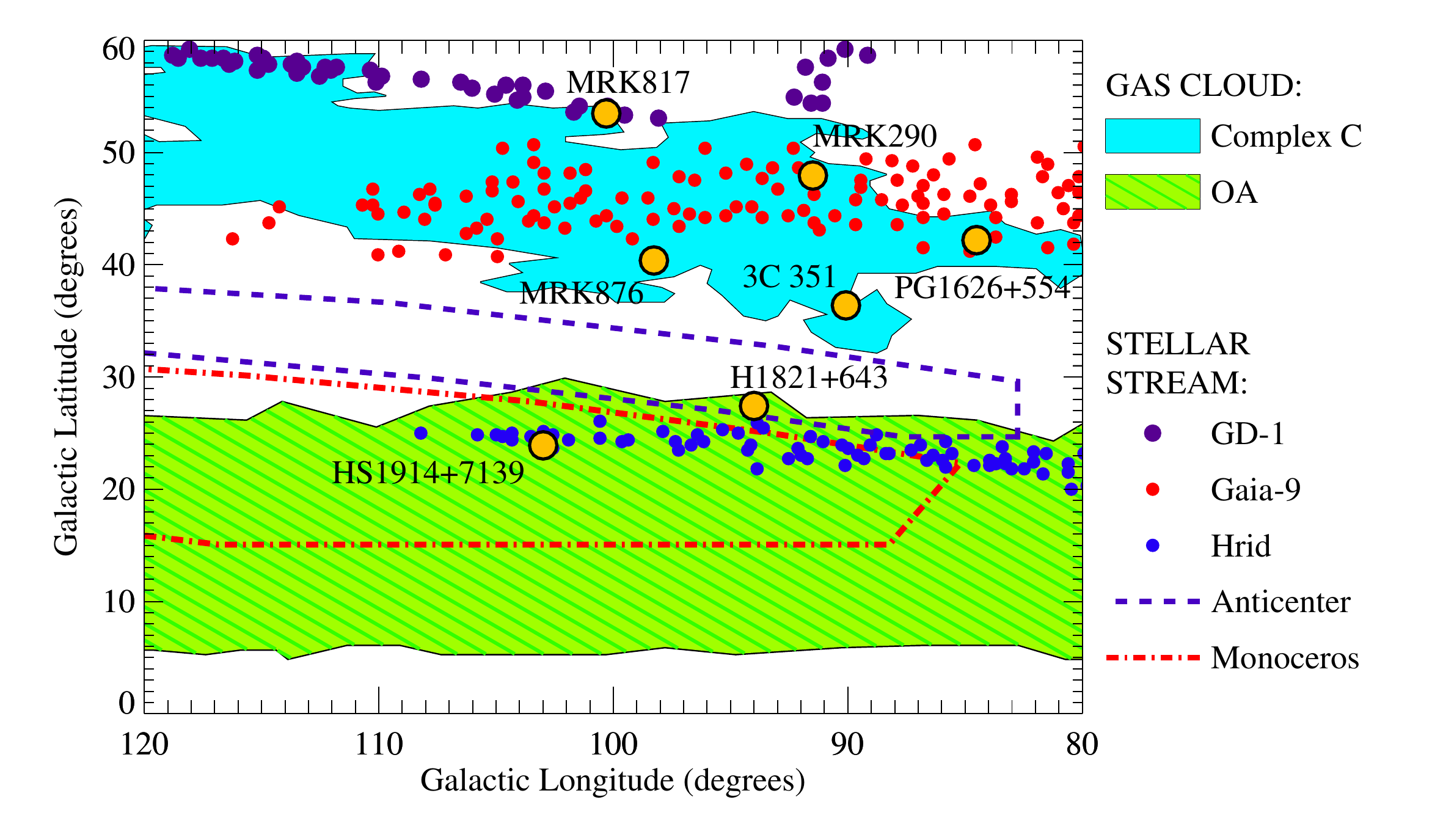}
\caption{Schematic map of the high-velocity clouds and stellar streams in the outer Galaxy region studied in this paper.  Filled regions indicate the $N$(\HI ) $\approx 10^{19}$ cm$^{-2}$ boundaries of the high-velocity gas clouds Complex C (cyan) and the ``Outer Arm'' (green) based on the Leiden-Agentine-Bonn (LAB) Survey \citep{kalberla05}. Red and blue dots mark the locations of stars in the Gaia-9 and Hr\'{i}d stellar streams identified by \citet{ibata21}, and the dashed purple and dot-dashed red lines schematically enclose the locations of the Anticenter Stream \citep{grillmair06} and the Monoceros Stream \citep{newberg02,penarrubia05} according to the recent study by \citet{laporte20}. Labeled yellow dots show the locations of the UV continuum sources (see Table~\ref{tab:targets}) used here to study the gas clouds in absorption. Some properties of these clouds and streams are summarized in Table~\ref{tab:clouds}.\label{fig:outergalmap}}
\end{figure*}

Complex C and the OA are dominating features in 21cm emission maps of HVCs in the the Milky Way halo \citep[see, e.g., Fig.~1 in][]{tripp12}.  There are a variety of hypotheses about the OA origin in the literature, as briefly reviewed in Section 2 of \citet{tripp12}.  The ``Outer Arm'' name is unfortunate because, for the reasons discussed in \citet{tripp12}, this gaseous object is probably not related to the spiral arm detected near the plane in the outer Galaxy that is also called the Outer Arm \citep[e.g.,][]{dame11}. HVCs are historically identified by 21cm emission and are typically defined as gas with velocities that deviate from Galactic rotation ($\equiv {\rm v}_{\rm dev}$) by more than 70 to 90 km s$^{-1}$ \citep[e.g.,][]{wakker91,putman12}, and on this basis the OA is often excluded from HVC maps and catalogs. However, this is a spurious omission because the OA exhibits abundant gas with $\mid {\rm v}_{\rm dev} \mid \ > \ 80$ km s$^{-1}$ \citep[see Fig.~26 of][]{marchal21}, and the OA kinematics cannot be reconciled with normal Galactic rotation.

% Goofy latex error
% \ 

Constraints on the location and metallicity of Complex C and the OA are summarized in Table~\ref{tab:clouds}. \citet{tripp12} have suggested that Complex C and the OA, as well as the smaller Complex G and Complex H HVCs, may be related since they have similar kinematics, similar metallicities, and are in the same spatial location of the Galaxy. \citet{tripp12} did not discuss Complex A, but it also has a similar distance and metallicity \citep{barger12}. Based on spectroscopy of stars with known distances, these HVCs are constrained to be at similar Galactocentric radii (see Tab.~\ref{tab:clouds}).  Given these distance constraints, Complex C and the OA are both $\approx$20 kpc across along their long axes. Absorption-line abundances also indicate that Complex C and the OA have similar metallicities (Tab.~\ref{tab:clouds}, but see S\ref{sec:1821lowionmodels} - \ref{sec:1821highionmodels} below).  The somewhat higher metallicity of the OA, which is closer to the plane than Complex C, may be due to mixing with ISM gas in the outer Milky Way, but within the systematic uncertainties of the photoionization models required to estimate ionization corrections, the OA and C metallicities could be essentially the same.  It is also interesting to note that Complex C exhibits a ``high-velocity ridge'' (hereafter referred to as the HVR) along its spine with a speed that is $\approx 50$ km s$^{-1}$ more negative than the lower-velocity portion of Complex C (see \citealp{tripp03} and Fig.6 in \citealp{wakker01}), and the OA is also clearly detected at the HVR velocity as well as the main Complex C velocity \citep{tripp03,tripp12}.

While these similar characteristics suggest that the OA and C could have a relationship, the nature of this relationship is not clear.  \citet{kawata03} used numerical simulations to investigate whether Complex C could have been produced by the passage of a satellite galaxy through the Milky Way disk, and while they found that such an event could produce a structure with the general properties of Complex C and the OA, they did not favor this hypothesis because they could not identify compelling evidence of the putative satellite.  However, subsequently several stellar streams and structures have been discovered in this region of the outer Galaxy, and the progenitor(s) of these stellar structures could potentially also explain the origin of Complex C and the OA in a scenario like the one investigated by \citet{kawata03}.   The most well-known stellar stream in this region is the Monoceros Ring \citep[e.g.,][]{newberg02,penarrubia05}, but surveys such as the Sloan Digital Sky Survey \citep[SDSS,][]{york00} and the Gaia Survey \citep{gaia16} have revealed additional stellar streams in this part of the Galaxy. Table~\ref{tab:clouds} summarizes some properties of stellar streams in this direction including Monoceros ``North'', the Anticenter Stream, Hr\'{i}d, Gaia 9, and GD1.  It is intriguing to note that some of the metallicities and locations of these stellar streams are similar to the published metallicities of Complex C and the OA, but as we will see in this paper, the metallicities of these HVCs are not securely measured -- there are degeneracies in the ionization models used to derive metallicities, and it is possible that these HVCs have significantly higher metallicities than the published abundances.  It is also worth noting that some of these stellar streams, including the northern Monoceros Ring, Hr\'{i}d, and GD1, have similar velocities to the HVCs. The Anticenter Stream, on the other hand, exhibits significantly different kinematics with mean velocities of $\emph{v}_{\rm LSR} = +48$ and $+78$ \kms\ in two directions studied by \citet{grillmair08}.  However, spatially overlapping streams with different kinematics could lead to a roiled, turbulent region, which in turn could make precipitation more likely \citep{voit21}.  To more visually show the spatial correspondence of the stellar structures and the gas clouds, Figure~\ref{fig:outergalmap} shows a schematic map of the projected locations of the streams and clouds from Table~\ref{tab:clouds}.  Note that Figure~\ref{fig:outergalmap} is centered and zoomed in on a set of active galactic nuclei employed in this paper (see below) and does not show the full extent of the HVCs and streams.

Various relationships between the outer-Galaxy stellar streams and the HVCs are possible.  Similarly to the \citet{kawata03} hypothesis about the origin of Complex C, some of the stellar streams could be extraplanar features elevated from the disk by the passage of a satellite galaxy \citep[e.g.,][]{laporte18,laporte21}, which could persist for significant amounts of time \citep{laporte21}. \citet{laporte18} favor the Sagittarius dwarf galaxy \citep{ibata94,law10} as the main perturber, but the Large Magellanic Cloud could also play a role \citep[see also][]{weinberg06}. The orbit of Sagittarius and the Cetus Stream are not shown in Figure~\ref{fig:outergalmap}; these structures also pass through this region -- see Fig.3 in \citet{yuan21} -- roughly orthogonally to the HVCs and streams shown in the figure. The gas could connect to a satellite by a classical mechanism such as ram-pressure or tidal stripping \citep[e.g.,][]{tonnesen21}, or it could be indirectly connected through the dynamical effects the satellite/streams have on the region.  If precipitation is more likely to occur in dynamically disturbed areas, then this region of multiple stellar streams is a promising place to search for examples of the process.

Currently, absorption spectroscopy using background QSOs/AGNs provides the most sensitive probe of gas in HVCs and the CGM. Fortunately, there are many UV-bright QSOs and active galactic nuclei (AGN) behind Complex C and the OA \citep{wakker01}, and extensive high-quality spectra of many of these AGN are available from space-telescope archives \citep[e.g.,][]{sembach03,collins07,richter17}.  Moreover, one QSO behind the OA (H1821+643) has been observed in the UV with very high spectral resolution (2.6 km s$^{-1}$ FWHM, see HST Program 15321). These high-resolution data provide a unique opportunity to search for clustered narrow absorption components (which would probably be unresolved in most available ultraviolet spectra through the CGM) as might be expected in rapidly cooling gas.

% TABLE: Targets
\ctable[
caption={High-Velocity Cloud Sightlines Near the Longitude of H1821+643},
label={tab:targets},
doinside=\footnotesize
]{llllcc}{
\tnote[a]{Galactic coordinates in degrees}
\tnote[b]{Velocity shift between the Heliocentric frame and the Local Standard of Rest (LSR): v$_{\rm LSR} = $ v$_{\rm Helio} + \Delta {\rm v}_{\rm LSR}$.}
\tnote[c]{Log of the total \textsc{H~i} column density (in cm$^{-2}$) from 21cm emission at v$_{\rm LSR}$ $< -90$ km s$^{-1}$ \citep[from][]{wakker01,wakker03}}
\tnote[d]{Star in the outer Galaxy that exhibits absorption lines similar to those observed toward H1821+643 \citep[see][and text]{lehner10}}}
{\FL
  \         &   \                     & \                                  &                & \                                  & HVC \NN
   \       &    \                     & Gal.                            &  Gal.      & $\Delta {\rm v}_{\rm LSR}$ \tmark[b]  & log \NN
Target & $z_{\rm AGN}$ & long.\tmark[a]        & lat.\tmark[a] & (km s$^{-1}$)               & $N$(\textsc{H~i})\tmark[c] \ML
PG1626+554 & 0.133     & 84.51                           & 42.19   & 16.8                              & 19.4  \NN
3C 351.0        & 0.372     & 90.08                           & 36.38   & 16.5                              & 18.6 \NN
MRK290         & 0.030     & 91.49                          & 47.95   & 15.4                               & 20.1 \NN
H1821+643   & 0.297      & 94.00                          & 27.42   & 16.1                              & 18.5 \NN
MRK876        & 0.129     & 98.27                           & 40.38    & 15.1                     & 19.4 \NN
MRK817        & 0.031      & 100.30                        & 53.48    & 13.8                     & 19.5 \NN
HS1914+7139 & ...\tmark[d] & 102.99               & 23.91        & 14.3               &  ... \LL
}

% TABLE: E140H Observation log
\ctable[
caption={Space Telescope Imaging Spectrograph E140H Echelle Observations of H1821+643},
label={tab:stis_log},
doinside=\footnotesize,
star
]{lllclll}{
\tnote[a]{The target was observed using two E140H grating tilts with central wavelength  $\lambda_{\rm cen}$.  The  $\lambda_{\rm cen}$ = 1343 \AA\ setting covers the 1242 -- 1440 \AA\ wavelength range, and the $\lambda_{\rm cen}$ = 1453 \AA\ tilt covers 1359 -- 1552 \AA .}
\tnote[b]{\footnotesize Identification codes for locating the observations in the Mikulski Archive for Space Telescopes (MAST) at https://archive.stsci.edu/ }.}
{ \FL
Visit & Obs. & Start & $\lambda_{\rm cen}$\tmark[a] & No. & Total Exp.       & MAST \NN
 \      & Date & UT (hh:mm) & (\AA)                          & Exposures & Time (ksec)    & ID\tmark[b] \ML
1 & 2018 May 10 & 13:57 & 1343 & 10 & 23.098 & ODMJ01010 - ODMJ010A0 \NN
2 & 2018 July 4   & 9:15 & 1343 & 10 & 23.098 & ODMJ02010 - ODMJ020A0 \NN
3 & 2019 July 19 & 15:16 & 1343 & 3 &  6.601   & ODMJ53060 - ODMJ53080 \NN
\  &    \                & 11:23 & 1453 & 5 & 11.391 & ODMJ53010 - ODMJ53050 \LL
}

This paper presents a set of ultraviolet spectroscopic observations of sightlines through Complex C and the OA, including the very high-resolution H1821+643 data, to investigate the physical conditions in these inner-CGM HVCs using absorption spectroscopy.  The sightlines studied are listed in Table~\ref{tab:targets} and were selected to probe a spatial stripe, roughly centered on H1821+643, extending from the OA up to higher latitudes through the full vertical extent of Complex C.  The locations of the pencil-beam sightlines are also marked with yellow dots in Figure~\ref{fig:outergalmap}. The data provide good constraints on the temperature, density, cloud size, dust content, and multiphase ionization of the HVC gas, and it is found that the gas cooling times and $t_{\rm cool}/t_{\rm ff}$ ratios are generally consistent with the idea that these clouds are caused by precipitation in a dynamically stirred region.  Other interpretations are certainly possible, but regardless, the overall set of measurements provides a coherent and unique suite of constraints that can be used to test theoretical work on gas physics in the inner CGM.

Broadly, this paper is organized into two parts. The primary branch of the paper presents the new very high-resolution (2.6 km s$^{-1}$) \textsl{HST} spectroscopy of H1821+643; these high-resolution data reveal complex component structure and uniquely constrain the gas physical conditions. The second branch of the paper considers several ancillary sightlines near H1821+643 (see Tab.~\ref{tab:targets}) that pass through both the OA and Complex C HVCs; these ancillary sightlines support the conclusions drawn from the high-resolution H1821+643 data. Section \ref{sec:dataredux} presents the UV spectroscopic observations and data reduction followed by the H1821+643 absorption-line measurements (Section \ref{sec:h1821measurements}) and analysis of the ionization and physical conditions implied by the H1821+643 data (Section \ref{sec:h1821ionization}).  The study then shifts to the ancillary sightlines in Section \ref{sec:ancmeas}, and discussion and summary of the findings appear in Sections \ref{sec:discussion} and \ref{sec:summary}.  The appendix presents a method for mitigating temporally variable hot pixels in Space Telescope Imaging Spectrograph data recorded with the far-UV Multianode Microchannel Array (MAMA) detector.

\section{Observations and Data Reduction}
\label{sec:dataredux}

The targets in Table~\ref{tab:targets} have been observed with a variety of UV spectrographs with good signal-to-noise (S/N), and the combined archival data from different instruments cover the line-rich region from $\approx$ 912 \AA\ to 1800 \AA.  This paper uses data from the following spectrographs, with (observed wavelength range, spectral resolving power $R = \lambda/\Delta \lambda$): \textsl{FUSE} (912 - 1185 \AA , $R$ = 20000), the \textsl{HST} Space Telescope Imaging Spectrograph (STIS) with the E140H echelle grating (1242 - 1444 \AA\ and 1352 - 1554 \AA, $R$ = 114000), STIS with the E140M echelle grating (1144 - 1710 \AA , $R$ = 45800), and the \textsl{HST} Cosmic Origins Spectrograph (COS, 1150 - 1800 \AA, $R$ = 12000 - 16000). Further information on the design and performance of these spectrographs can be found in: (1) \textsl{FUSE}: \citet{moos00,moos02,sahnow00,dixon07}, (2) STIS: \citet{woodgate98,kimble98,branton21}, and (3) COS: \citet{green12,hirschauer21}.

The observations of the objects in Table~\ref{tab:targets} were made for various purposes by various investigators over the course of many years, and aspects of these data have been published in many previous papers.  The program abstracts, instrument setups, observation dates, exposure times, etc., as well as tabulations of publications that use the data, are available from the Mikulski Archive for Space Telescopes (MAST).\footnote{See the MAST web page at https://archive.stsci.edu/} 

The STIS E140H observations of H1821+643 are new and have not been presented in previous papers, however, so some remarks on these new data are warranted.  As summarized in Table~\ref{tab:stis_log}, the E140H observations were obtained over the course of three visits in 2018 and 2019 using the E140H grating tilts with central wavelengths $\lambda_{\rm cen}$ = 1343 and 1453 \AA, which cover the 1242 - 1444 and 1352 - 1554 \AA\ ranges, respectively.  When observing extragalactic targets with the E140H mode, long exposures are required to attain good signal-to-noise ratios.  Fortunately, it was possible to observe H1821+643 in the \textsl{HST} continuous viewing zone (CVZ), which doubled the exposure time per orbit.  However, this efficiency came with a tradeoff: the E140H detector heats up more in the CVZ, which in turn leads to more hot pixels (see Appendix). The observations were primarily designed to probe extragalactic \OVI\ absorption systems (Tripp et al., in prep.) and consequently required much higher S/N in the 1242 - 1444 \AA\ range where weak redshifted \OVI\ absorbers are detected (the longer-wavelength tilt was obtained to record corresponding \HI\ lines, which are often stronger and thus require lower S/N).  This program design has consequences for the present paper because the E140H S/N ratios for Galactic lines such as the \SII\ triplet, \SiII\ $\lambda$1304.37, and the \SiIV\ doublet are much better than the E140H S/N ratios for \SiII\ $\lambda$1526.71 and the \CIV\ doublet.  Some important transitions, notably the \FeII\ line at 1608.45 \AA , the \OVI\ doublet, and weaker lines of \OI, \SiII, and \FeII, are not covered in the new STIS E140H data at all, so for these important lines it is necessary to rely on (lower-resolution) STIS E140M and \textsl{FUSE} spectra.  

Ideally, the E140H observations would have used the $0.2'' \times 0.09''$ aperture for the cleanest line-spread function (LSF).  Unfortunately, due to concern about potentially significant throughput losses stemming from an offset between the STIS best focus and the \textsl{HST} best focus \citep{proffit17}, STScI advised our team to use the $0.2'' \times 0.2''$ aperture, which does not suffer from appreciable throughput reduction due to this problem.  The larger aperture does not catastrophically degrade the data, so we elected to use the $0.2'' \times 0.2''$ slit.  The LSF with the $0.2'' \times 0.2''$ aperture does have broader wings, and we account for these wings in our absorption-line measurements (see the following section).
   
For the initial data reduction steps, the pipeline software from the STScI and \textsl{FUSE} archives provide excellent results, and this paper relies on the pipeline extractions of one-dimensional (1-d) spectra from individual exposures using CalSTIS (version 3.4.2),  CalFUSE (v. 3.2.2), and CalCOS (v. 3.1.7). The various reduction steps applied to produce calibrated 1-d spectra are described in the STIS Data Handbook \citep{sohn19}, the COS Data Handbook \citep{rafelski18}, and the CalFUSE paper \citep{dixon07}.  The 1-d spectra from individual exposures were aligned and coadded using the procedures of \citet{tripp01,tripp08}, \citet{meiring11}, and \citet{tumlinson13} for the STIS E140M, \textsl{FUSE}, and COS data.  Finally, the spectra were binned to Nyquist sampling. 

When these procedures were applied to the new STIS E140H spectra, it became apparent that the new data are adversely affected by a large number of hot pixels that significantly impede analysis.  Fortunately, the hot pixels are temporally variable, so it was possible to mitigate this problem by identifying and masking out hot pixels in individual exposures before coaddition.  Examples of the hot-pixel behaviour, and the procedure used for hot-pixel mitigation, are presented in the Appendix.

\begin{figure*}
\includegraphics[width=15.8cm]{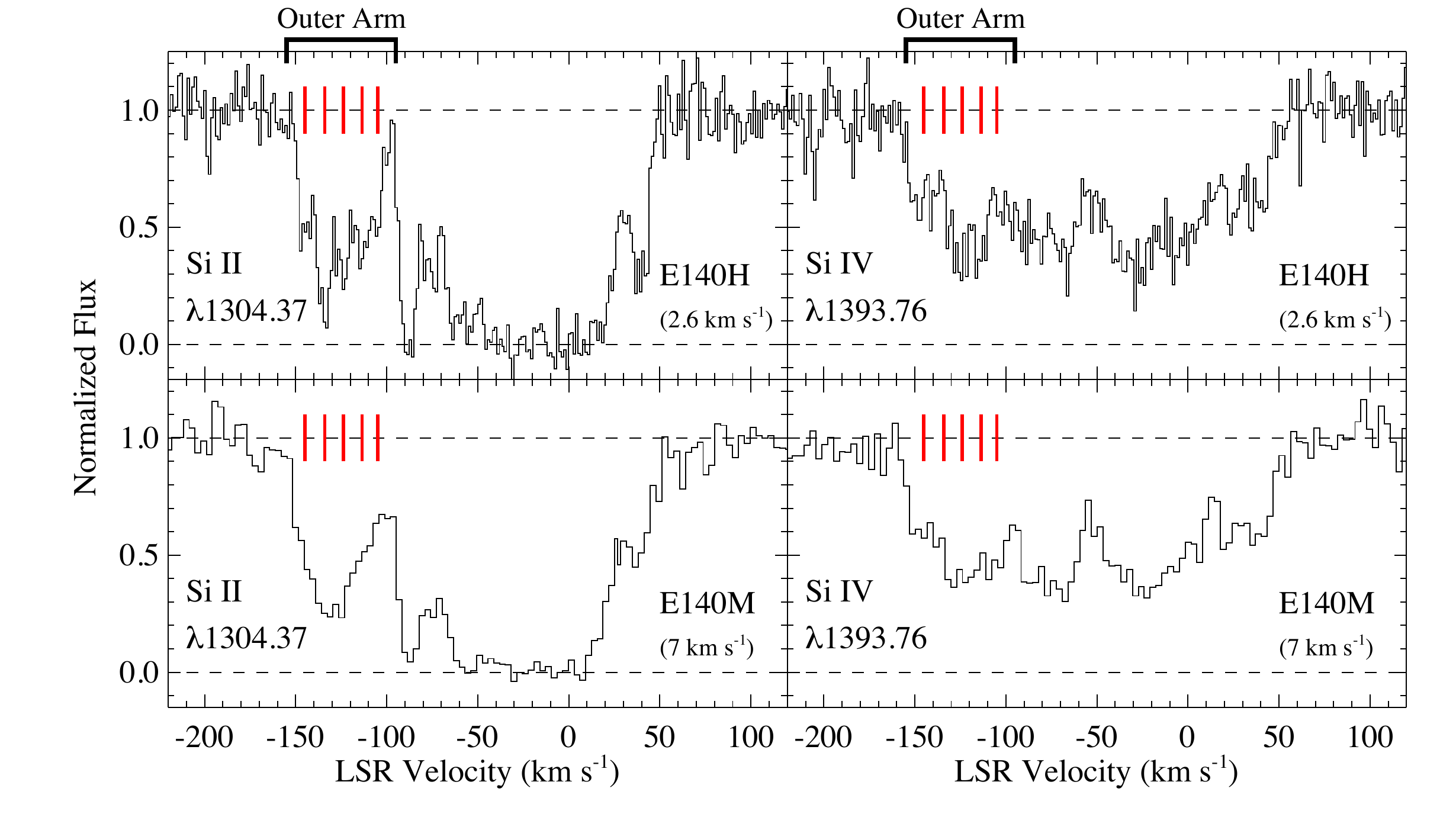}
\caption{Comparison of the new STIS E140H echelle spectra of H1821+643 (upper panels) to the previous lower resolution STIS E140M spectra (lower panels) obtained by \citet{tripp00b,tripp01}. The left and right columns compare the absorption profiles of the \SiII\ 1304.37 \AA\ and \SiIV\ 1393.76 \AA\ transitions, respectively, plotted vs. velocity in the Local Standard of Rest (LSR).  The E140H spectra provide a spectral resolution of 2.6 km s$^{-1}$, and the E140M spectra have a resolution of 7 km s$^{-1}$.  While the E140H spectra are noisier than the earlier E140M observations, the higher spectra resolution reveals the precise shapes of many narrow absorption components in these profiles; this is particularly evident in the ``Outer Arm'' feature in the \SiII\ 1304.37 \AA\ line (compare upper left and lower left panels).  The E140M data, despite the lower resolution, corroborate the component structure in the E140H spectra. The red tick marks indicate the five components identified in the \SiII\ profile. \label{fig:e140h_vs_e140m}}
\end{figure*}

\section{H1821+643 Absorption-Line Measurements}
\label{sec:h1821measurements}

Figure~\ref{fig:e140h_vs_e140m} shows examples of Galactic-absorption profiles from the new STIS E140H spectrum of H1821+643 (upper panels) compared to the previous STIS E140M spectra (lower panels).  The left-hand panels in Figure~\ref{fig:e140h_vs_e140m} compare the Milky Way \SiII\ 1304.37 \AA\ lines, and the right-hand panels compare the Galactic \SiIV\ 1393.76 \AA\ lines.  The LSR velocity range of the OA HVC is indicated above the upper panels.  

The E140H data are noisier than the E140M spectra, but nevertheless the benefits of the improved resolution are striking.  The \SiII\ 1304.37 \AA\ line detected from the OA, which at first glance appears to be a single absorption component at the 7 \kms\ resolution of the E140M spectrum (lower-left panel of Fig.~\ref{fig:e140h_vs_e140m}), resolves into \textbf{five} significantly detected components in the 2.6 \kms\ E140H spectrum (upper-left panel of Fig.~\ref{fig:e140h_vs_e140m}). The five \SiII\ components are marked with red tickmarks in each panel.  However, careful inspection of the E140M data reveals edges and inflections in the profiles that hint at multiple blended components, and indeed \citet{tripp12} fitted the E140M data with a three-component model.   Nevertheless, it is clear that the properties of most of the individual \SiII\ components (velocity centroids, line widths, and individual column densities) cannot be robustly measured in the E140M profile, which at 7 \kms\ resolution is generally considered to have very good resolution. Measurements from the E140H and E140M data are compared below.

Turning to the \SiIV\ profile comparison on the right side of Figure~\ref{fig:e140h_vs_e140m}, we see that in contrast to the \SiII\ data, most of the \SiIV\ components evident in the higher-resolution spectra are (marginally) discernible in the earlier E140M observations.  However, even though the E140M spectra show the complex structure of the profiles, some components are still seriously blended at E140M resolution, and the component properties are better constrained at the higher resolution.  The overall good agreement of the detailed component structure in the E140H vs. E140M profiles underscores a crucial benefit of this multi-instrument dataset: the good-quality E140M data \textsl{corroborate} the complicated component structure indicated by the higher-resolution spectra.   The STIS E140M observations use the same MAMA detector as the STIS E140H observations, but the echelle spectrographs are independent, and the spectra at particular wavelengths are recorded on different regions of the detector and on very different dates.  Thus the consistent evidence of the same component structure verifies that the components are real.  This is valuable in light of the high number of hot pixels in the new E140H data (see the Appendix).

\subsection{Apparent Column Density Comparisons}
\label{sec:navmethod}

To study the properties of individual absorption components, this paper employs two measurement methods. The analyses start with apparent column-density profiles \citep{savage91,jenkins96} of the absorption data.  In this technique, after fitting the continuum in the vicinity of an absorption lines with a low-order polynomial \citep[this paper uses the continuum-fitting method of][]{sembach92}, the ``apparent'' optical depth $\tau_{\rm a}({\emph v})$ in a pixel at velocity \textit{v} is calculated from the observed intensity $I({\emph v})$ and the fitted continuum intensity $I_{\rm c}({\emph v})$ in that pixel: $\tau_{\rm a}({\emph v}) = {\rm ln}[I_{\rm c}({\emph v})/I({\emph v})]$. In turn, the apparent column density per unit velocity, $N_{\rm a}({\emph v})$, is calculated from $\tau_{\rm a}({\emph v})$,
\begin{equation}
N_{\rm a}({\emph v}) = \frac{m_{\rm e}c}{\pi e^{2}} \frac{\tau_{\rm a}({\emph v})}{f\lambda}
\end{equation}
where $f$ is the oscillator strength and $\lambda$ is the wavelength of the transition, and the other symbols have their usual meanings.  If $N_{\rm a}({\emph v})$ is not affected by saturation, then it can be integrated to obtain the total column density, $N_{\rm tot} = \int N_{\rm a}({\emph v}) d{\emph v}$. These quantities are referred to as ``apparent'' because the true optical depth is smeared out by the LSF of the spectrograph. If the lines are well resolved, then $\tau_{\rm a}({\emph v})$ is a good approximation of the true optical depth profile, but even if the profiles are not optimally resolved, $N_{\rm a}({\emph v})$ profiles have a number of virtues. Comparison of the $N_{\rm a}({\emph v})$ profiles of two or more transitions that differ in $f\lambda$ by $\gtrsim$0.3 dex will reveal velocity ranges where the data are affected by unresolved saturation, and by converting the exponential absorption profiles into linear $N_{\rm a}({\emph v})$ profiles, different species can be scaled and overlaid to compare their detailed component structures and kinematics. 

\subsection{Voigt-Profile Fitting}

With guidance from the \nav\ profiles regarding the component structure, the absorption features are then fitted with Voigt profiles using a modified version of the profile-fitting software originally developed by \citet{fitzpatrick97}.  Each component in a profile model has a column density $N$, a line width expressed as a $b-$value, and a velocity centroid \textit{v}.  The full set of (often overlapping) components are convolved with the appropriate LSF for the instrument before comparison to the data, and the code iteratively adjusts the component parameters to find the best fit \citep[see][for additional details]{spitzer95,fitzpatrick97}. Different profiles from different instruments can be fitted simultaneously with the appropriate resolution and LSF for each instrument.  When multiple lines are fitted, the data are weighted by their inverse variances, with the exception that higher resolution data are required to have equal or better weight than the lower resolution data (so that the component structure revealed by better resolution impacts the fit even if the lower-resolution data have higher S/N).

\subsection{Line Deblending}
\label{sec:deblending}

When using QSOs and AGNs as background continuum sources, blending with interloping lines from various redshifts can occur, especially if the continuum source has an appreciable redshift.  In addition, as the observed wavelength decreases, the number of foreground lines from the nearby ISM goes up.  The \textsl{FUSE} bandpass in particular covers a large number of Lyman- and Werner-band H$_{2}$ absorption lines\footnote{See \citet{jenkins97} for wavelengths and $f\lambda$ data for H$_{2}$ Lyman and Werner lines.} from the ISM, and these H$_{2}$ lines can contaminate higher-velocity absorption features of interest. Fortunately, these blends can often be rectfied by modeling other transitions from the same interloper and then using the models to estimate the strength of, and ultimately remove, the offending blend \citep[see, e.g., Fig.~S1 and accompanying text in][]{tripp11}.  However, sometimes insufficient information (e.g., no other transitions from the same species are covered) prevents removal of blends. In the H1821+643 dataset, there are three important blends that can be modeled and removed:

First, the \SII\ 1259.52 \AA\ line at OA velocities is blended with \OVI\ $\lambda$1037.62 from $z_{\rm abs}$ = 0.2133 \citep{tripp08,savage14}. The other two lines of this \SII\ triplet are too weak to detect in the OA, so the \SII\ 1259.52 \AA\ line provides unique and valuable information (see below).  To deblend this important line, the strength of the blended \OVI\ 1037.62 \AA\ line was estimated based on the corresponding 1031.93 \AA\ transition of the \OVI\ doublet, which is well constrained \citep[see Fig.~10 on page 49 of][]{savage14}.  After fitting the 1031.93 \AA\  line, we use the fit to estimate the strength of  \OVI\ 1037.62 \AA\ and divide it out of the the \SII\ 1259.52 \AA\ line.

Second, the \OI\ 1039.23 \AA\ absorption is close to the ISM H$_{2}$ L5R2 and L5P2 lines at 1038.69 and 1040.37 \AA, respectively; we are only concerned with the H$_{2}$ L5R2 line in this paper since it is on the negative-velocity side of  \OI\ $\lambda$1039.23 where the OA absorption is recorded.  For this blend, seven H$_{2}$ transitions from the $J$ = 2 level, with a range of $f\lambda$ values, were fitted: L7R2, L7P2, L4R2, L4P2, L3R2, L3P2, L2R2, and the fit was then used to predict and remove the H$_{2}$ L5R2 line from the data.

Finally, the \OVI\ 1031.93 \AA\ line is blended with the H$_{2}$ L6P3 transition at the OA/HVR velocity \citep[see Fig.1 in][]{wakker03}.  To deal with this blend, the H$_{2}$ L7P3, L4P3, and L3R3 transitions from $J$ = 3 were fitted, and then the fit was used to predict and rectify the H$_{2}$ L6P3 interloper.  The same method was used to model and remove the H$_{2}$ L6P3 line from the ancillary sightlines.

\subsection{H1821+643 Low-Ionization Absorption Lines}
\label{sec:h1821_lowion_meas}

\begin{figure}
\includegraphics[width=9.1cm]{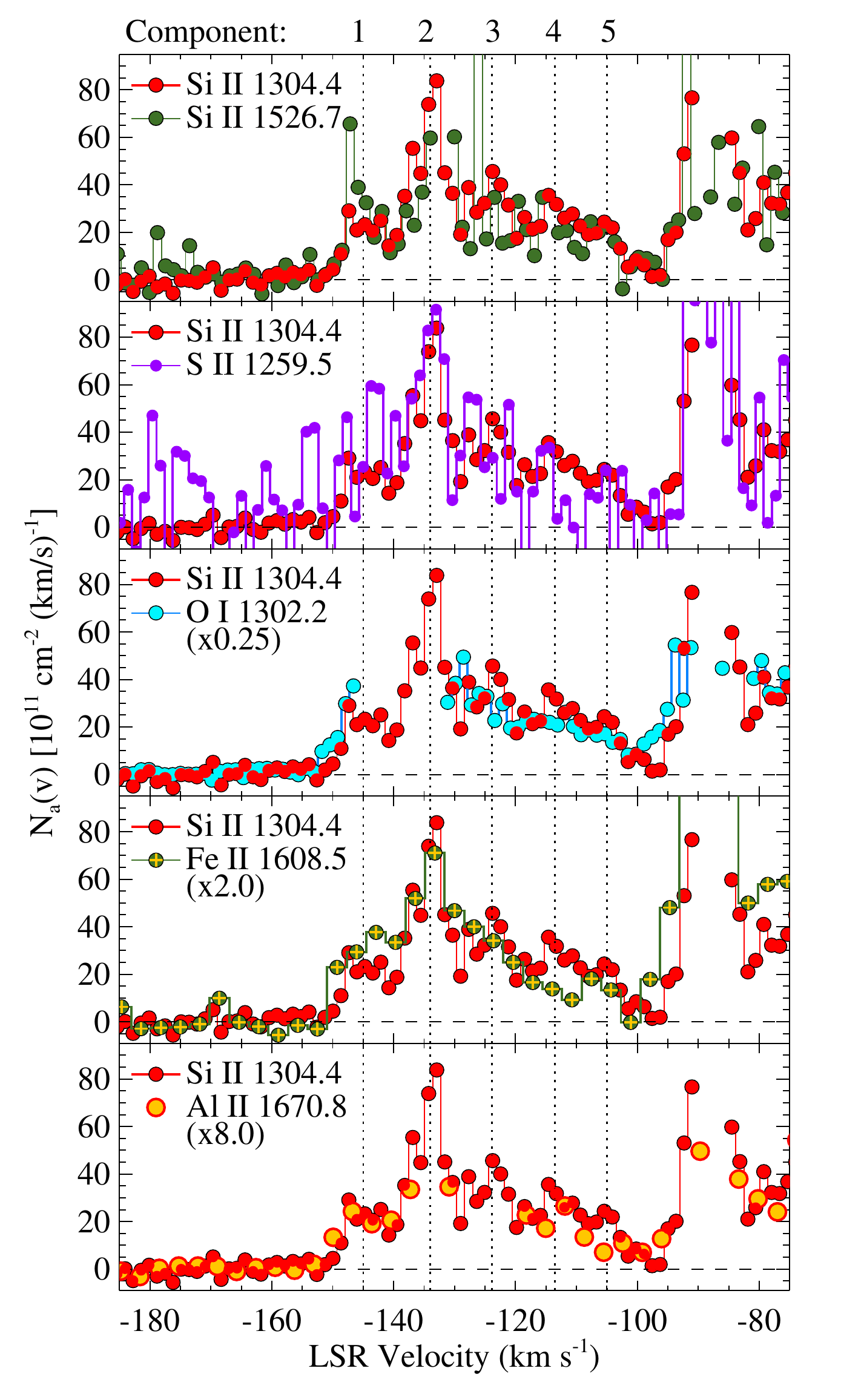}
\caption{Apparent column-density profiles of the low-ionization lines detected in the Outer Arm toward H1821+643.  The species plotted are indicated by the legend at upper left in each panel. In some panels, the \nav\ profiles are scaled (see legend) so that the detailed shapes of the profiles can be compared. When noise drives a pixel to have a negative value in the core of a strong component, \nav\ cannot be evaluated, which causes some of the stronger-line profiles to have gaps. Components $1 - 5$ are marked with vertical dotted lines. \label{fig:low_nav}}
\end{figure}

Figure~\ref{fig:low_nav} compares the \nav\ profiles of \SiII, \SII, \OI, \FeII, and \AlII\ detected in the Outer Arm toward H1821+643.  For convenience, throughout this paper the five components revealed in the \SiII\ E140H spectra will be referred to, from most-negative to least-negative velocity, as Components $1 - 5$, as labeled at the top of Figure~\ref{fig:low_nav}.  The \SiII, \SII, and \OI\ profiles in Figure~\ref{fig:low_nav} are constructed from the E140H data, while the \FeII\ and \AlII\ lines are only covered in the the E140M spectrum.  The good agreement of the \SiII\ 1304.37 and 1526.71 \AA\ profiles \citep[see also Fig.7 in][]{tripp12} indicates that the \SiII\ data are mostly unaffected by unresolved saturation; some very weak saturation is possible in the core of component 2, but this should be adequately modeled in profile fits to the high-resolution spectra.\footnote{In general, the weaker \SiII\ 1020.70 \AA\ transition in the \textsl{FUSE} spectrum can be used to further probe for weak saturation, but in the H1821+643 spectrum the \SiII\ 1020.70 \AA\ line cannot be used for this due to an uncorrectable blend with an \textsc{O~iii} 832.93 \AA\ line from $z_{\rm abs}$ = 0.2250.}   As expected based their ionization potentials, the detailed shapes of the \SiII, \SII, \OI, \FeII, and \AlII\ profiles are very similar over the full velocity range of the OA.  This is important because it indicates that the lower resolution and/or noisier data can be modeled using the high-quality \SiII\ profile as a template (see below).  The \FeII\ and \AlII\ profiles are somewhat smeared out by the lower resolution provided by the E140M grating, but it is nevertheless clear from Figure~\ref{fig:low_nav} that the \FeII\ \nav\ profile rises and falls in the same way as the \SiII\ 1304.37 \AA\ profile, and moreover components 1 and 2 are evident in the \FeII\ data.  The usefulness of the \AlII\ is, unfortunately, severely limited by the combination of possibly strong saturation and the lower E140M resolution in the only available \AlII\ line at 1670.79 \AA.\footnote{\citet{tripp12} reported \AlII\ measurements.  However, due to improvements in the CALSTIS data reduction procedures, a newer reduction of the E140M data reveals that the \AlII\ line, which is recorded near the edge of an echelle order, is more strongly saturated than previously indicated, and consequently the \AlII\  profile mostly provides lower limits.} 

\begin{figure*}
\includegraphics[width=16cm]{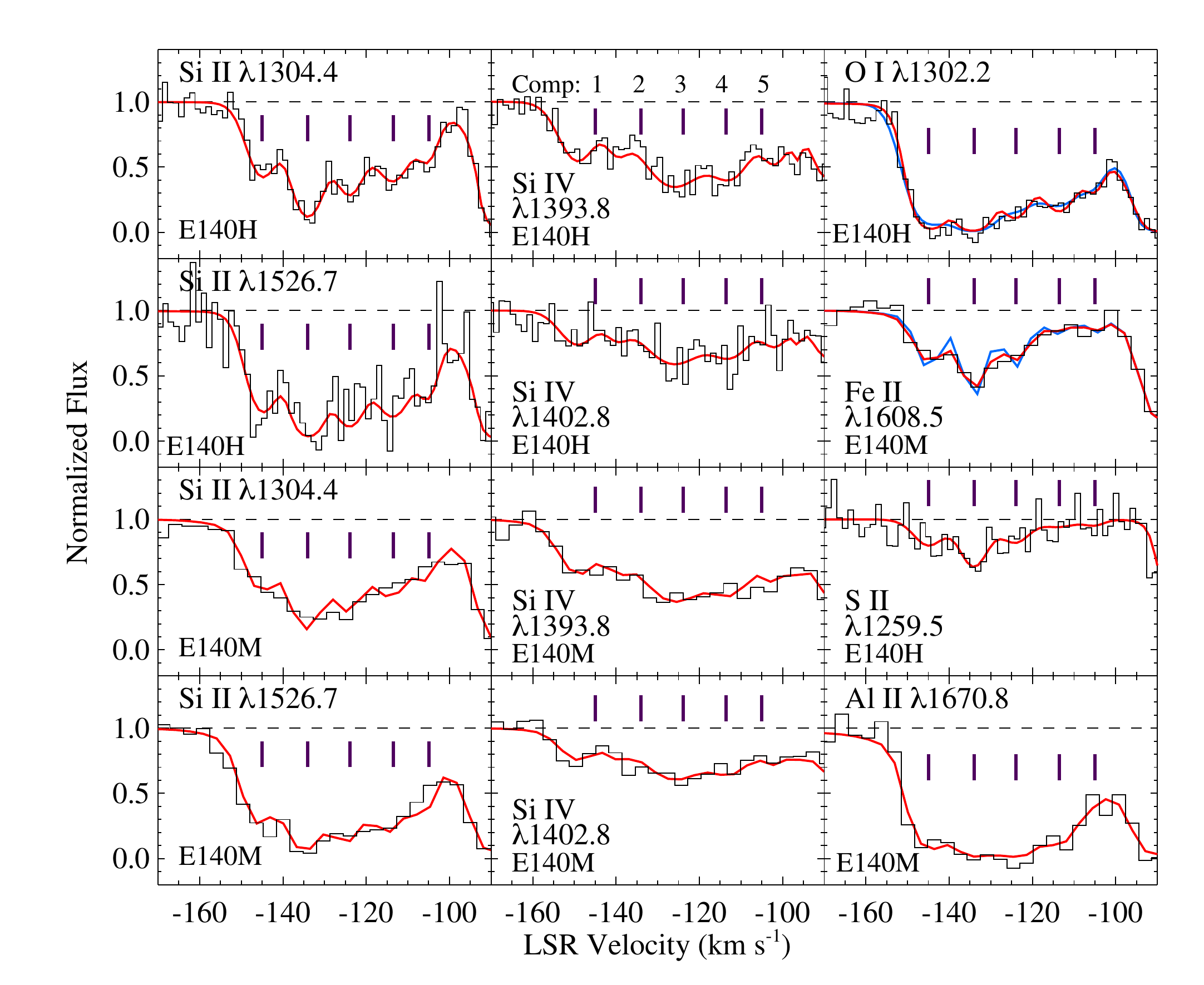}
\caption{Voigt-profile fits (colored lines) overplotted on the continuum-normalized absorption profiles (black histograms) of species detected in the Outer Arm in the spectrum of H1821+643. Each panel is labeled with the species shown and the STIS spectroscopic mode (E140H or E140M) used to record the data.  For the \SiII\ and \SiIV\ lines, the E140H and E140M data were fitted simultaneously. As explained in the text, in the \OI\ and \FeII\ panels, the blue line shows the fit obtained assuming the lines are predominantly thermally broadened, and the red line plots the fit assuming that non-thermal broadening dominates. The tick marks indicate the velocity centroids of the the five components identified in the low-ionization detections in the Outer Arm. These components are referred to in the main text by the numbers indicated in the \SiIV\ 1393.8 \AA\ E140H panel. \label{fig:vpstack}}
\end{figure*}

Motivated by the \nav\ comparison in Figure~\ref{fig:low_nav}, a two-pass procedure was used to fit Voigt profiles to the low-ionization lines in the H1821+643 data:

In the first pass, the \SiII\ 1304.37 and 1526.71 \AA\ profiles from both the E140H and the E140M spectra were fitted simultaneously with \emph{v}, $b$, and $N$ all allowed to freely vary for all components. A five-component model was adopted based on the components evident in Figure~\ref{fig:e140h_vs_e140m}. The resulting fit is overlaid on the continuum-normalized E140H and E140M \SiII\ absorption profiles in the left-most column of Figure~\ref{fig:vpstack}, and the \emph{v}, $b$, and $N$ of each component from the best fit are listed in the upper half of Table~\ref{tab:vp_si2_si4}.  This procedure worked well for the well-constrained \SiII\ dataset. The E140H and E140M profiles consistently support the component model (Fig.~\ref{fig:vpstack}); small discrepancies in some regions are likely due to imperfect continuum placement.

\begin{figure}
\includegraphics[width=8.0cm]{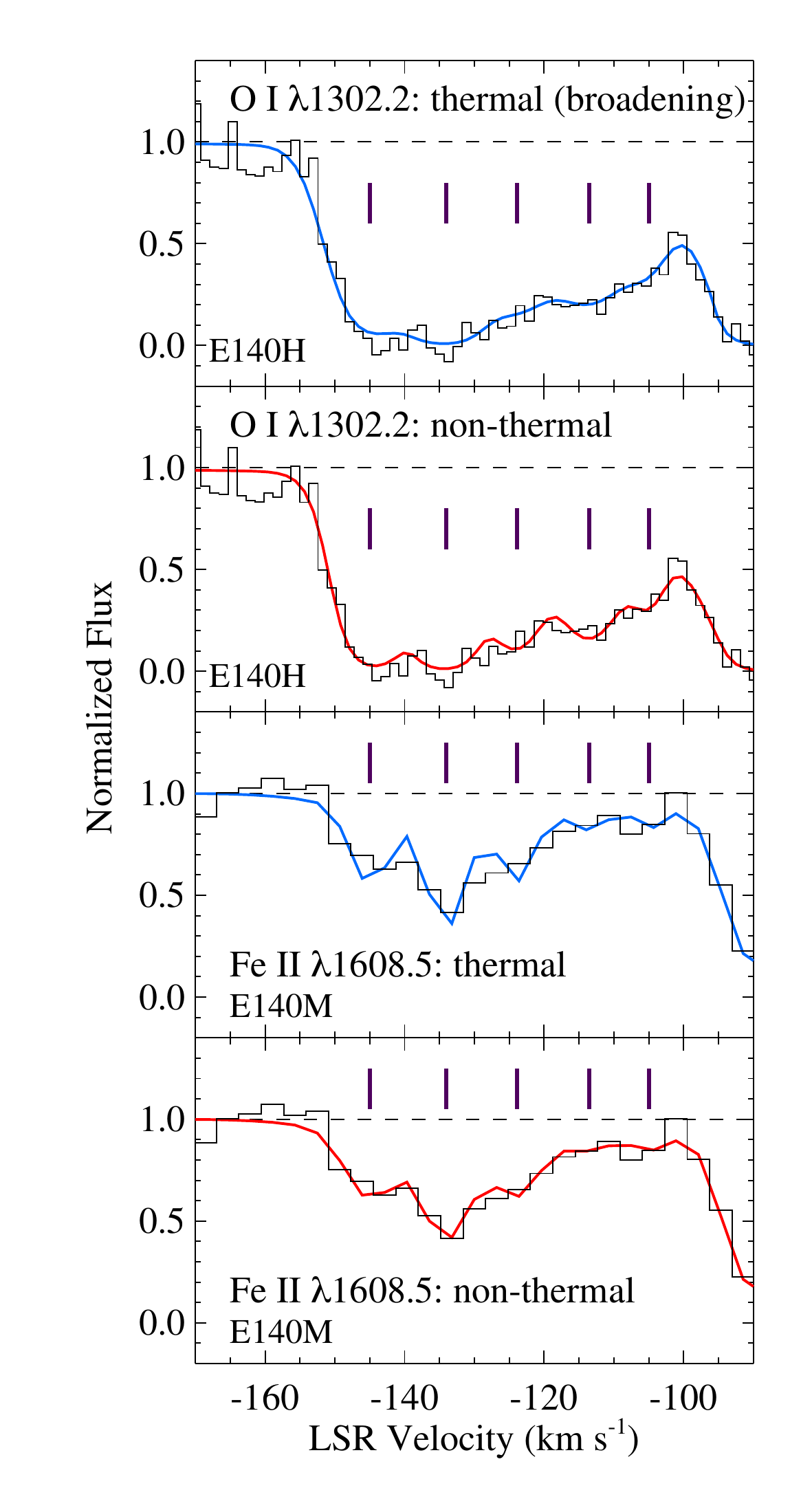}
\caption{Fits to the H1821+643 Outer Arm \OI\ 1302.169 \AA\ absorption profile (upper two panels) and the \FeII\ 1608.451 profile (lower two panels) assuming predominantly thermal or non-thermal broadening, as labeled in each panel.  The tick marks indicate the velocity centroids of the the five components identified and fitted in the Outer Arm.\label{fig:vp_therm_nontherm}}
\end{figure}

In the second pass, the \OI, \SII, and \FeII\ data were fitted using the \SiII\ results from the first pass as a template.  This was done for several reasons.  First, while the STIS E140H grating delivers the best spectra resolution, it also covers a smaller wavelength range in a single exposure, and HST Program 15321 did not observe all of the key transitions with the E140H mode.  Consequently, to measure \FeII\ column densities in the OA, the lower-resolution E140M \FeII\ $\lambda$1608.45 data were used combined with the \textsl{FUSE} recordings of the \FeII\ lines at 1121.98, 1125.45, 1142.37, 1143.23, and 1144.94 \AA\ (all lines were fitted simultaneously and provide a range of $f\lambda$ values). These \FeII\ data do not have sufficient resolution to successfully constrain a five-component fit if \emph{v}, $b$, and $N$ are all allowed to vary freely; there isn't enough structure in the lower-resolution absorption profiles for the profile-fitting algorithm to recognize the centroids and line widths of the five components.  However, the overall optical depths and shapes of this set of six \FeII\ transitions still constrain the component column densities if prior constrains on the component centroids and line widths can be applied.  Given the very similar ionization potentials and behavior of \SiII\ and \FeII\ in ionization models \citep[e.g.,][]{tripp03}, it is reasonable to assume that \SiII\ and \FeII\ will originate in the same gas cloud and therefore will have the same velocity centroids; this argument is supported by the very similar \SiII\ and \FeII\ \nav\ profiles shown in Figure~\ref{fig:low_nav}.   Line widths are more complicated because \SiII\ and \FeII\ have significantly different atomic masses, so two limiting cases can be assumed to bracket the possible column densities.  In one case, the lines are assumed to be predominantly thermally broadened, and the \FeII\ $b-$values are set equal to the \SiII\ $b-$values scaled according to the Si and Fe masses.\footnote{Since $b = \sqrt{2kT/m}$, in this model $b$(\FeII ) = $\sqrt{m_{\rm Si}/m_{\rm Fe}} \ b$(\SiII).}  This is referred to as the ``themally-broadened'' fit.  The second limiting case occurs if non-thermal broadening dominates to the extent that $b$(\FeII) = $b$(\SiII). This assumption was adopted for the ``non-thermally broadened'' fit. The \FeII\ column densities obtained for the five components assuming the two limiting cases are presented in Table~\ref{tab:lowion_columns}. The best-fit \FeII\ models are overplotted on the \FeII\ 1608.45 \AA\ absorption profile in Figures~\ref{fig:vpstack} and \ref{fig:vp_therm_nontherm}.

The same procedure was applied to fit the \SII\ 1259.52 \AA\ profile, for a different reason.  The \SII\ line is weak and the profile is noisy, and while the shape of the \SII\ \nav\ profile tracks the \SiII\ 1304.37 \AA\ profile (Fig.~\ref{fig:low_nav}), the \SII\ profile (see also Figure~\ref{fig:vpstack}) is too noisy to usefully constrain a fit with five components.  So, to measure the \SII\ column densities in a way that would allow a meaningful component-to-component comparison with \SiII\ columns, the \SiII\ fit was again used as a template with the \SII\ \emph{v} and $b$ fixed to the same values as found for \SiII .  Sulfur and silicon have similar atomic masses, so the thermally vs. non-thermally broadened cases yield nearly identical results.  The resulting \SII\ column densities are also in Table~\ref{tab:lowion_columns}, and the fit is shown in Figure~\ref{fig:vpstack}.  

Finally, this procedure was also used to fit the \OI\ 1302.17 and \OI\ 1039.23 \AA\ lines.  Components $1 - 3$ are very strong in the \OI\ 1302.17 \AA\ profile, and consequently the component structure is not well-constrained by the \OI\ 1302.17 \AA\ profile even though it is in the E140H spectrum.  However, with the addition of (\emph{v},$b$) constraints from the \SiII\ fit, the \OI\ component column densities can be measured. Inclusion of the weaker \OI\ 1039.23 \AA\ line helps to ensure that the \OI\ column densities are not underestimated due to saturation effects on the 1302.17 \AA\ profile. Other weak \OI\ lines in the \textsl{FUSE} were considered for use in the fit but were found to be too noisy and confused by blending. The atomic mass of oxygen is enough lower than the mass of silicon that the two limits cases discussed above were used to fit \OI .  Table~\ref{tab:lowion_columns} summarizes the \OI\ results from the two limiting cases, and the \OI\ fits are plotted in Figures~\ref{fig:vpstack} and \ref{fig:vp_therm_nontherm}.

Whether the data favor thermal- or non-thermal broadening is an interesting question in its own right.  To enable the reader to closely inspect the \OI\ and \FeII\ fits in the two limiting cases, Figure~\ref{fig:vp_therm_nontherm} compares the results from the two models.  In the \OI\ fits, the non-thermal broadening model does qualitatively appear to follow the undulations of the 1302.17 \AA\ slightly better, but the fits residuals are not necessarily smaller, and at any rate the superiority of the non-thermal fit is modest.  However, the non-thermal fit to the \FeII\ data appears to be clearly better: some of the individual \FeII\ components are too narrow in the predominantly thermally broadened model (compare the bottom two panels in Fig.~\ref{fig:vp_therm_nontherm}).  Considered together, the reduced $\chi_{\nu}^{2}$ for the \OI\ and \FeII\ fits is $\chi_{\nu}^{2}$ = 1.36 for the thermally broadened model and $\chi_{\nu}^{2}$ = 1.24 for the non-thermally broadened model, so the two models provide similarly good fits, with a modest preference for the non-thermally broadened fits.  The remainder of this paper will use the \OI\ and \FeII\ columns from the non-thermal broadening case, but this choice has only minor impacts on the main findings below because the column densities are mostly similar in either case (see Table~\ref{tab:lowion_columns}). 

\subsection{H1821+643 High-Ionization Absorption Lines}
\label{sec:h1821highion}

\begin{figure*}
\includegraphics[width=16cm]{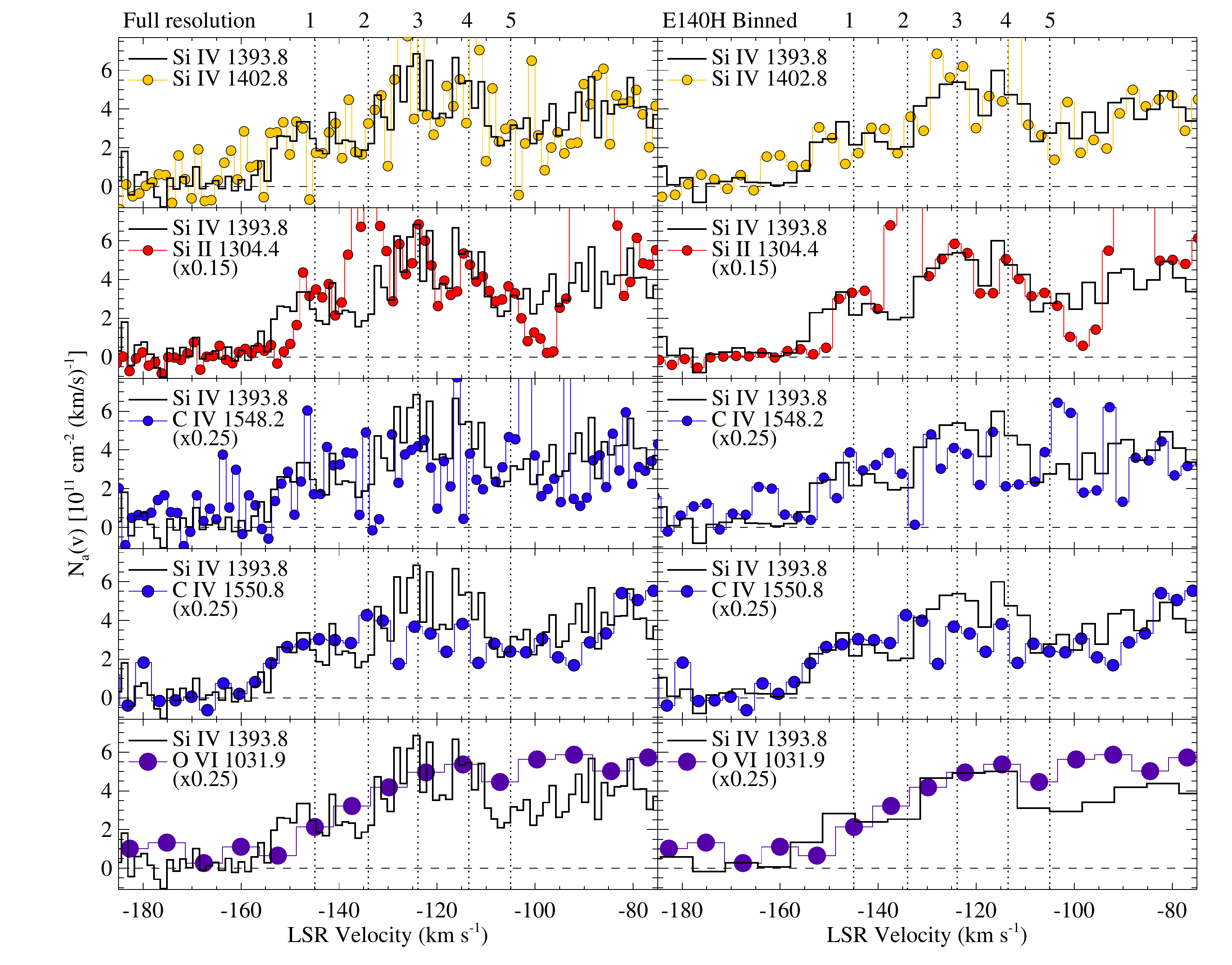}
\caption{Apparent column-density profiles of high-ionization lines in the spectrum of H1821+643; the species and transitions that are compared are labeled in each panel.  The \SiII, \SiIV, and \CIV\ 1548.2 \AA\ profiles are from the STIS E140H spectrum, the \CIV\ 1550.8 \AA\ data are from the STIS 140M spectrum, and the \OVI\ 1031.9 \AA\ profile is from the \textsl{FUSE} spectrum.  For reference, the vertical dotted lines mark the velocities of the five components identified in the low-ionization \SiII\ profile (see Figs.~\ref{fig:vpstack} and \ref{fig:low_nav}).  The panels in the left stack show the STIS E140H data at full (unbinned) spectral resolution.  In the right-hand panels, the STIS E140H data are binned two pixels into one, so that the resolution is more similar to that of the E140M data, in the upper four panels. In the lowest panel on the right, the E140H data are binned 5 pixels to one for resolution comparable to the \textsl{FUSE} data. Only the E140H data are binned in the right-hand panels. \label{fig:high_nav}} 
\end{figure*}

Highly significant absorption lines of the \SiIV, \CIV, and \OVI\ doublets are also detected at OA velocities in the H1821+643 dataset.\footnote{Absorption by the strong \textsc{C~iii} 977.02 \AA\ and Si~\textsc{iii} 1206.50 \AA\ lines are also detected at OA velocities, but they are so strongly saturated and blended that they provides only very loose limits and are not useful for most purposes of this paper.} Figure~\ref{fig:high_nav} shows the \nav\ profiles of these highly-ionized species, including a comparison to the  \SiII\ 1304.37 \AA\ \nav\ profile.  It must be borne in mind that the data in Figure~\ref{fig:high_nav} have various spectral resolutions ranging from 2.6 \kms\ (STIS E140H) to $\approx$15 \kms\ (FUSE \OVI ).  To compare the data at similar resolution (and also show the profiles with higher S/N), the E140H data are binned two pixels into 1 (leading to resolution comparable to E140M) in the upper four panels on the right side of Figure~\ref{fig:high_nav}, and five pixels into one in the lowest right panel, which shows the E140H data with a resolution similar to the \textsl{FUSE} data shown in the same panel.  The panels on the left show the \nav\ profiles at full resolution.

% TABLE: fits to Si II and Si IV
\ctable[
caption={Profile-Fitting Measurements of Si II and Si IV in the Spectrum of H1821+643},
label={tab:vp_si2_si4},
doinside=\footnotesize
]{llccl}{
\tnote[a]{STIS E140H and E140M data simultaneously fitted; see text.}}
{\FL
 \            & Fitted          & \emph{v}$_{\rm LSR}$ & $b$                    & \ \NN
Species & Lines (\AA)\tmark[a] & (km s$^{-1}$) & (km s$^{-1}$) & log [$N$ (cm$^{-2}$)] \ML
Si II & 1304.370,          & $-145.0\pm0.5$ & $4.0\pm0.7$ & 13.39$\pm$0.05 \NN
   \   &  1526.707          & $-134.0\pm0.4$ & $3.7\pm0.7$ & 13.82$\pm$0.06 \NN
   \   &    \                      & $-123.9\pm0.6$ & $3.8\pm1.3$ & 13.52$\pm$0.10 \NN
   \   &    \                      & $-113.5\pm0.8$ & $4.8\pm2.1$ & 13.48$\pm$0.13 \NN
   \   &    \                      & $-105.0\pm1.0$ & $2.5\pm1.3$ & 13.03$\pm$0.20 \ML
Si IV & 1393.755,        & $-149.3\pm0.7$ & $5.1\pm1.0$ & 12.51$\pm$0.06 \NN
 \       & 1402.770         & $-139.7\pm0.9$ & $3.1\pm2.0$ & 12.14$\pm$0.28 \NN
 \       &   \                     & $-126.1\pm1.4$ & $9.1\pm2.8$ & 12.97$\pm$0.12 \NN
  \       &   \                    & $-112.0\pm1.2$ & $5.7\pm2.6$ & 12.67$\pm$0.24 \NN
  \       &   \                    & $-101.7\pm0.9$ & $3.5\pm2.0$ & 12.37$\pm$0.22 \LL
}

% TABLE: Low-ion column densities
\ctable[
caption={Profile-Fitting Measurements of O~I, S~II, Fe~II, and Al~II in the Spectrum of H1821+643$^{\ a}$},
label={tab:lowion_columns},
doinside=\footnotesize
]{ccccc}{
\tnote[a]{The combination of signal-to-noise, blending, and (for E140M data) lower spectral resolution compromises the column densities obtained when v, b, and N are allowed to freely vary, so we have used the results from the Si~II fits (see Tab.~\ref{tab:vp_si2_si4}) as a template for the (v,b) values used in these fits. For the non-thermally broadened model, the b-values for all species are set equal to the Si~II b-values.  For the thermally broadened model, the b-values are assumed to be predominantly thermally broadened ($b = \sqrt{2kT/m}$), and the assumed b-values of the various species are scaled from the measured Si~II b-values accordingly. Since the atomic weights of Al, Si, and S are not that different, the two models yield very similar results for Al and S.} 
\tnote[b]{Line is strongly saturated; we conservatively report lower limits based on direct integration of the optical depth over the velocity range of the component.}
\tnote[c]{This component is strongly blended and weak, and the column density is not reliably constrained.}}
{\FL
 \emph{v}$_{\rm LSR}$ & log $N$(O~I)      & log $N$(S~II)     & log $N$(Fe~II)     & log $N$(Al~II) \NN
(km s$^{-1}$) & (E140H)             & (E140H)             & (E140M)              & (E140M) \ML
\multicolumn{5}{l}{Non-thermally broadened model:} \ML
 $-145.0$       & 14.40$\pm$0.06  & 13.54$\pm$0.13 & 13.27$\pm$0.06 & $>12.63$\tmark[b] \NN
 $-134.0$       & 14.52$\pm$0.07  & 13.83$\pm$0.08 & 13.53$\pm$0.05 & $>12.59$\tmark[b]  \NN
 $-123.9$       & 14.02$\pm$0.05 & 13.45$\pm$0.16 & 13.21$\pm$0.07 & $>12.61$\tmark[b]  \NN
 $-113.5$       & 14.00$\pm$0.04 & 13.01$\pm$0.43 & 12.76$\pm$0.17 & 12.43$\pm$0.13 \NN
 $-105.0$       & 13.45$\pm$0.09 & 12.71$\pm$0.67 & 12.52$\pm$0.25 &  Blnd.\tmark[c]  \ML
\multicolumn{5}{l}{Thermally broadened model:} \ML
$-145.0$        & 14.25$\pm$0.04 & 13.54$\pm$0.13 & 13.28$\pm$0.06 & $>12.63$\tmark[b]  \NN
$-134.0$        & 14.50$\pm$0.06 & 13.82$\pm$0.08 & 13.59$\pm$0.05 & $>12.59$\tmark[b] \NN
$-123.9$        & 13.95$\pm$0.05 & 13.43$\pm$0.16 & 13.22$\pm$0.07 &  $>12.61$\tmark[b] \NN
$-113.5$        & 14.03$\pm$0.04 & 13.04$\pm$0.39 & 12.75$\pm$0.15 &  12.43$\pm$0.13 \NN
$-105.0$        & 13.40$\pm$0.10 & 12.87$\pm$0.46 & 12.56$\pm$0.22 & Blnd.\tmark[c] \LL
}

Several important aspects of the high-ionization gas in the OA are revealed by Figure~\ref{fig:high_nav}: (1) The \SiIV\ 1393.76 and 1402.77 \AA\ \nav\ profiles agree well, so the \SiIV\ profiles are not saturated. Similar comparison of the \CIV\ $\lambda \lambda$ 1548.20,1550.78 doublet likewise shows that the \CIV\ lines are not significantly affected by unresolved saturation.  Unfortunately, the \OVI\ 1037.62 \AA\ line at $z = 0$ is seriously degraded by blending with many ISM lines \citep[see Fig.~4 in][]{bowen08} as well as uncertain continuum placement, so the \OVI\ 1037.62 \AA\ data are not used here. However, considering its overall strength and similar shape to the other high ions, the \OVI\ 1031.93 \AA\ absorption is not likely to be significantly saturated. (2) With the exception of Component 2, there is a striking similarity between the \SiIV\ and (scaled) \SiII\ \nav\ shapes: there is a strong kinematical correlation between the \SiIV\ and \SiII\ components, and the \SiIV/\SiII\ column-density ratio ($\approx 0.15$) is similar in most of the components.  The similarity is particularly notable in Components 3 and 4.  On the other hand, the \SiIV/\SiII\ column-density ratio is much lower in Component 2 (see also Figure~\ref{fig:e140h_vs_e140m}). \SiIV\ absorption is present near Component 2, but profile fitting (see below) indicates that the \SiIV\ velocity centroid is offset from the \SiII\ centroid in this component. Likewise, there are clearly shape differences in Component 1.  Nevertheless, the general correspondence of these ions suggests some type of physical relationship, as discussed further below. (3) Less surprisingly, the \SiIV\ and (scaled) \CIV\ profiles have similar shapes. However, while the \SiIV\ and \CIV\ \nav\ profiles match nicely in Component 1, the component correspondence in Comps. 3 and 4 is less clear in the \SiIV\ vs. \CIV\ comparison than in the \SiIV\ vs. \SiII\ comparison. This may be partially due to the fact that the E140H spectra covering \CIV\ are much noisier, but even with the higher S/N E140M \CIV\ data, the component correspondence is not as apparent.  This could be due to line broadening: if the high-ionization components are predominantly thermally broadened, then the \CIV\ components will be broader (and more smeared together) than the \SiIV\ components (more on this below). (4) Finally, although the \OVI\ profile has somewhat lower resolution, the bottom panels of Figure~\ref{fig:high_nav} clearly show that the \OVI\ profile generally follows the \SiIV\ profile over the OA velocity range ($\emph{v}_{\rm LSR} < -100$ \kms ). This is interesting since 113.9 eV is required to create \OVI\ (i.e., to ionize \textsc{O~v} to \textsc{O~vi}), while it only takes 45.1 eV to destroy \SiIV\ (by ionizing \SiIV\ $\rightarrow$ Si~\textsc{v}).  This suggests that the \SiIV\ and \OVI\ absorption lines originate in distinct phases despite the kinematical correspondence, a situation that has been shown to occur in QSO absorbers with detailed constraints on many adjacent ionization stages \citep[e.g.,][]{charlton03,ding05,tripp11,haislmaier21}. For convenient reference in the ionization modeling and discussion below, Table~\ref{tab:IPS} summarizes the ionization energies required to create and destroy the metal species considered in this paper.  

\subsection{Si IV Measurements}

\ctable[
caption={Species Ionization Potentials $^{a}$},
label={tab:IPS},
doinside=\footnotesize
]{lll}{
\tnote[a]{Ionization potentials of the next-lower ionization stage ($X^{i-1}$) and of the ion listed in column 1 ($X^{i}$) \citep{verner94}.}}
{\FL
Species \ \ \ & $X^{i-1}$ (eV) \ \ \ & $X^{i}$ (eV) \ \ \ \ML
\OI         & ...                     & 13.62 \NN
\SiII       & 8.15                  & 16.35 \NN
\SII        & 10.36               & 23.33 \NN
\FeII      & 7.87                 & 16.19 \NN
\SiIV     & 33.49               & 45.14 \NN
\CIV      & 47.89               & 64.49 \NN
\OVI      & 113.90              & 138.12 \LL
}

With regard to Voigt-profile fitting, the \SiIV\ data are similar to the \SiII\ data, i.e., the \SiIV\ 1393.76 and 1402.77 \AA\ lines are independently recorded in both the E140H and E140M spectra.  Consequently, the E140H and E140M \SiIV\ spectra were treated the same way as the \SiII\ data; all four \SiIV\ absorption profiles were simultaneously fitted with the \emph{v}, $b$, and $N$ parameters allowed to freely vary with a five-component fit.  The results of this fit are plotted in Figure~\ref{fig:vpstack} and summarized in the lower half of Table~\ref{tab:vp_si2_si4}.

\begin{figure}
\includegraphics[width=7.8cm]{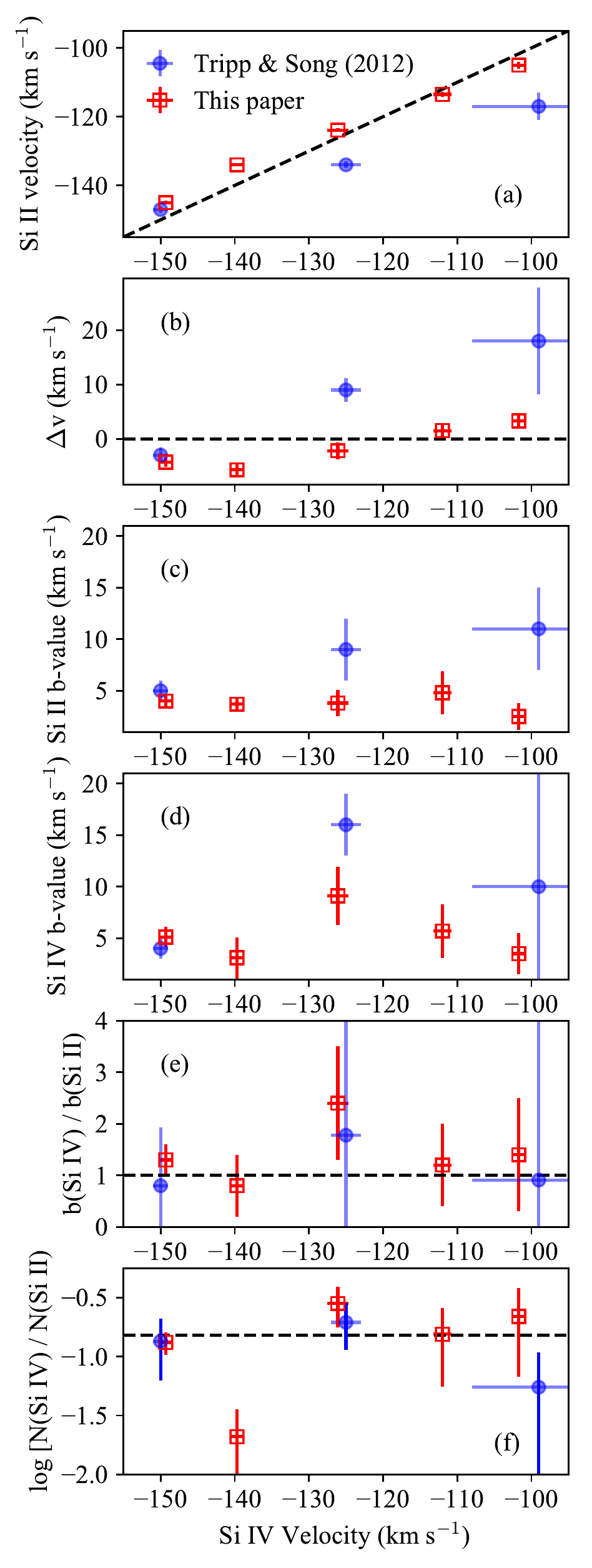}
\caption{Comparison of the H1821+643 Outer Arm profile-fitting measurements obtained by \citet{tripp12} using the STIS E140M data (blue dots) to the new measurements using the STIS 140H spectra (open red squares) reported in this paper.  The panels show (a) the \SiII\ vs. \SiIV\ velocity centroids, (b) the velocity centroid differences $\Delta$v = v(\SiIV) - v(\SiII), (c) \SiII\ $b-$values, (d) \SiIV\ $b-$values, (e) the \SiII / \SiIV\  $b-$value ratios, and (f) log of the \SiII / \SiIV\  column-density ratios. Here each \SiII\ component is matched to the \SiIV\ component that is closest in velocity, and all measurements are shown with $1\sigma$ error bars.\label{fig:si2si4fitparams}}
\end{figure}

Since the \SiIV\ and \SiII\ profiles were freely and independently fitted, we can compare all of the \SiIV\ and \SiII\ parameters from the fits.  Figure~\ref{fig:si2si4fitparams} graphically compares the fitting results in a variety of ways. For this figure, each \SiII\ component has been matched to the \SiIV\ component that is closest in velocity, and the results in the various panels are all plotted vs. the velocity centroid of the matched \SiIV\ component.  To investigate how spectral resolution affects individual-component measurements, Figure~\ref{fig:si2si4fitparams} also compares the results obtained here using the combined E140H+E140M data (red squares) to the measurements reported previously by \citet{tripp12} using only the E140M data (blue circles).

Beginning with the \SiII\ and \SiIV\ velocity centroid comparison (panels a and b in Figure~\ref{fig:si2si4fitparams}), several features are notable.  Firstly, analysis of only the lower-resolution E140M spectra created, effectively, an illusion of very different kinematics in the \SiII\ and \SiIV\ velocity centroids: the \citet{tripp12} measurements (blue points) indicated velocity differences of 9 and 18 \kms\ in two (out of their three) components.  However, the new higher-resolution E140H centroids (red points) reveal that all of the five components are much better aligned than this.  The velocity differences, $\mid \Delta \emph{v}\mid = \mid$\emph{v}(\SiIV) - \emph{v}(\SiII)$\mid$, are $2 - 6$ \kms\ in the new measurements.  The E140M results are off because at E140M resolution, components 2 and 3 cannot be securely recognized and distinguished, and likewise the existence of component 5 was not clear, so \citet{tripp12} elected to use only three components in their profile fits.  Even if they had decided to use five components, there would not be sufficient information in the E140M data to reliably constrain all five components. Interestingly, the \citet{tripp12} measurements and the new results agree better in component 1, largely because this component is at one edge of the multicomponent absorption profile and is less badly affected by blending.  Secondly, while the new \emph{v}(\SiII) and \emph{v}(\SiIV) measurements show velocity correspondence to 6 \kms\ or better, the formal error bars from the profile-fitting software nevertheless indicate that the \SiII\ and \SiIV\ velocity centroids in components 1 and 2 are different at the $5.0\sigma$ and 2.7$\sigma$ levels, respectively, as one might expect from the profile comparison in Figure~\ref{fig:high_nav}.  For components $3 - 5$, the measurements do not indicate any statistically significant differences between \emph{v}(\SiIV) and \emph{v}(\SiII).

Turning to the $b-$value comparison (panels c, d, and e in Figure~\ref{fig:si2si4fitparams}), we see that not surprisingly, the E140M-only $b-$values are too large since features that we now know to be multiple components were fitted as single components. Focusing on the new measurements (red squares), we see that within the 1$\sigma$ error bars, the line widths could be the same, i.e., $b$(\SiII) $\approx$ $b$(\SiIV).  However, since some of the velocity centroids are different,  the \SiIV\ likely comes from a different gas phase, and the error bars allow this phase to be appreciably hotter than the \SiII -bearing phase.  We will return to this question below.  The individual-component column densities in the older fits are confused by component blending, but the total column densities in the OA, summed over all components, are in excellent agreement in the previous and new measurements.

But is the OA absorption well resolved even at E140H resolution?  Would higher-resolution observations reveal even more component structure?  Given the \HI\ column density of the OA in the H1821+643 direction, the gas is not fully shielded against photoionization by the ultraviolet background, and photoionization from this background light will heat the gas to $\gtrsim 10^{4}$ K \citep[see, e.g., top axis of Fig.12 in][]{tripp03}.  This floor on the gas temperature, in turn, requires $b \gtrsim$ 2.4 \kms\ for silicon ions.  Therefore it is likely that the STIS E140H spectra have adequately resolved the component structure in the H1821+643 silicon data.

The systematic issues revealed by the comparisons in Figure~\ref{fig:si2si4fitparams} are a general concern for QSO absorption-line observations, which are usually obtained with a resolution comparable to the STIS E140M resolution.  If it is expected that an absorption system will contain low-ionization gas analogous to the OA, it could be valuable to obtain higher resolution spectra of at least some suitable species to use as a template for the analysis of that system. %However, it is useful to note that the \textsl{total} column densities from the E140M vs. E140H data measurements, summed over all components in the OA, are generally in excellent agreement. 

\subsection{C IV and O VI Measurements}

Like the \OI, \SII, and \FeII\ measurements presented in Section \ref{sec:h1821_lowion_meas}, the available \CIV\ data are not able to support a five-component fit with all parameters allowed to freely vary.  However, analogously to the procedure employed in Section \ref{sec:h1821_lowion_meas}, we can use the \SiIV\ results as a template for the \CIV\ component structure, again assuming either predominantly thermally broadened lines with the $b-$values for \CIV\ scaled from the \SiIV\ $b-$values based on their respective masses, or assuming non-thermal broadening with $b$(\CIV) = $b$(\SiIV).  The results obtained with the \SiIV\ template under these two assumptions are listed in Table~\ref{tab:civ_comp}.

\ctable[
caption={H1821+643 \CIV\ Component Column Densities Based on the \SiIV\ Template $^{a}$},
label={tab:civ_comp},
doinside=\footnotesize
]{cll}{
\tnote[a]{\CIV\ column densites obtained by Voigt-profile fitting using the (\emph{v},$b$) values from the \SiIV\ fit as a template.  Column 2 adjusts the $b-$values for \CIV\ assuming the lines are predominantly thermally broadened, and column 3 lists the results obtained assuming the lines are predominantly non-thermally broadened with $b$(\CIV) = $b$(\SiIV).}}
{\FL
\emph{v}$_{\rm LSR}$  & log [$N$(\CIV) cm$^{-2}$] & log [$N$(\CIV) cm$^{-2}$] \NN
(\kms)    & (Thermal brd.)                   & (Non-thermal brd.) \ML
$-149.3$  & 13.10$\pm$0.08 & 13.04$\pm$0.08 \NN
$-139.7$  & 12.74$\pm$0.10 & 12.96$\pm$0.10 \NN
$-126.1$  & 13.54$\pm$0.05 & 13.43$\pm$0.05 \NN
$-112.0$  & 12.87$\pm$0.18 & 13.06$\pm$0.08 \NN
$-101.7$  & 13.13$\pm$0.09 & 13.02$\pm$0.12 \LL
}

Unlike the low ions, \SiIV\ and \CIV\ do not have overlapping ionization potentials (see Table~\ref{tab:IPS}), so there is less confidence that the component structure from \SiIV\ applies to \CIV.  However, physical conditions can often lead to \SiIV\ and \CIV\ coexisting in the same phase. This situation is not assured, but the similarity of the \SiIV\ and \CIV\ profiles (see Figure~\ref{fig:high_nav}) indicates that it is worthwhile to fit the \CIV\ using the \SiIV\ results as a template. On the other hand, in the case of \OVI\ the differences in ionization potentials between \SiIV\ and \OVI\ are much larger (Tab.~\ref{tab:IPS}), so in this case we do not use the \SiIV\ ($b,\emph{v}$) as a template for fitting the \OVI\ data.  Nevertheless, the \nav\ profiles are similar, so for \OVI, a simpler approach is adopted: we integrate the \nav\ profiles of \SiIV, \CIV, and \OVI\ across the velocity range of the OA, and then we compare the ratios from the \nav\ integration to ratios from various theoretical models (see below).  With this approach, we find log [$N$(\SiIV)/$N$(\CIV)] = $-0.52\pm0.02$ and log [$N$(\CIV)/$N$(\OVI)] = $-0.17\pm0.02$. Note that these ratio uncertainties reflect the random noise in the \nav\ profiles, and the intrinsic component-to-component ratio variation could be somewhat larger than these error bars.

\section{H1821+643 Ionization Models}

\begin{figure*}
\includegraphics[width=5.8cm]{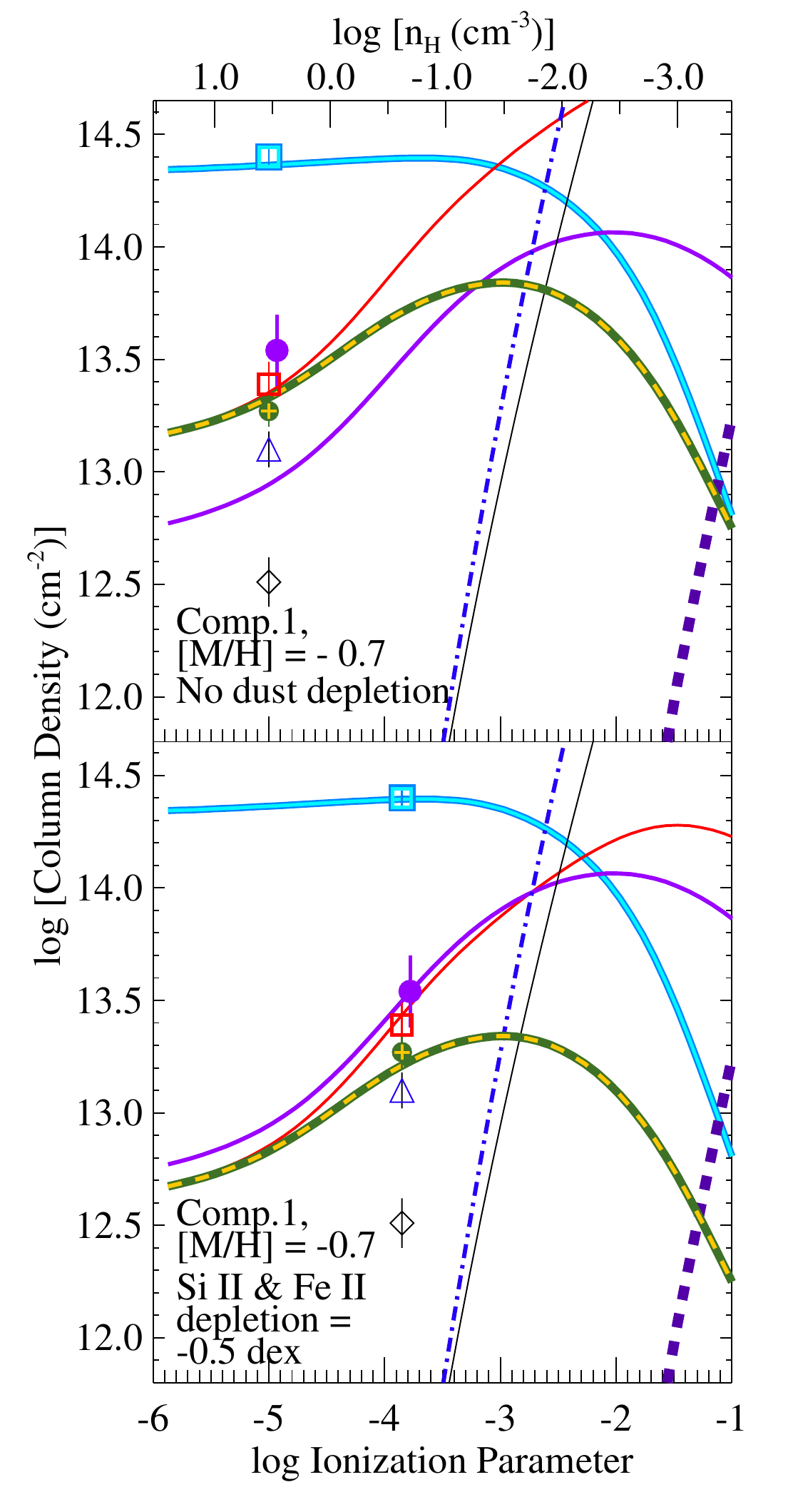}
\includegraphics[width=5.8cm]{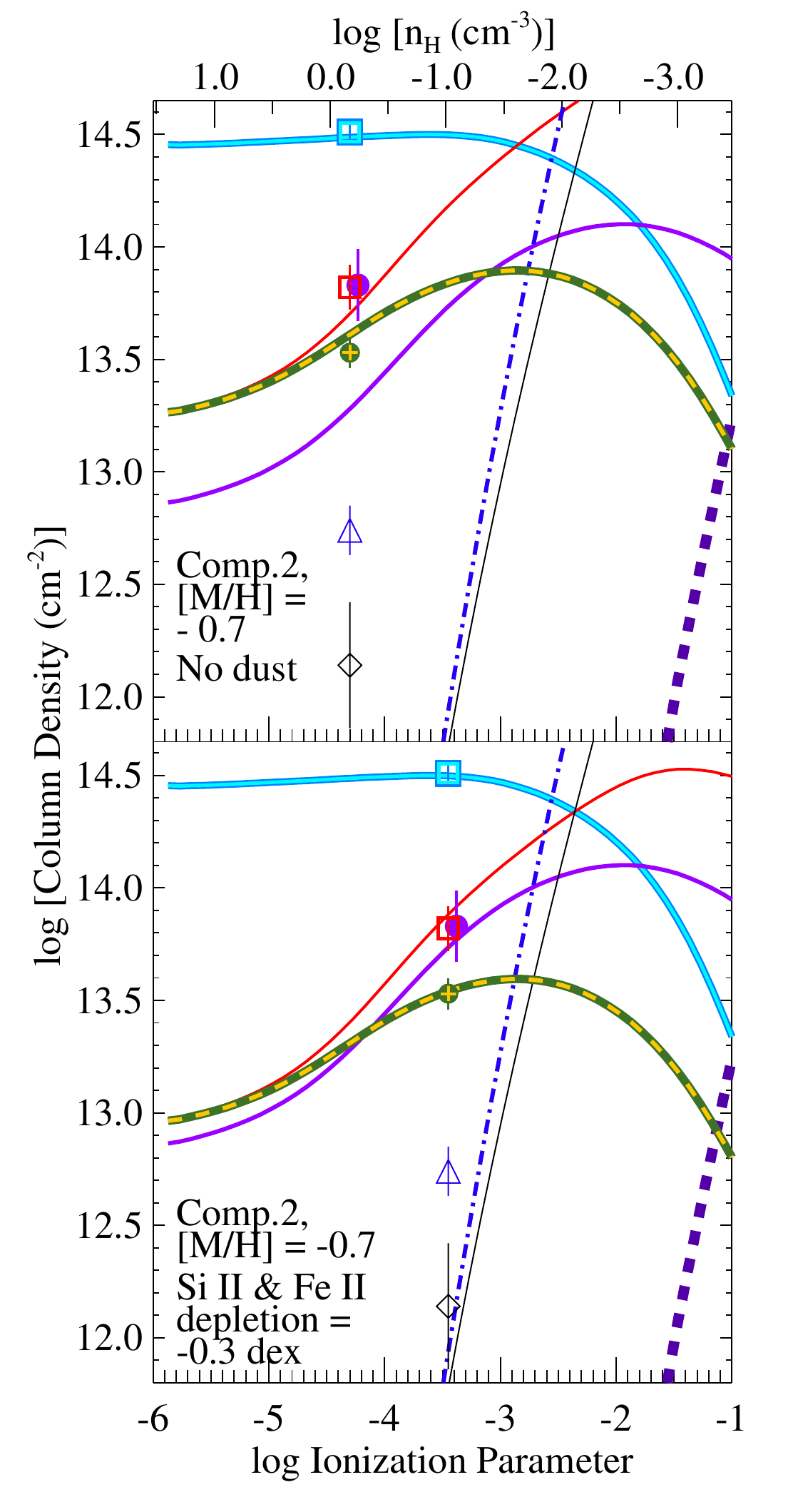}
\includegraphics[width=5.8cm]{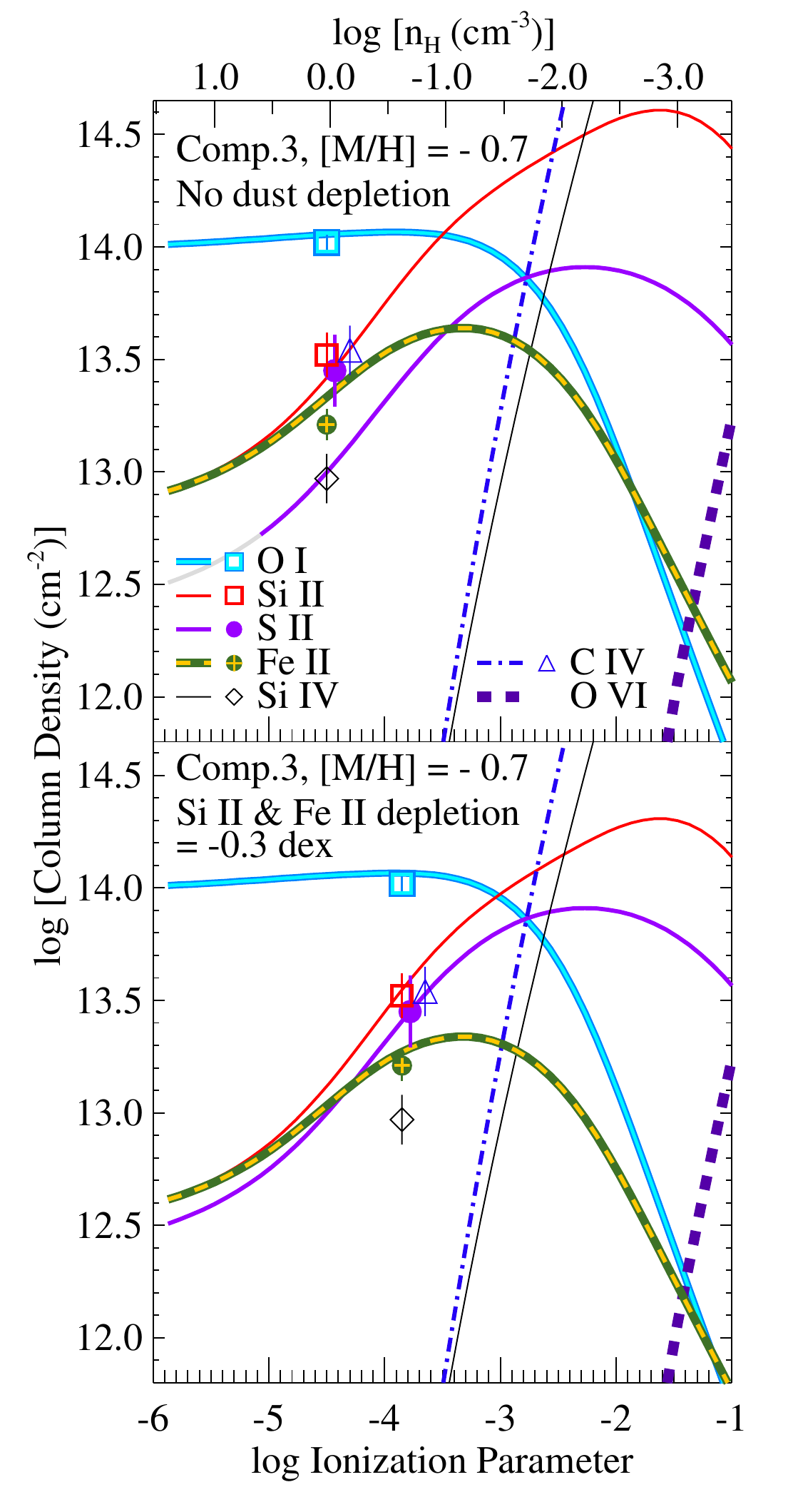}
\caption{Photoionization models calculated with \textsc{cloudy} \citep{ferland17} for components $1 - 3$ detected in the Outer Arm toward H1821+643. The logarithmic column densities of various species predicted by the models, as a function of the log of the ionization parameter ($U \equiv n_{\gamma}/n_{\rm H}$), are shown with smooth curves following the legend in the upper-right panel; the $n_{\rm H}$ corresponding to the $U$ values are indicated on the top axes.  The observed column densities with $1\sigma$ error bars are plotted with discrete points, also following the legend at upper right.  The left, middle, and right stacks show models for components 1, 2, and 3 respectively.  The upper panels show models computed assuming that \underline{none} of the species are depleted by dust, and the observed points are plotted at an ionization parameter that provides the best fit to the \OI, \SiII, and \FeII\ measurements (some points are offset slightly for clarity).  These dust-free models always provide a poor fit to the observed \SII\ column densities (compare the purple lines and purple dots). The lower panels show models in which the silicon and iron is depleted by 0.3 to 0.5 dex, as noted by the annotations within each panel.  With some depletion of the Si and Fe, the models achieve much better agreement with the observed low-ionization stages.  On the other hand, at the ionization parameters that match the measured columns of low-ionization stages, the higher ions are underpredicted by orders of magnitude.  The high-ionization stages must arise from a separate gas phase. \label{fig:photomodcol}}
\end{figure*}

The detailed contextual information on the Milky Way HVCs, and in particular the OA and Complex C (S\ref{sec:hvc_summary}), provides an opportunity to probe the physics of gas in the disk-halo interface and the inner CGM.  To investigate the physical origin of the five narrow components detected in the OA toward H1821+643, this section derives observational constraints, starting with the low-ionization absorption lines in Section~\ref{sec:1821lowionmodels} and then turning to the highly-ionized line in Section~\ref{sec:1821highionmodels}. Considering the low- and high-ionization lines together, we will find that the data can be explained by either low-metallicity or high-metallicity gas, but the implied physical conditions are similar in either case. In this section and throughout the paper, we adopt the solar abundances measured by \citet{asplund09}.

\label{sec:h1821ionization}

\subsection{Low-Ionization Phases}
\label{sec:1821lowionmodels}

\subsubsection{Photoionization Model 1: Low Metallicity}
\label{sec:dust1821}

To study abundance patterns and physical conditions, we can use the photoionization code \textsc{Cloudy} \citep[v17.02,][]{ferland17} to model the ionization of the low-ionization gas. The known location of the OA (S\ref{sec:hvc_summary}) is valuable for this purpose. At the OA's location, the ionizing flux emerging from the Milky Way should dominate over the UV flux from the extragalactic background \citep[see Fig.~8 in][]{fox05}.  Therefore, we initially assume the gas is mainly photoionized by flux from the Milky Way.  \citet{fox05} publish models of the flux at $z-$height = 0, 10, 50, and 100 kpc in their Figure~8; for these models we use the ionizing radiation field shape $z$ = 10 kpc and interpolate to the approximate intensity at $z$ = 7 kpc, the estimated height of the OA gas in the direction of H1821+643. Each absorbing ``cloud'' is modeled as a constant density slab illuminated by this radiation field. In this section the OA gas is assumed to have a metallicity\footnote{Logarithmic metallicity in the usual notation, [M/H] = log (M/H) - log (M/H)$_{\odot}$, where M generically represents metals.} [M/H] = $-0.7$ based on previous studies of the outer-Galaxy HVCs \citep{richter01,tripp03,collins07,tripp12,putman12}; we will examine high-metallicity models in the next section.  With the radiation field from \citet{fox05}, the first step in the modeling is to adjust the gas density (and hence the ionization parameter) to fit the measured \OI/\SII\ column-density ratio.  Therefore good measurements of \OI\ and \SII\ are needed for these models; since \SII\ is not well-measured in components 4 and 5, this section will concentrate on components $1 - 3$. If [M/H] = $-0.7$, components $1 - 3$ are required to have log $N$(\HI) = 18.3, 18.4, and 18.0, respectively, to fit \OI/\SII, assuming oxygen and sulfur suffer little depletion by dust.  Some of the detected low-ionization stages, particularly \SiII\ and \FeII, can be significantly depleted from the gas phase by incorporation into dust \citep{savage96,jenkins09}, so while we will initially assume no dust depletion of any species, we will subsequently consider models that allow for some depletion of refractory elements below.  

The higher Lyman series lines in the FUSE spectrum of H1821+643 \citep[see Fig. 21 in][]{french21} are strongly saturated and can easily accommodate the \HI\ column densities derived for components $1 - 3$ with [M/H] = $-0.7$. The constraint placed on $N$(\HI) by the observed 21cm emission toward H1821+643 is more stringent, but some caveats about the 21cm emission measurement should be noted.  The 21cm emission was recorded with a single-dish telescope \citep{wakker01}. As such, the measured $N$(\HI) is vulnerable to various systematic errors due to the large single-dish beam at 21cm: there could be material in the 21cm beam that is not actually in front of the QSO, there could be beam-dilution effects, and the measurements can be systematically affected by ripples in the baseline or the removal of sidelobe stray radiation \citep[e.g.,][]{wakker01}. Moreover, the H1821+643 21cm spectrum cannot discern the individual components seen in the E140H data, so the apportionment of $N$(\HI) among the components was set by assuming all components have [M/H] = $-0.7$, and then $N$(\HI) was adjusted by small amounts to fit the column densities of \OI\ and \SII .  The log $N$(\HI) values found in this way are in reasonable agreement with the \citet{wakker01} measurement (see Tab.~\ref{tab:targets}) considering possible systematic errors, but there is some degeneracy between [M/H] and $N$(\HI), so various combinations of [M/H] and $N$(\HI) lead to equally acceptable fits.  In addition, even though the \citet{fox05} study provides a reasonable estimate of the ionizing flux impinging on the absorbing gas, there is still uncertainty in the impinging flux, partly because the location of the gas, while usefully constrained by the studies in Table~\ref{tab:clouds}, still allows for a range of $z-$heights, and of course the \citet{fox05} model details have their own uncertainties.  

Figure~\ref{fig:photomodcol} shows the column densities of detected ions predicted by the photoionization models as a function of ionization parameter (lower axes) and gas density $n_{\rm H}$ (upper axis).  From left to right, the three columns in Figure~\ref{fig:photomodcol} show predictions for components 1, 2, and 3 respectively; the upper panels in each column show models that assume no dust depletion, while the lower panels show the same models but with logarithmic depletions of silicon and iron by $-0.3$ to $-0.5$ dex, as indicated by the annotations in each panel. The model predictions are shown with smooth curves, and the the observed column densities are indicated with discrete points plotted at an ionization parameter that provides agreement between the models and the data (see below). The species indicated by the curves and discrete points are indicated by the legend in the upper-rightmost panel.  This section is focused on the low-ionization gas, but it turns out that the \SiIV\ and \CIV\ column densities will rule out some models (see below), so the observed $N$(\SiIV) and $N$(\CIV) are also plotted in Figure~\ref{fig:photomodcol}.  We immediately see that photoionization models that match the low-ion column densities significantly underpredict the high-ion column densities, a point we will return to in S\ref{sec:1821highionmodels}.

In the dust-free models (upper panels of Fig.~\ref{fig:photomodcol}), the observed column densities are plotted at an ionization parameter that provides the best fit to the \OI, \SiII, and \FeII\ measurements for components $1-3$.  However, for all three components, optimizing the fit of \OI, \SiII, and \FeII\ causes the dust-free models to underpredict the observed log $N$(\SII) by $0.4 - 0.6$ dex.  The \SII\ discrepancies in the dust-free models are significant at the 4.6, 6.3, and 2.5$\sigma$ levels based on the error bars from the profile-fitting code.  The difference cannot be attributed to the \SII\ deblending discussed in S\ref{sec:deblending}; the deblending removed a maximal amount of the interloper optical depth, and any plausible alternative deblending model (including ignoring the blend entirely) would lead to higher measured $N$(\SII) thereby exacerbating the discrepancy.  Two possible explanations of this discrepancy are (1) some \SII\ arises in a more highly ionized phase, and this boosts the \SII/\SiII\ ratio, or (2) some of the Si (and Fe) is depleted by dust grains.  We can show that the dust-depletion explanation is strongly preferred by the data.

\subsubsection{Extra S II From Ionized Gas?} Since \SII\ has a higher ionization potential than \OI, \SiII, and \FeII\ (see Table~\ref{tab:IPS}), it is plausible that \SII\ can persist in more highly ionized material where the other low-ion column densities are reduced due to ionization. Indeed, using the non-equilbrium collisional ionization models of \citet{gnat07}, or using the photoionization models ionized by the \citet{fox05} radiation field but with the gas temperature fixed at $T \approx 10^{4.5}$ K (which adds collisional ionization), we can find models of a more ionized phase that boost \SII\ relative to \SiII\ and contribute negligible amounts of additional \OI\ and \FeII.  When the additional column densities from this more highly ionized phase are added to the lower-ionization phases shown in Figure~\ref{fig:photomodcol}, we can adequately fit all of the low-ionization stages with small adjustments of the ionization parameters.  However, the additional ionized gas creates a serious problem: the \SiIV\ and \CIV\ column densities from the more ionized phase substantially exceed the observed $N$(\SiIV) and $N$(\CIV), and the disagreement is much greater than the measurement uncertainties.  On this basis, we can rule out this explanation.

\subsubsection{Evidence of Dust \label{sec:dustdepletion}}
A more likely explanation is that the \SII\ discrepancy indicates that Si and Fe are mildly depleted by dust in the OA in the direction of H1821+643.  In the lower halo and the disk-halo interface, previous studies have shown that Si and Fe are depleted from the gas phase by $-0.3$ to $-0.7$ dex due to incorporation into dust, while the sulfur depletion is negligible \citep{savage96}.  The oxygen depletion should also be minimal in this regime of light depletion. If we allow the Si and Fe to be depleted by $0.3 - 0.7$ dex in the photoionization models, then we can find fits at somewhat higher ionization parameters that do accommodate the observed \SII\ column densities.  This is shown in the lower panels of Figure~\ref{fig:photomodcol}: with Si and Fe depletions of $-0.5$, $-0.3$, and $-0.3$ dex in components 1, 2, and 3 respectively, the \OI, \SII, \SiII, and \FeII\ columns in all of the components can be fitted with log $U$ ranging from $-4.2$ to $-3.7$.

\begin{figure}
\includegraphics[width=7.5cm]{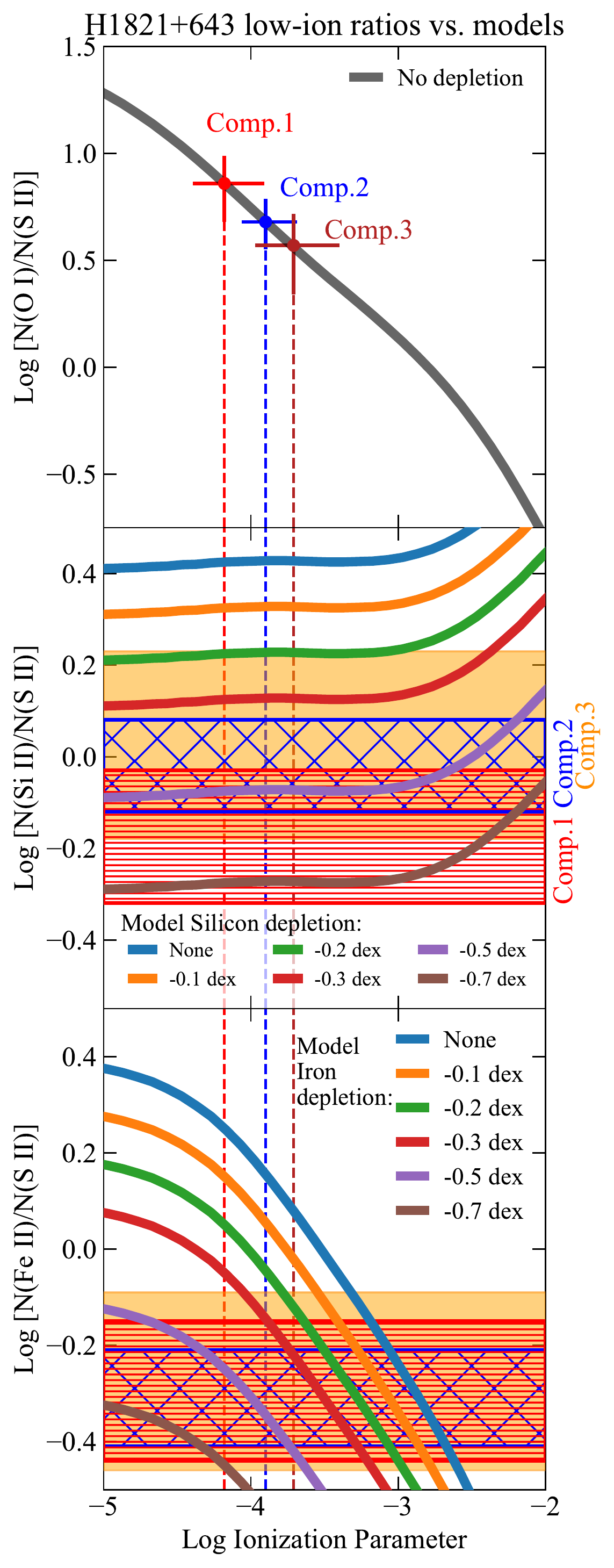}
\caption{Logarithmic column-density ratios predicted by the photoionization models from Fig.~\ref{fig:photomodcol}. The upper, middle, and lower panels show the predicted $N$(\OI)/$N$(\SII), $N$(\SiII)/$N$(\SII), and $N$(\FeII)/$N$(\SII) ratios, respectively, as a function of log $U$. In the upper panel, only the ratio assuming no depletion is shown.  In the middle and lower panels, the colored curves assume various amounts of Si and Fe depletion (see legends).  The observed \OI/\SII\ ratios are shown with labeled points with $1\sigma$ error bars for components $1-3$ in the upper panel.  The observed $\pm 1\sigma$ ranges for the \SiII/\SII\ and \FeII/\SII\ ratios are shown as red hatched, blue hatched, and solid orange bands for components 1, 2, and 3, respectively.  At any ionization parameter, the \SiII/\SII\ ratio cannot be matched by the ionization model unless the Si is depleted by at least a factor or 2 or 3.\label{fig:photomodrat}}
\end{figure}

While the evidence of dust depletion in the lower panels of Figure~\ref{fig:photomodcol} is predicated on simultaneously fitting \OI, \SII, \SiII, and \FeII, the presence of dust in the OA is robustly indicated by the \SiII/\SII\ ratio alone.  This is illustrated in Figure~\ref{fig:photomodrat}. In this figure, the upper, middle, and lower panels plot the \OI/\SII, \SiII/\SII, and \FeII/\SII\ ratios predicted by the models, with the different colored curves assuming no depletion up to $-0.7$ dex of depletion, as indicated in the legend. Focusing first on the middle panel, we see that the \SiII/\SII\ ratio is almost constant in the ionization parameter range where \OI\ is the dominant oxygen ion, and this ratio increases at higher $U$ values.  The solid or hatched red, blue, and orange bands indicate the observed $\pm 1\sigma$ ranges of the \SiII/\SII\ ratios in comps. 1-3.  Comparing the model and observed ratios, we see that silicon must be depleted by at least $-0.2$ to $-0.4$ dex if log $U \lesssim -3.0$, and the Si depletion must be even greater at higher ionization parameters. Some amount of Si depletion is unavoidable.  The \FeII/\SII\ ratio (lower panel of Fig.~\ref{fig:photomodrat}) is less cooperative; to constrain the Fe depletion, the ionization parameter range must be constrained, e.g., by using the \OI/\SII\ ratio as illustrated in the top panel.  Using Figure~\ref{fig:photomodrat} and the constraints on log $U$ from \OI/\SII, we find the minimum and maximum Si and Fe depletions (that are consistent with the 1$\sigma$ measurement uncertainties) listed in the upper half of Table~\ref{tab:h1821_dustdepl}.

\ctable[
caption={Dust Depletion in Low-Ionization Outer-Arm Components in the Spectrum of H1821+643},
label={tab:h1821_dustdepl},
doinside=\footnotesize
]{lcccc}{
\tnote[a]{Logarithmic depletion by dust required to achieve agreement with the ionization models and the $\pm 1\sigma$ measurement uncertainties.} 
\tnote[b]{Component number, as indicated in Figure~\ref{fig:vpstack}, with the component velocity in km s$^{-1}$ in parentheses.}
}
{\FL \ & \multicolumn{2}{c}{\underline{Si Depletion (dex)}\tmark[a]} & \multicolumn{2}{c}{\underline{Fe Depletion (dex)}\tmark[a]} \NN
Comp.\tmark[b] & Min.   & Max.   & Min. & Max. \ML
\multicolumn{5}{l}{Model 1: Low Metallicity ($Z = 0.2 \ Z_{\odot}$)} \NN
   1 ($-145$)      & $-0.4$ & $-0.8$ & $-0.3$ & $-0.7$ \NN 
   2 ($-134$)      & $-0.3$ & $-0.6$ & $-0.3$ & $-0.6$ \NN 
   3 ($-124$)      & $-0.2$ & $-0.6$ & $-0.1$ & $-0.6$ \NN \hline
\multicolumn{5}{l}{Model 2: High Metallicity ($Z = 1.0 \ Z_{\odot}$)} \NN
   1 ($-145$)      & $-0.5$ & $-0.8$ & $-0.6$ & $-0.9$ \NN 
   2 ($-134$)      & $-0.4$ & $-0.7$ & $-0.6$ & $-0.9$ \NN 
   3 ($-124$)      & $-0.3$ & $-0.8$ & $-0.5$ & $-0.9$  \LL
}

\subsubsection{Photoionization Model 2: High Metallicity}
\label{sec:lowion_highZ_photo}

The photoionization models in the previous section assumed that the vast majority of the \HI\ is located in the same phase as the \OI, \SII, \SiII, and \FeII.  This is a standard assumption in the literature, and it is logical to suppose that \HI\ is mainly in the lowest-ionization phase.  However, \HI\ can persist and be detectable in more highly ionized gas, and given the kinematical alignment of the low- and high-ionization phases, we should ask if some of the \HI\ absorption could come from the more highly ionized gas.  Indeed, in the next section we will find that non-equilibrium models \textsl{favor} roughly equal contributions to $N$(\HI) from the low- and high-ionization phases of the H1821+643 OA cloud.

What are the consequences of relaxing this assumption about the origin of the \HI \ ?  In photoionization models, $N$(\HI) and metallicity are largely degenerate, so if we decrease $N$(\HI) and increase [M/H] by the same amount, the predicted metal-ion column densities will remain mostly the same, i.e., the observed metal columns can be explained by various combinations of $N$(\HI) and [M/H].  However, as $N$(\HI) increases above $\approx 10^{17}$ cm$^{-2}$, self shielding becomes increasingly important, and this leads to changes in ionization structure at higher $N$(\HI) values, so it is necessary to explicitly compute models with different $N$(\HI) rather than just scaling a single model to different values of $N$(\HI) and [M/H].  Of course, if the low-ionization phase does not account for the large majority of the observed \HI, then some other location must be identified with enough \HI\ to make up the difference between the part in the low-ionization gas and the observed total.  In S\ref{sec:noneq_highions} we will show that in some models, the highly ionized phase can provide sufficient additional \HI\ to make an important contribution to the total \HI\ budget.

If the outer-Galaxy HVCs are gaseous structures elevated from the plane by the passage of a perturber through the disk \citep[as in the model of][]{kawata03}, then (at least some portions of) the HVC gas could have originated in the disk and therefore could have roughly solar metallicity.  Alternatively, if these HVCs are part of a Galactic fountain \citep[e.g.][]{fraternali15}, high-metallicity material from the supernova ejecta that drives the fountain flow could be present.  To explore these scenarios, we have calculated a second set of photoionization models for the low-ionization phases in components $1 - 3$ using solar metallicity ($Z = 1.0 \ Z_{\odot}$) and reduced $N$(\HI) values.

\begin{figure}
\includegraphics[width=7.3cm]{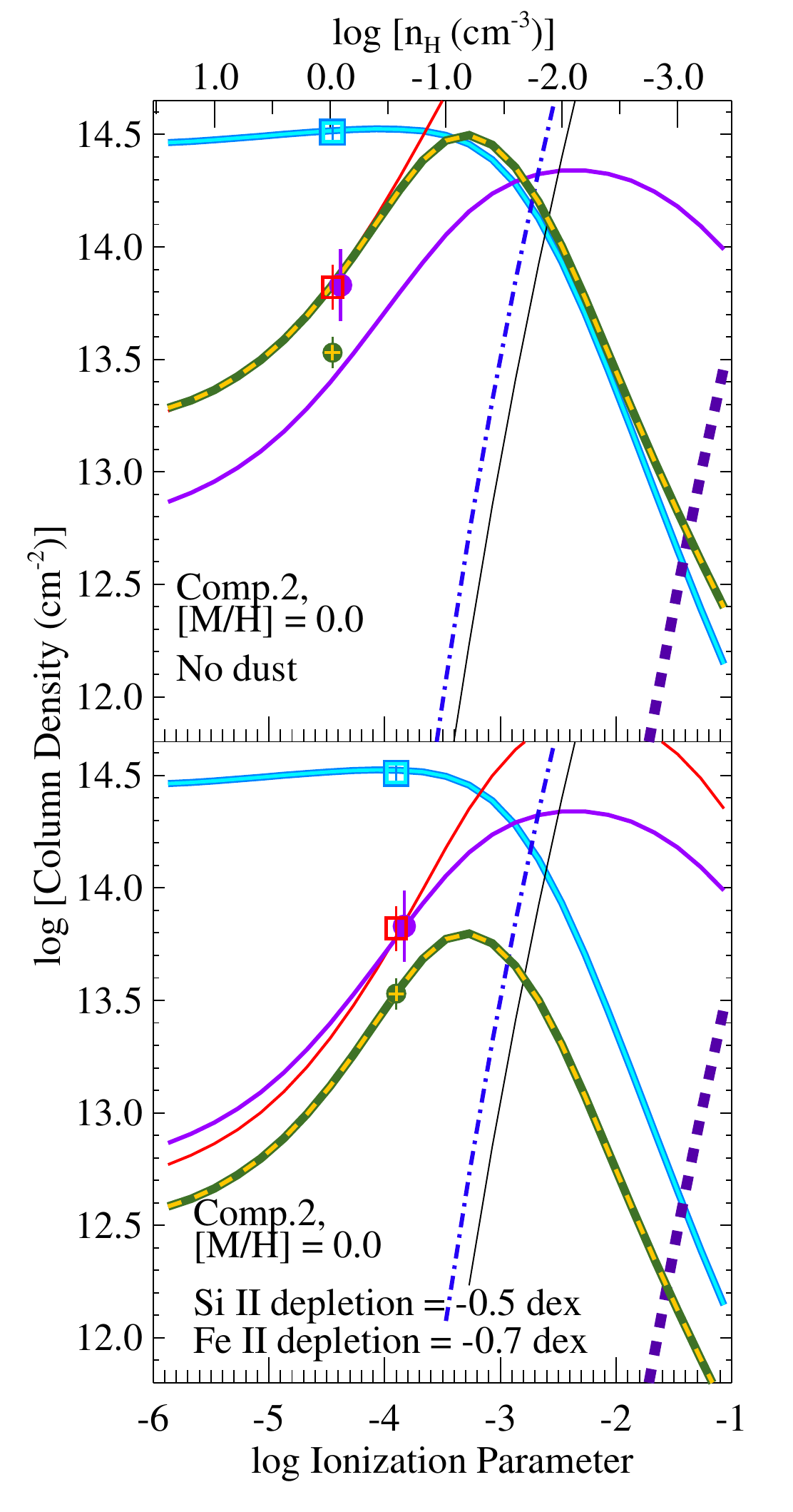}
\caption{Photoionization models, as in Figure~\ref{fig:photomodcol} but with $Z = 1.0 \ Z_{\odot}$ and log $N$(\HI) = 17.7, for component 2 of the OA toward H1821+643. This high-metallicity model yields a comparably good fit to the observations, but like the low-metallicity model in Figure~\ref{fig:photomodcol}, Si and Fe depletion is necessary (compare upper and lower panels). 
\label{fig:photomod_1Z}}
\end{figure}

Figure~\ref{fig:photomod_1Z} shows a $Z = 1.0 \ Z_{\odot}$ model that fits the columns measured in component 2.  Comparing Figure~\ref{fig:photomod_1Z} to the middle column of Figure~\ref{fig:photomodcol}, we see that there are some important differences in the $Z = 0.2 \ Z_{\odot}$ model versus the $Z = 1.0 \ Z_{\odot}$ case, as expected since the higher-metallicity run has a significantly lower \HI\ column density and less self shielding. Nevertheless, the calculation with solar metallicity leads to qualitatively similar conclusions:  First, the $Z = 1.0 \ Z_{\odot}$ model still requires appreciable dust depletion of refractory species; in fact, the $Z = 1.0 \ Z_{\odot}$ run actually requires slightly more depletion. In the upper panel of Figure~\ref{fig:photomod_1Z}, which does not include dust depletion, it is impossible to find an ionization parameter that simultaneously fits \OI, \SII, \SiII, and \FeII.  On the other hand, if \SiII\ is depleted by 0.5 dex and \FeII\ is depleted by 0.7 dex, as shown in the lower panel, then all of these low-ionization stages can be fitted.  This pattern of Fe depletion that is slightly greater than the Si depletion is more similar to depletion patterns observed in the Milky Way disk \citep{jenkins09} than the equal Si and Fe depletion required by the $Z = 0.2 \ Z_{\odot}$ models (see Fig.~\ref{fig:photomodcol}).  Second, this high-metallicity model requires a small shift (compared to the low-$Z$ run) in the ionization parameter/gas density to optimally fit the observed columns, but the best-fitting ionization parameter is close to the best value in the $Z = 0.2 \ Z_{\odot}$ model, so the implications for the physical conditions of the gas (next section) are essentially the same regardless of whether we chose one metallicity or the other. 

Models for components 1 and 3 with $Z = 1.0 \ Z_{\odot}$ (not shown in Figure~\ref{fig:photomod_1Z}) show very similar results.  For components $1 - 3$, the minimum and maximum depletions of Si and Fe required in the high$-Z$ models are summarized in the lower half of Table~\ref{tab:h1821_dustdepl}.

\ctable[
caption={Physical Conditions of Low-Ionization Outer-Arm Components in the Spectrum of H1821+643},
label={tab:h1821_physcon},
doinside=\footnotesize,
star
]{llclllccll}{
\tnote[a]{Component number, as indicated in Figure~\ref{fig:vpstack}, with the component velocity in km s$^{-1}$ in parentheses.}
%\tnote[b]{Logarithmic depletion by dust required to achieve agreement with the ionization models and the $\pm 1\sigma$ measurement uncertainties. For each component, the upper number is the minimum depletion and the lower number is the maximum depletion that is consistent with the data at the $1\sigma$ level.}
\tnote[b]{Upper limit on the component temperature derived from the Si~II line width (see Table~\ref{tab:vp_si2_si4}).}
\tnote[c]{Ionization parameter $U \equiv n_{\gamma}/n_{\rm H}$.  Error bars indicate the range in $U$ that is consistent with the \OI/\SII\ ratio at the 1$\sigma$ level.}
\tnote[d]{As discussed in the text, there are systematic uncertainties that affect the gas density, total H column density, and cloud size.  Nevertheless, even allowing for these uncertainties, $t_{\rm cool}/t_{\rm ff}$ is robustly indicates that this gas is able to cool rapidly.}
\tnote[e]{Ratio of the cooling time to the sound-crossing time, $t_{\rm sc} = L/c_{\rm s}$.}}
%{\FL               \   &  \                       & \                           & \                             & log                        & \              &     \                  & \                                    \NN
%                     \   &                         & \                           &                                          &                                                                 & Size       & $t_{\rm cool}$ &  \  \NN
{\FL    \                            & log                              & \            & log                         & log                                & log & \ & \ & \ & \NN                                                     
Comp.\tmark[a] & [T (K)]\tmark[b] & log $U$\tmark[c] & [$n$ (cm$^{-3}$)]  & [$N$(\HI) (cm$^{-2}$)]  & [$N_{\rm tot}$(H) (cm$^{-2}$)]         & Size (pc)        & $t_{\rm cool}$ (years)    & $t_{\rm cool}/t_{\rm ff}$\tmark[d] & $t_{\rm cool}/t_{\rm sc}$\tmark[e] \NN
 \ \ \ (1) & \ \ \ (2) & (3) & \ \ \ (4) & \ (5) & (6) & (7) & \ (8) & (9) & (10) \ML
 \multicolumn{10}{l}{Model 1: Low Metallicity ($Z = 0.2 \ Z_{\odot}$)} \NN
        \                  & \                      &\                             & \                            &  \                           &    \           &  \                  & \    & \ & \NN
   1 ($-145$)      & $\leq 4.43$      & $-4.18^{+0.26}_{-0.20}$  & $\approx -0.3$           & 18.3 & 19.0            & $\approx 6$  & 5$\times 10^{4}$ & 5$\times 10^{-4}$ & 0.1 \NN
       \                  & \                      &\                             & \                            &  \                           &    \           &  \                    & \  & \ & \NN
   2 ($-134$)      & $\leq 4.36$      & $-3.90^{+0.20}_{-0.15}$  & $\approx -0.6$           & 18.4 & 19.2            & $\approx 19$ & 1$\times 10^{5}$ & 1$\times 10^{-3}$ & 0.05 \NN
          \                  & \                      & \                             & \                            &  \                           &    \           &  \                 & \     & \ & \NN
   3 ($-124$)      & $\leq 4.39$      & $-3.70^{+0.30}_{-0.25}$  & $\approx -0.8$           & 18.0 & 19.3            & $\approx 37$ & 2$\times 10^{5}$ & 2$\times 10^{-3}$ & 0.05 \NN
             \                  & \                      & \                             & \                            &  \                           &    \           &  \                 & \     & \ & \NN \hline
 \multicolumn{10}{l}{Model 1: High Metallicity ($Z = 1.0 \ Z_{\odot}$)} \NN
         \                  & \                      & \                             & \                            &  \                           &    \           &  \                  & \    & \ & \NN
   1 ($-145$)      & $\leq 4.43$      & $-4.14^{+0.25}_{-0.19}$  & $\approx -0.3$           & 17.6 & 18.5            & $\approx 2$  & 3$\times 10^{4}$ & 3$\times 10^{-4}$ & 0.1 \NN
       \                  & \                      & \                             & \                            &  \                           &    \           &  \                   & \   & \ & \NN
   2 ($-134$)      & $\leq 4.36$      & $-3.88^{+0.18}_{-0.15}$  & $\approx -0.6$           & 17.7 & 18.8            & $\approx 8$ & 3$\times 10^{4}$ & 3$\times 10^{-4}$ & 0.03 \NN
          \                  & \                      & \                             & \                            &  \                           &    \           &  \                  & \    & \ & \NN
   3 ($-124$)      & $\leq 4.39$      & $-3.90^{+0.28}_{-0.20}$  & $\approx -0.6$           & 17.3 & 18.5            & $\approx 4$ & 3$\times 10^{4}$ & 3$\times 10^{-4}$ & 0.06 \LL
}

\subsubsection{Physical Conditions in the Low-Ionization Gas}
\label{sec:lowphase_physcond}

The ionization models in Figures~\ref{fig:photomodcol} and \ref{fig:photomod_1Z} constrain the physical conditions of the OA, and in turn these physical conditions can be compared to predictions from circumgalactic physics theory (S\ref{sec:cgm_physics}). The conditions of the low-ionization phase are particularly interesting when compared to those of the more highly ionized gas at the same velocity, which are discussed in the next section (S\ref{sec:1821highionmodels}).  In this section we concentrate on the models that include dust depletion of Si and Fe.

One of the most straightforward physical-conditions measurements is provided by the component line widths: in the case of purely thermal broadening, $T = mb^{2}/2k = A(b/0.129)^{2}$, where $A$ is the atomic mass number and the rightmost expression is for $T$ in K and $b$ in \kms. Since other broadening mechanisms may contribute (and indeed some evidence of non-thernal broadening is shown in Figure~\ref{fig:vp_therm_nontherm}), these temperature constraints should be treated as upper limits.   Temperature limits derived this way, assuming the best-fit $b-$values from the \SiII\ measurements, are listed in Table~\ref{tab:h1821_physcon} for components $1 - 3$.

With the ionizing flux and photon density fixed by the \citet{fox05} calculations, the photoionization models also constrain the gas density (see top axes in Figs.~\ref{fig:photomodcol} and \ref{fig:photomod_1Z}). The ionization parameters that best fit the low-ionization absorption lines, and the corresponding gas densities, are summarized in columns 3 and 4 of Table~\ref{tab:h1821_physcon}.  This table summarizes the results for the low$-Z$ and high$-Z$ models in the upper and lower halves, respectively. The sizes and total column densities of clouds are also useful in various contexts, so the table also lists the derived total H column densities (\HI\ + \textsc{H~ii}) in column 6 and an estimate of the absorbing cloud thickness ($L = N_{\rm H}/n_{\rm H}$) in column 7. The error bars on log $U$ indicate the ranges of $U$ that are consistent with the observed \OI/\SII\ ratios within $\pm 1\sigma$.  Uncertainties in the ionizing flux lead to uncertainties in the gas density, so the log $n$ values are indicated as rough estimates.  For example, if the absorbing gas is at a $z-$height of 10 kpc instead of 7 kpc, then the reduction of the ionizing flux, according to the \citet{fox05} calculations, would lead to gas densities that are lower by a factor of $\approx$ 3 and cloud thicknesses that are larger by a factor of $\approx$ 3. The total column densities have an additional uncertainty from the apportionment of the \HI\ among the components (see \S\ref{sec:dust1821}), but given the relative low-ionization metal column densities in the components, the relative amount of \HI\ in the components is reasonable and is likely correct to within $\approx$ 0.3 dex.

Even with the various uncertainties, the results in Table~\ref{tab:h1821_physcon} robustly suggest that the OA low-ionization gas is in a rapidly cooling situation.  To show this, we can calculate the cooling time for components $1 - 3$ using Eqn.~\ref{cooling_eqn} with the gas densities, temperatures,\footnote{The \textsc{cloudy} model gas temperatures are consistent with the observational temperatures implied by the line widths.} and $\Lambda$ taken from the \textsc{cloudy} model at the best-fitting $U$.  These cooling times are tabulated in column 8.  To calculate the diagnostic $t_{\rm cool}/t_{\rm ff}$ ratio that is frequently discussed in CGM theory papers (S\ref{sec:cgm_physics}), we can evaluate $t_{\rm ff}$ using $t_{\rm ff} = \sqrt{2} R_{\rm G}/\emph{v}_{\rm cir}$, or we can estimate $t_{\rm ff}$ using the vertical gravitational acceleration in and beyond the solar neighborhood \citep[e.g.,][]{boulares90} and the $z-$height of the gas.   For the H1821+643 sightline, these two methods yield very similar results for $t_{\rm ff}$; column 9 in Table~\ref{tab:h1821_physcon} summarizes the $t_{\rm cool}/t_{\rm ff}$ calculated using the first method.  We see that these estimates indicate that $t_{\rm cool}/t_{\rm ff} \ll 1$.  While some uncertainties could lead to longer  $t_{\rm cool}$ times, these uncertainties are not sufficient to increase $t_{\rm cool}/t_{\rm ff}$ enough to indicate that it is in the stable regime. Table~\ref{tab:h1821_physcon} pertains to the cool phase; we will examine constraints on the $t_{\rm cool}/t_{\rm ff}$ ratio in the more ionized \textsl{mixing} gas in the next section.

We can also use the sound-crossing time, $t_{\rm SC} = L/c_{\rm s}$, where $L$ is the pathlength through the cloud and $c_{\rm s}$ is the sound speed, as an estimate of the dynamical timescale of the gas. Column (10) in Table~\ref{tab:h1821_physcon} summarizes $t_{\rm cool}/t_{\rm sc}$ calculated this way. In components $1 - 3$, $t_{\rm cool}/t_{\rm sc} \ll 1$ in either the low$-Z$ or high$-Z$ models.    

These low $t_{\rm cool}/t_{\rm sc}$ values suggest that the OA may be in the process of ``shattering'' (see S\ref{sec:cgm_physics}).  Using the calculation of \citet{mccourt18}, %\citeauthor{mccourt18} (2018; see also \citealt{voit90}), 
that the ``cloudlet'' size $\approx$ (0.1pc)($n$/cm$^{-3}$)$^{-1}$, we find that the \textsc{Cloudy} absorber sizes in the $Z = 0.2 \ Z_{\odot}$ ($Z = 1.0 \ Z_{\odot}$) models are $30 - 60$ ($7 - 20$) times larger than the size of a stable cloudlet, so components $1-3$ are expected to shatter into a larger number of cloudlets (if they have not done so already).  \citet{voit90} and \citet{mccourt18} deduce that the stable cloudlets should have a total column density $\approx 10^{17}$ cm$^{-2}$.  If so, the total $N$(H) inferred for components $1 - 3$ indicate that each of these components comprise $\approx 100 - 200$ cloudlets in the $Z = 0.2 \ Z_{\odot}$ model or $\approx 30 - 60$ cloudlets with $Z = 1.0 \ Z_{\odot}$. It is not trivial to distinguish between a single monolithic cloud with log $N_{\rm tot}$(H) = 19 vs. a mist of 100 cloudlets with log $N_{\rm tot}$(H) = 17; if the cloudlets coherently move at the same bulk velocity in the parent cloud, they would produce a single absorption line similar to the line from a monolithic cloud.  However, it is possible that in the shattering scenario, line broadening due to the turbulent/non-thermal motions of the cloudlets would dominate over thermal line broadening, so the fact that the line widths in the OA favor non-thermal broadening (see Fig.~\ref{fig:vp_therm_nontherm} and S\ref{sec:h1821_lowion_meas}) is intriguing in this context.  Moreover, analyses of the high-ionization absorption in the OA components also favors a misty arrangement of $\approx$100 cloudlets, as we discuss in the next section.

\subsection{High-Ionization Phase}
\label{sec:1821highionmodels}

The kinematic similarity of the low-ionization and high-ionization absorption lines in the OA (see S\ref{sec:h1821highion}) suggests that there is a relationship between the low-ionization and high-ionization gas.  This similarity would be unlikely to occur if the highly-ionized material has no relationship whatsoever to the less-ionized gas.  However, the small-but-significant kinematical differences indicate that the low-ion and high-ion absorption lines do not arise in a single-phase gas cloud, and this conclusion is corroborated by ionization models.  From Figures~\ref{fig:photomodcol} and \ref{fig:photomod_1Z} we see that in purely photoionized gas, when the ionization parameter is adjusted to best fit the low-ionization lines, the models fall vastly short of producing the observed column densities of \SiIV, \CIV, and \OVI\ in the OA.  Likewise, equilibrium and non-equilibrium collisional ionization models \citep{gnat07} cannot simultaneously match the low ions and high ions, even if the joint effects of photoionization plus non-equilibrium collisional ionization are considered \citep{oppenheimer13}.  We can confidently conclude that the OA absorption arises in a multiphase entity.  The following sections examine ionization models that could produce the highly-ionized phase in this multiphase context.

\subsubsection{Photoionization}
\label{sec:photohighion}

It is possible to tune purely photoionized models to fit the observed column densities with (1) a low-ionization phase (as in S\ref{sec:1821lowionmodels}), (2) an intermediate photoionized phase that agrees with the \SiIV/\CIV\ ratio and the \SiIV\ and \CIV\ column densities, and (3) a highly-ionized phase that only contributes \OVI.  These additional phases can have a similar metallicity to the low-ionization gas and do not violate the \HI\ constraints -- the \HI\ profiles can accommodate the \HI\ is the additional phases.  The more ionized phases also do not contribute appreciable low-ionization metal absorption, so they do not conflict with the low-ion metal measurements either.  

However, this multiphase but purely photoionized model suffers from two flaws.  First, similarly to the findings of \citet{haislmaier21} in extragalactic absorbers, in this tuned multiphase photoionization model, the thermal pressures in the \SiIV+\CIV\ and the \OVI\ phases are orders of magnitude lower than the thermal pressure in the low-ionization phase.  Therefore the low-ionization phase is not pressure-confined by the more highly-ionized phases and is highly unstable unless it is confined by non-thermal pressure sources.\footnote{An even hotter phase could be present that is too hot to be detected in UV absorption (i.e., a phase that can only be detected in X-rays).  Such a hotter phase could have a high enough thermal pressure to confine the low-ionization phase, but the stability problem would still persist because some of the phases would still have much lower pressures than the (unseen) hot phase.} Second, in this model the spatial dimensions of the more ionized phases are awkward and unlikely.  In component 1, for example, the required thickness of the \SiIV\ phase is $\approx$40 times larger than the thickness of the low-ionization phase, and the \OVI\ phase is $>10^{5}$ larger if the gas is photoionized by the \citet{fox05} radiation field.  Highly-ionized photoionized phases require very low densities and substantial $N$(H), which in turn implies a large size.  The size of the \SiIV\ phase is marginally workable, but considering the cross section of the \SiIV\ phase vs. the cross section of the low-ionization phase, it is surprising that five components would be intercepted in the OA that are all detected in both \SiIV\ and low ions.  Given its larger cross section in this model, we would expect to frequently see the \SiIV\ without any low-ionization absorption at similar velocities.  To make this work, a large number of small low-ionization droplets would need to fill the \SiIV\ phase with a low volume filling factor, i.e., this requires a \textsl{misty} multiphase configuration.  But this begs a question: how would the low-ionization and \SiIV-phases, with their very different thermal pressures, interact and evolve? How would physical processes at the boundaries of the phases change the column densities?   

Considering the constraints on the location of the OA discussed in S\ref{sec:hvc_summary} and the fact that similar low-ion and \SiIV+\CIV\ absorption is observed toward HS1914+7139, a star at a heliocentric distance of 14.9 kpc \citep{lehner10}, this model mandates a contrived situation where the low-ion and \SiIV\ phases are nearby and the \OVI\ arises in a very long tail extending almost directly away from the observer.  These model dimensions come from a model assuming photoionization by the \citet{fox05} radiation field, which has a large drop in flux at the He~\textsc{ii} edge (see their Fig.~8).  Switching to an ionizing flux field that does not have the He~\textsc{ii} edge does not solve this problem.  Using the \citet{haardt12} UV background flux, for example, reduces the size of the \OVI\ phase, but it is still $7000\times$ larger than the low-ionization phase.  While non-thermal pressure support is a viable and interesting solution to the confinement problem, given the large sizes required by the photoionization models and the possibility that photoionization is not the only process affecting the columns, it is worthwhile to explore whether other ionization models might more naturally explain the observations.

\subsubsection{Non-equilibrium Collisional and Hybrid Ionization}
\label{sec:noneq_highions}

\begin{figure}
\includegraphics[width=8cm]{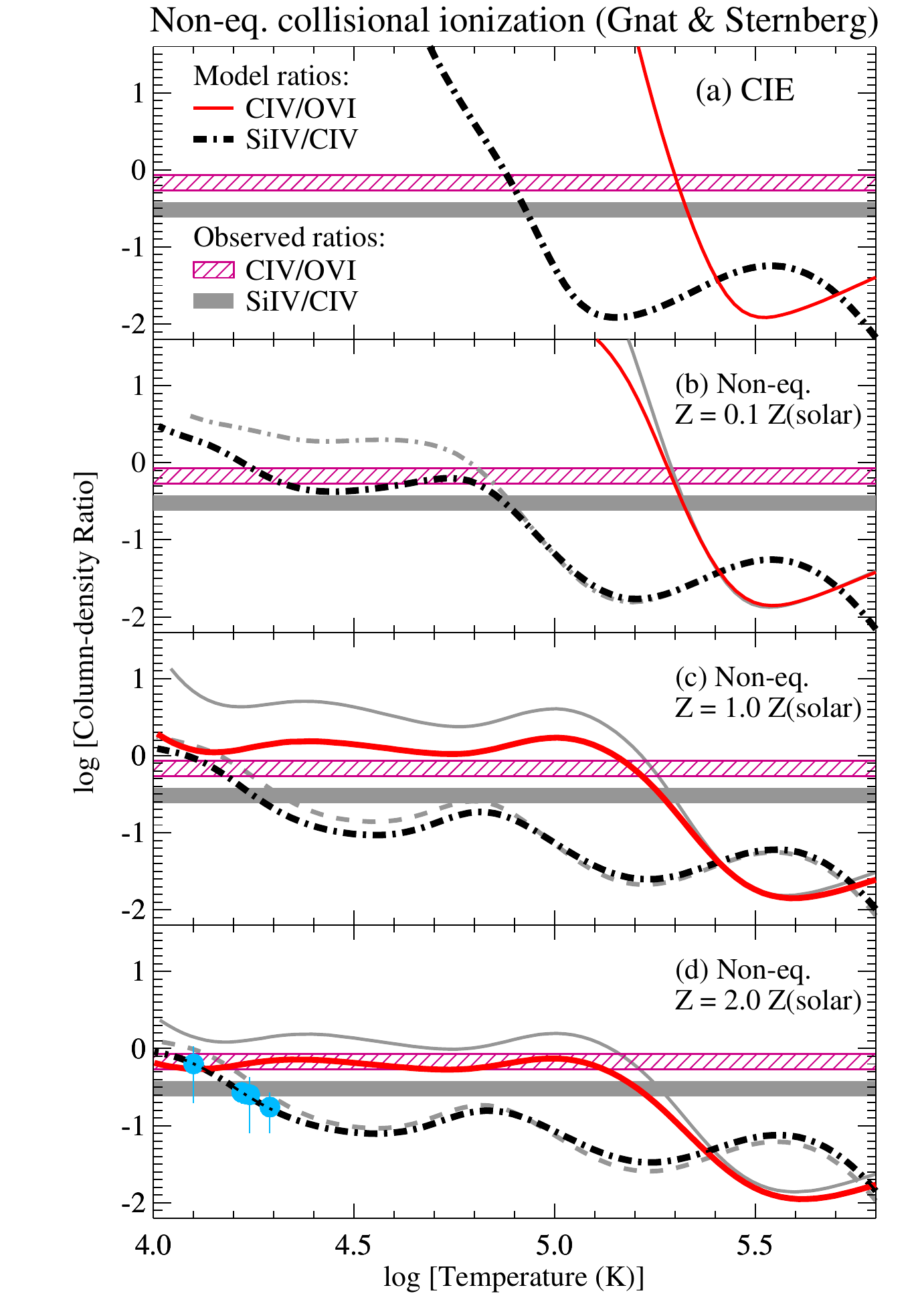}
\caption{Comparison of the \SiIV/\CIV\ and \CIV/\OVI\ ratios vs. temperature predicted by the collisionally ionized gas models of \citet[][solid curves]{gnat07} to the observed OA ratios in the H1821+643 spectrum (horizontal bands). The legend in panel (a) identifies the curves and bands. Several models from \citet{gnat07} are compared, including (a) gas in collisional ionization equilibrium, (b) non-equilibrium cooling gas with $Z = 0.1 \ Z_{\odot}$, (c) non-equilibrium cooling gas with $Z = 1.0 \ Z_{\odot}$, and (d) non-equilibrium cooling gas with $Z = 2.0 \ Z_{\odot}$. In panels (b) -- (d), isochorically cooling models are plotted with dark lines, and the corresponding isobarically cooling models are indicated with lighter gray lines.  In panel (d), the blue dots show the measured \SiIV/\CIV\ ratios in the five individual OA components, each plotted at the temperature that best fits the ratio in that component. \label{fig:gnat07compare}}
\end{figure}

Collisional ionization can alleviate the size problems encountered in the photoionized models because collisional ionization does not require low gas densities.  Can we find a two-phase model that includes a photoionized low-ionization phase and a second hotter phase that produces the \SiIV, \CIV, and \OVI? A two-phase model with the hot phase in collisional ionization equilibrium (CIE) can be immediately ruled out because, as shown in panel (a) of Figure~\ref{fig:gnat07compare}, the CIE temperatures required by the \SiIV/\CIV\ ratio (i.e., $T \approx 10^{4.9}$ K) and the \CIV/\OVI\ ratio ($T \approx 10^{5.3}$ K) are inconsistent at high significance (the discrepancy cannot be attributed to uncertainties in the measured ratios). 

\begin{figure}
\includegraphics[width=8cm]{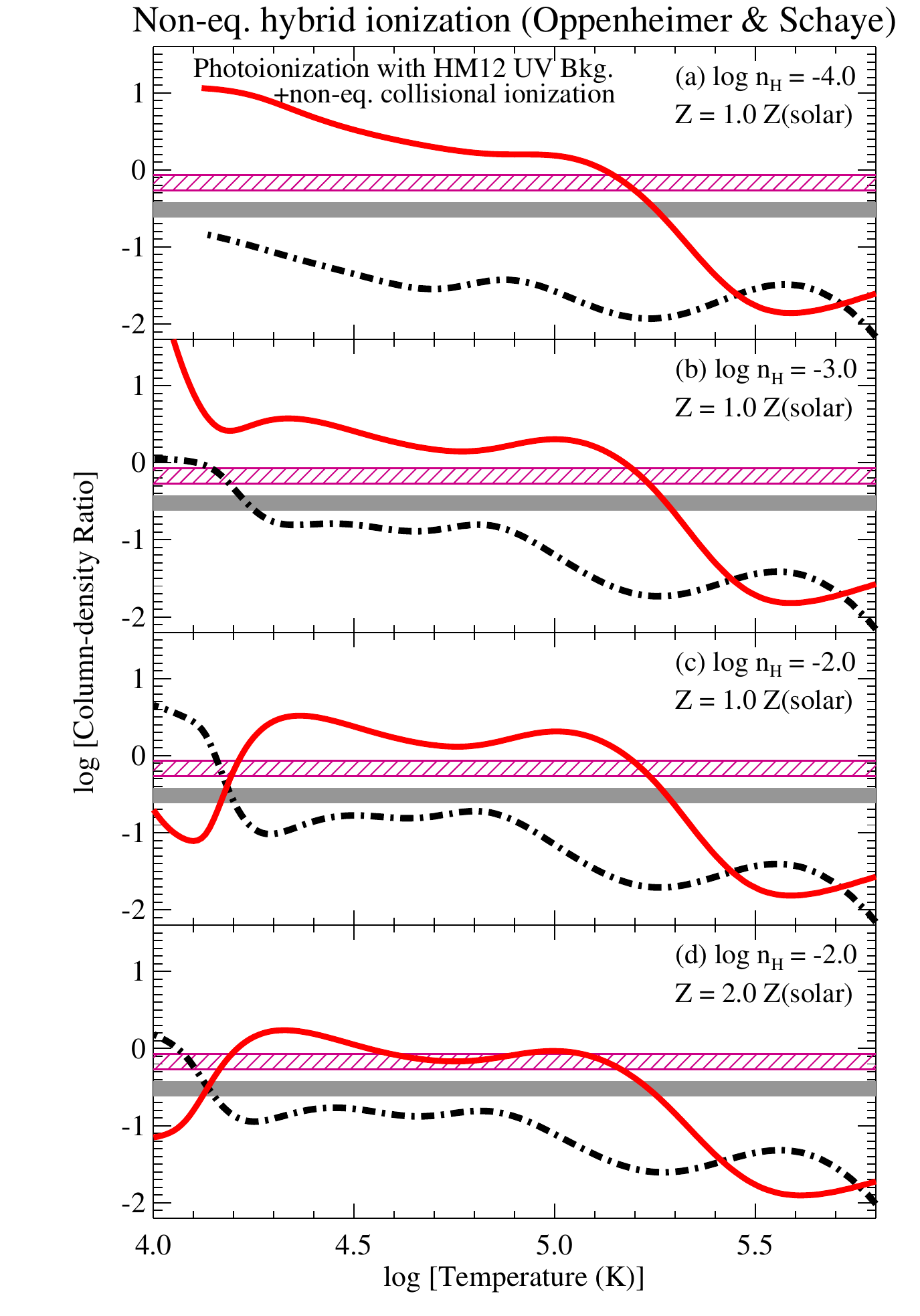}
\caption{Same as Figure~\ref{fig:gnat07compare} but compared to the non-equilibrium cooling gas models of \citet{oppenheimer13}, which include photoionization by the UV background light (this figure shows the models using the UV background from \citealp{haardt12}) in addition to the rapidly cooling collisionally ionized gas. The inclusion of photoionization introduces a density dependence. The panels show the models with (a) log $n_{\rm H} = -4.0$ and $Z = 1.0 \ Z_{\odot}$, (b) log $n_{\rm H} = -3.0$ and $Z = 1.0 \ Z_{\odot}$, (c)  log $n_{\rm H} = -2.0$ and $Z = 1.0 \ Z_{\odot}$, and (d) log $n_{\rm H} = -2.0$ and $Z = 2.0 \ Z_{\odot}$. This figure only shows the isochorically cooling models, which generally fit the observed ratios better than the isobaric runs. \label{fig:opp13compare} }
\end{figure}

However, it is possible to reconcile all of the H1821+643 OA data with a two-phase model if the highly-ionized absorption lines arise in collisionally ionized gas that is out of ionization equilibrium because it is cooling more rapidly than it can recombine, i.e., overionized gas that is colder than expected given its degree of ionization.  To show this, Figures~\ref{fig:gnat07compare} and \ref{fig:opp13compare} compare the observed \SiIV/\CIV\ ratio (horizontal gray band) and  \CIV/\OVI\ ratio (purple-hatched horizontal band) to the rapidly cooling non-equilibrium models of \citet{gnat07} and \citet{oppenheimer13}. 

\subsubsection{Gnat \& Sternberg Non-equilibrium Cooling.}  Starting with the computed ratios from \citet{gnat07} in Figure~\ref{fig:gnat07compare}, we see that the cooling gas can simultaneously match both \SiIV/\CIV\ and \CIV/\OVI\ if the gas has cooled isochorically to log $T \leq$ 4.3 and the gas has relatively high metallicity, $Z \approx 2 \ Z_{\odot}$.  This cool temperature requirement is consistent with the \SiIV\ line widths: the blue dots in Fig.~\ref{fig:gnat07compare}(d) show the \SiIV/\CIV\ ratios plotted at the temperatures that match the theoretical ratio for the five components; the fits require log $T$ = 4.1 -- 4.3.  The isochoric $Z = 1.0 \ Z_{\odot}$ \citet{gnat07} model can also agree with the \SiIV/\CIV\ ratio but predicts a \CIV/\OVI\ ratio that is slightly too high; this small excess could be due to various systematic uncertainties, so this model is also viable.  Likewise, the isobaric $Z = 2.0 \ Z_{\odot}$ model fits \SiIV/\CIV, and \CIV/\OVI\ is close enough to be in contention.  The other isobaric models and the lower-metallicity \citet{gnat07} models fail to simultaneously match both ratios.

\subsubsection{Oppenheimer \& Schaye Non-equilibrium Cooling with Photoionization. \label{sec:opp13model}} Perhaps the most uncomfortable aspect of the successful \citet{gnat07} models in Figure~\ref{fig:gnat07compare} is the high metallicity required for the cooling overionized gas, which is $\gtrsim 5$ times higher than the metallicity usually reported for the low-ionization phases of the outer-Galaxy HVCs \citep{richter01,tripp03,collins07,tripp12}.  As shown in S\ref{sec:lowion_highZ_photo}, such high metallicities are possible as long as all of the \HI\ can be explained, but before embracing high$-Z$ models, it is important to examine how photoionization, in addition to non-equilibrium collisional ionization with rapid cooling, could affect these ratios. We will refer to this combination as ``hybrid'' ionization hereafter.  The addition of photoionization will further alter the ion ratios and radiative gas cooling \citep{efstathiou92,wiersma09}.  

\citet{oppenheimer13} have calculated ion ratios (and other quantities) in the hybrid-ionization scenario assuming photoionization with the UV background models of \citet{haardt01} or \citet{haardt12}.  Figure~\ref{fig:opp13compare} presents the \SiIV/\CIV\ and \CIV/\OVI\ results from \citet{oppenheimer13} for isochorically cooling gas\footnote{The isobarically cooling models from \citet{oppenheimer13} fit the H1821+643 observations poorly, so we focus on the isochoric calculations here.} with photoionization by the \citet{haardt12} radiation field, $Z = 1.0 \ Z_{\odot}$ or 2.0 $Z_{\odot}$, and log $n_{\rm H} = -4.0, \ -3.0$ or $-2.0$. As in the collisionally ionized models of \citet{gnat07}, the lower-metallicity hybrid models (not shown) from \citet{oppenheimer13} cannot simultaneously fit the \SiIV/\CIV\ and \CIV/\OVI\  ratios, so the addition of photoionization does not remove the need for a metal-rich high-ionization phase.  However, the hybrid models do somewhat alleviate the degree to which the gas must be metal enriched: Figure~\ref{fig:opp13compare} shows that the hybrid models attain good fits to both \SiIV/\CIV\ and \CIV/\OVI\ with $Z = 1.0 \ Z_{\odot}$, whereas the \citet{gnat07} collisional models formally fit the data better with $Z = 2.0 \ Z_{\odot}$.

Figure~\ref{fig:opp13compare} provides another insight about the OA ionization in the hybrid scenario.  The addition of photoionization introduces a density dependence since the impact of photoionization depends on the ionization parameter.  From Figure~\ref{fig:opp13compare} we see that the lower-density hybrid models fit the data poorly; only the log $n_{\rm H} = -2.0$ models simultaneously match both ratios.  The non-equilibrium collisional ionization process is therefore predominant -- increasing the importance of photoionization (by lowering the gas density) degrades the fits.  This does not necessarily indicate that photoionization can be neglected entirely (compare Figs.~\ref{fig:gnat07compare} and \ref{fig:opp13compare}), but the gas density is constrained to be relatively high in this model.  The analysis in this section is somewhat grainy because \citet{gnat07} and \citet{oppenheimer13} have only published models with discrete (and somewhat widely separated) values of model parameters such as $Z$, $n_{\rm H}$, and ionizing radiation field shape.  Thus we cannot investigate whether models with $n_{\rm H} > 10^{-2}$ would fit the observations better (or worse).  Likewise, it is not possible to examine how these models would change if the \citet{fox05} ionizing flux field were used instead of \citet{haardt12}.  Nevertheless, the available models in Figure~\ref{fig:opp13compare} show that hybrid/non-equilibrium ionization does provide compelling agreement with the observations.  It would be interesting to carry out a more detailed parameter study with this model in the future.

\begin{figure}
\includegraphics[width=8cm]{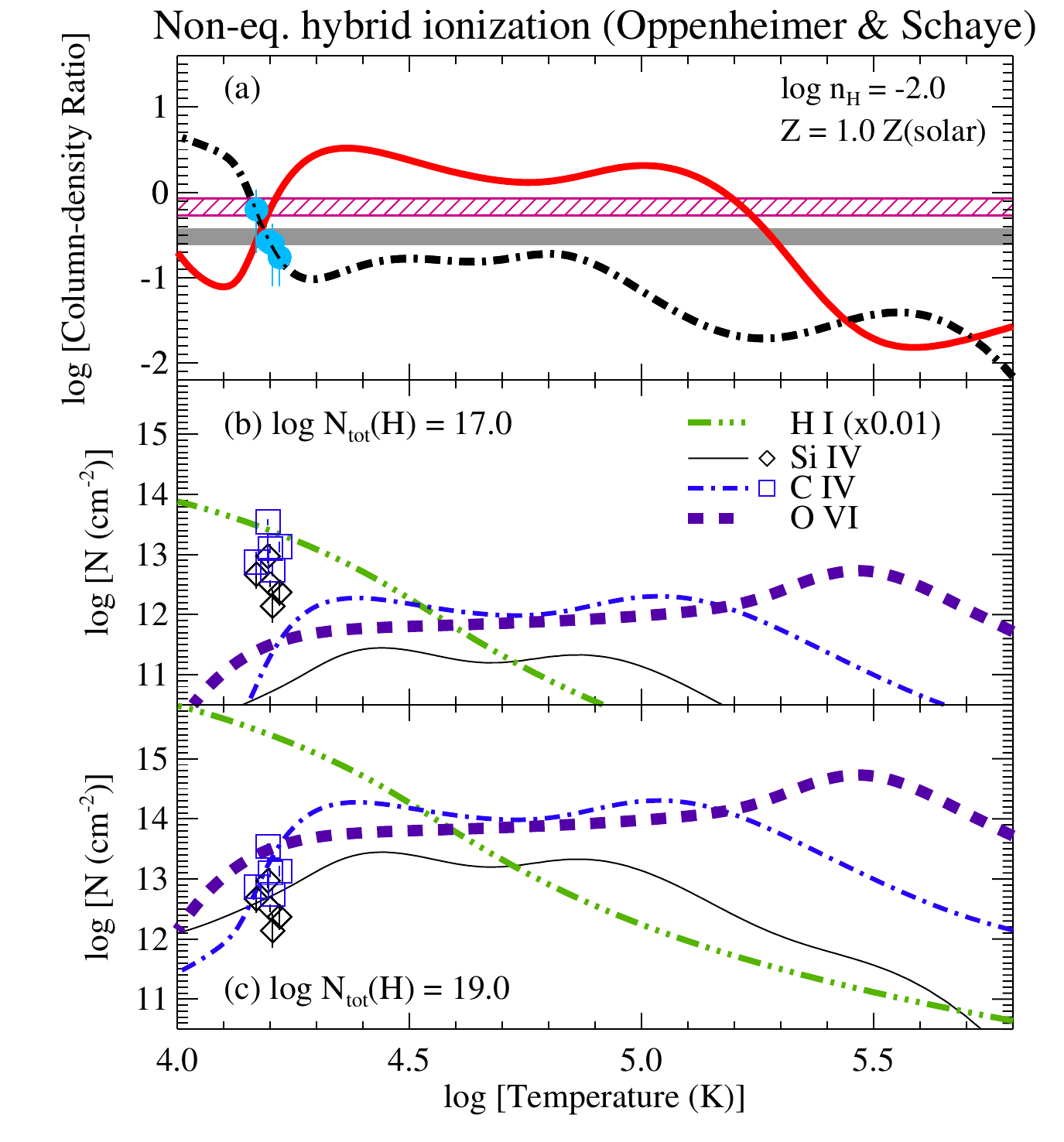}
\caption{ Comparison of the hybrid model from \citet{oppenheimer13}, with log $n_{\rm H} = -2.0$ and $Z = 1.0 \ Z_{\odot}$, to the H1821+643 high-ionization stage measurements: (\textsl{a}) The  \SiIV/\CIV\ and \CIV/\OVI\ ratios (as in Fig.~\ref{fig:opp13compare} but with the measured \SiIV/\CIV\ ratios in the five individual OA components indicated with blue dots at the best-fitting temperature). The lower panels show the column densities of \SiIV, \CIV, and \OVI\ (see panel-\textsl{b} legend) with (\textsl{b}) log $N_{\rm tot}$(H) = 17.0 and (\textsl{c}) log $N_{\rm tot}$(H) = 19.0.  The observed  $N$(\SiIV) and $N$(\CIV) are plotted at the same temperatures as the corresponding blue dots in (\textsl{a}).  \label{fig:bestopp}}
\end{figure}

So far this discussion has focused on the high-ion ratios, but it is also useful to consider just the column densities predicted by these models.  There are several parameters that can be adjusted to fit the data with these models (e.g., metallicity, temperature, density, and isochoric vs. isobaric cooling). The ratios constrain these parameters. To predict the high-ion column densities, an additional parameter must be set: the total hydrogen column density.  Figure~\ref{fig:bestopp} shows the \SiIV, \CIV, and \OVI\ column densities precicted by the best-fitting \citet{oppenheimer13} model with $Z = 1.0 \ Z_{\odot}$ and log $n_{\rm H} = -2.0$, assuming two different values for the total hydrogen column: log $N_{\rm tot}$(H) = 17.0 (panel b) and log $N_{\rm tot}$(H) = 19.0 (panel c). The ratios are shown in panel (a) of Fig.~\ref{fig:bestopp} with blue dots showing the temperatures that best fit the ratios in the five OA components.  

From Figure~\ref{fig:bestopp}b, we see that if log $N_{\rm tot}$(H) = 17.0 in a single parcel of highly-ionized gas from the model, then $\approx 10 - 200$ individual parcels of non-equilibrium highly ionized gas are required to match the observed $N$(\SiIV) and $N$(\CIV) in each OA component (i.e., the model column densities in one parcel are factors of $10-200$ lower than the observed columns).  Similar numbers of parcels are also needed to produce the observed \CIV/\OVI\ ratio. This number of individual parcels of highly ionized gas is close to the number of cloudlets required for the low-ionization phase in the \citet{mccourt18} shattering model (see S\ref{sec:lowphase_physcond}), so if the high ions originate in interface layers on the surface of the low-ionization cloudlets, then this model can naturally explain both the low-ionization and high-ionization phases with a single coherent concept.  This model also explains the narrow width of the observed \SiIV\ components.

With this model for the high ions, the combined \HI\ from the low- and high-ionization phases (using the $Z = 1.0 \ Z_{\odot}$ model for the low-ionization gas, which makes the metallicity of the low- and high-ionization phases the same) is roughly in agreement with the $N$(\HI) indicated by the 21cm emission from the OA.   The green triple-dot-dash line in Fig.\ref{fig:bestopp} shows the $N$(\HI) predicted by this model (scaled by 0.01 for plotting convenience).  We see that at the temperature indicated by the \SiIV/\CIV\ ratios, roughly 200 high-ion parcels would lead to a total $N$(\HI) (summed over all highly-ionized parcels) that is comparable to the $N$(\HI) in the low-ionization phase, and the total low phase + high phase $N$(\HI) is close to that indicated by the 21cm data.  Less than $\approx 200$ highly-ionized parcels would fall somewhat short of the 21cm $N$(\HI), but this could be due to various limitations of the available models.  For example, we have used a $Z = 1.0 \ Z_{\odot}$ model from \citet{oppenheimer13} because it is available, but a somewhat lower metallicity, say $Z = 0.8 \ Z_{\odot}$, might fit comparably well but require a higher $N$(\HI), which would be more in line with the 21cm emission.  We cannot check this because a $Z = 0.8 \ Z_{\odot}$ run is not publicly available currently. Again, it would be useful to further explore these \citet{oppenheimer13} models with a finer grid of parameters and different radiation fields.

\begin{figure}
\includegraphics[width=8cm]{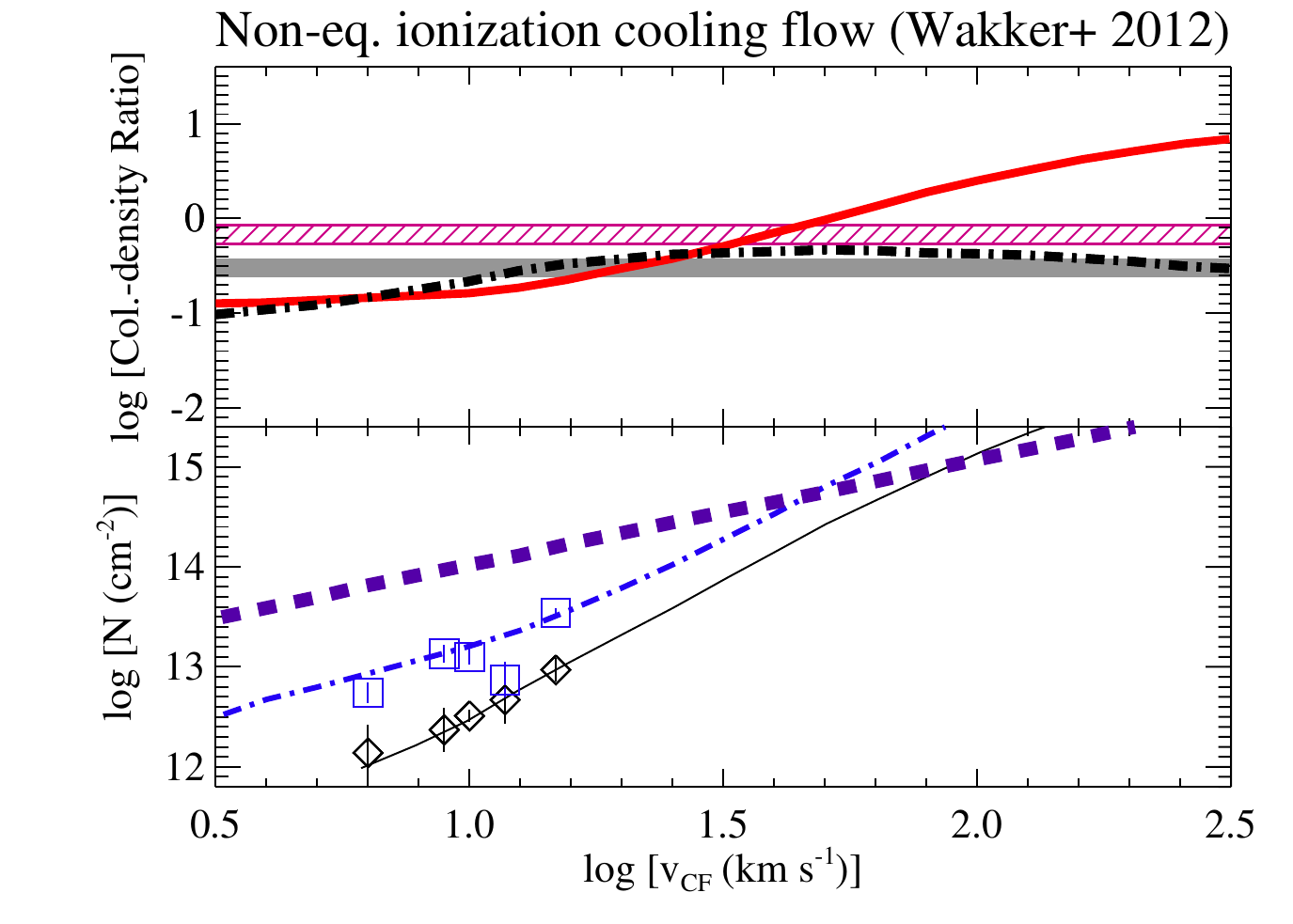}
\caption{Comparison of the non-equilbrium isochoric cooling flow model from \citet{wakker12}, including self-photoionization, to the observed \SiIV/\CIV\ and \CIV/\OVI\ ratios (upper panel) and the observed individual component column densities (lower panel).  The curves and symbols represent the same quantities as in Figures~\ref{fig:gnat07compare} and \ref{fig:bestopp}.  In this model, the column densities are predicted as a function of the flow velocity $\emph{v}_{\rm CF}$. In the lower panel, the observed component column densities of \SiIV\ and \CIV\ are plotted at the $\emph{v}_{\rm CF}$ value that matches the observations; with the exception of component 4, this model is in reasonable agreement with $N$(\SiIV) and $N$(\CIV).  As shown in the upper panel, at the $\emph{v}_{\rm CF}$ values that agree with the \SiIV\ and \CIV, the model predicts too much \OVI\ by a factor of $\approx 5$.  \label{fig:wak12cf} }
\end{figure}

The possibility that highly-ionized species like \SiIV, \CIV, \textsc{N~v}, and \OVI\ arise in interface layers on the surfaces of low-ionization clouds has been studied for many years with a variety of scenarios such as turbulent mixing layers or conductive interfaces \citep[][and references therein]{werk16}.  These theoretical models have often been criticized or dismissed because they predict column densities (in an \textsl{individual} surface layer) that are much lower than the observed columns \citep[e.g.,][]{wakker12,werk16}.  However, this bug may actually be a feature: if an absorption ``component'' is composed of hundreds of cloudlets, as suggested above, then many of these theoretical interface models might achieve good agreement with the observations when the surface layers on all of the cloudlets are added up.  

The model in Figure~\ref{fig:bestopp}b has considerable appeal, but as shown in Figure~\ref{fig:bestopp}c, it is also possible to fit the observed \SiIV\ and \CIV\ columns with a single monolithic gas cloud with log $N_{\rm tot}$(H) $\approx$ 19.0.  As in the photoionization models for the highly-ionized gas, this monolithic cloud would have a large size compared to the low-ionization gas phase, but with two important differences. First, while the purely photoionized models of S\ref{sec:photohighion} required an additional (and awkwardly much larger) phase to explain the \OVI, the non-equilibrium collisionally ionized model in Figure~\ref{fig:bestopp}c produces about the right amount \OVI\ in the same phase as the \SiIV\ and \CIV; an additional very large region of \OVI\ without \SiIV\ and \CIV\ is not required. Second, since the densities in the highly-ionized phase are significantly higher in the non-equilibrium collisional model, the low-ionization and high-ionization phases could be in thermal pressure equilibrium, so this model does suffer from the stability problems encountered in the purely photoionized scenario.

\subsubsection{Wakker et al. Non-equilbrium Cooling Flows. \label{sec:wakkerflows}}  Building on the work of \citet{benjamin94}, a different rapidly cooling and non-equilibirum cooling flow model for the highly-ionized UV species is found in \citet[][]{wakker12}. In this model, gas flows from a hotter (unseen) phase with log $T > 6$ into a thermally unstable region with velocity $\emph{v}_{\rm CF}$ and cools as the material flows away from the hot point of origin.  Interestingly, in this model, the predictions are insensitive to density or gas metallicity (although the conditions must enable thermal instability in the first place) and largely depend on $\emph{v}_{\rm CF}$.  Figure~\ref{fig:wak12cf} compares the \citet{wakker12} isochorically cooling model, including self-photoionization, to the H1821+643 OA measurements (using the same curves and symbols as in Fig.\ref{fig:bestopp}). In the lower panel of Figure~\ref{fig:wak12cf}, the observed \SiIV\ and \CIV\ column densities are plotted at the values of $\emph{v}_{\rm CF}$ that best fit the \SiIV/\CIV\ in each component, with the exception of component 4.  This cooling flow fits the observed \SiIV\ and \CIV\ columns and ratios reasonably well in components 1, 2, 3, and 5; in component 4, the measured \SiIV/\CIV\ ratio requires a much higher $\emph{v}_{\rm CF}$ which, in turn, predicts $N$(\SiIV) and $N$(\CIV) that are much too high.  However, at the $\emph{v}_{\rm CF}$ values that fit \SiIV\ and \CIV, the model predicts \OVI\ columns that are a factor of $\approx$ 5 too high. But, this comparison has a caveat: while the model accounts for ``self-photoionization'' by flux from the hot gas upstream, it does not include photoionization by the external ionizing flux (i.e., flux emerging from the Milky Way and flux from the cosmic UV background light).  The addition of photoionization from the external flux impinging on the cooliing flow could significantly alter the column densities and ratios.  It would be useful to revisit this model including this additional photoionization source. While the \citet{wakker12} model is usually referred to as a ``cooling flow'', in this context it is effectively an implementation of precipitation. The model begins with a region where hot gas is precipitating into a cooler region, and as it cools, it is likely to fall down onto the disk.

\subsubsection{$t_{\rm cool}/t_{\rm ff}$ in the Mixing Gas.} As highlighted in previous sections, the $t_{\rm cool}/t_{\rm ff}$ ratio is a useful diagnostic of circumgalactic gases.  The physical conditions derived in S\ref{sec:lowphase_physcond} indicate that the low-ionization phases have low $t_{\rm cool}/t_{\rm ff}$ values, but what about the putative mixing gas that is transitioning from the hot phase into the cool phase and thereby potentially driving precipitation?  The OA absorption data in the H1821+643 spectrum provide an opportunity to study this mixing gas.  The \SiIV\ absorption is correlated with the \SiII\ and other low-ionization lines. As shown in Figs.~\ref{fig:high_nav} and \ref{fig:si2si4fitparams}, with the exception of component 2, the \SiIV\ and \SiII\ data have a similar, though not identical, component structure with similar $b-$values, velocity centroids, and \SiIV/\SiII\ ratios.  And yet, the \SiIV\ clearly comes from a different phase.  The \CIV\ and \OVI\ profiles have the same overall shape as the \SiIV\ profile (Fig.\ref{fig:high_nav}) and likewise must arise from a more highly ionized phase.  The \CIV\ and \OVI\ appear to be smoother than the \SiIV\ profiles; this could be due to thermally broadening (the components could be more smeared together in the profiles of lower-mass species), but it is also possible that this is simply a resolution effect since the \SiIV\ is recorded at the highest (E140H) spectral resolution and has better S/N than the \CIV\ E140H data.  At any rate, the overall similarity of these ions suggests that these high ions originate in the mixing gas.  Of course, there could be ionization gradients in the mixing regions.

\begin{figure*}
\includegraphics[width=16cm]{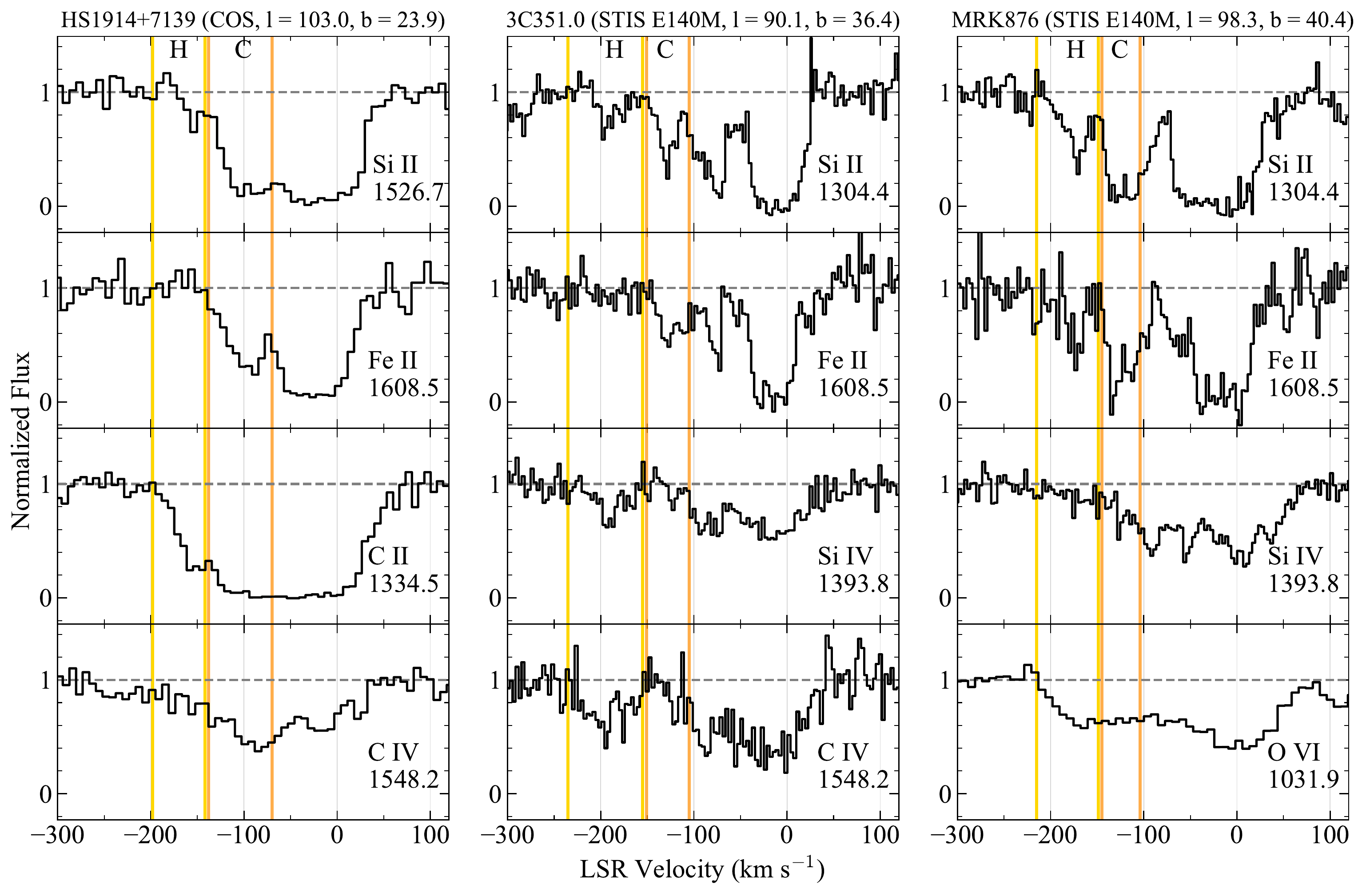}
\includegraphics[width=16cm]{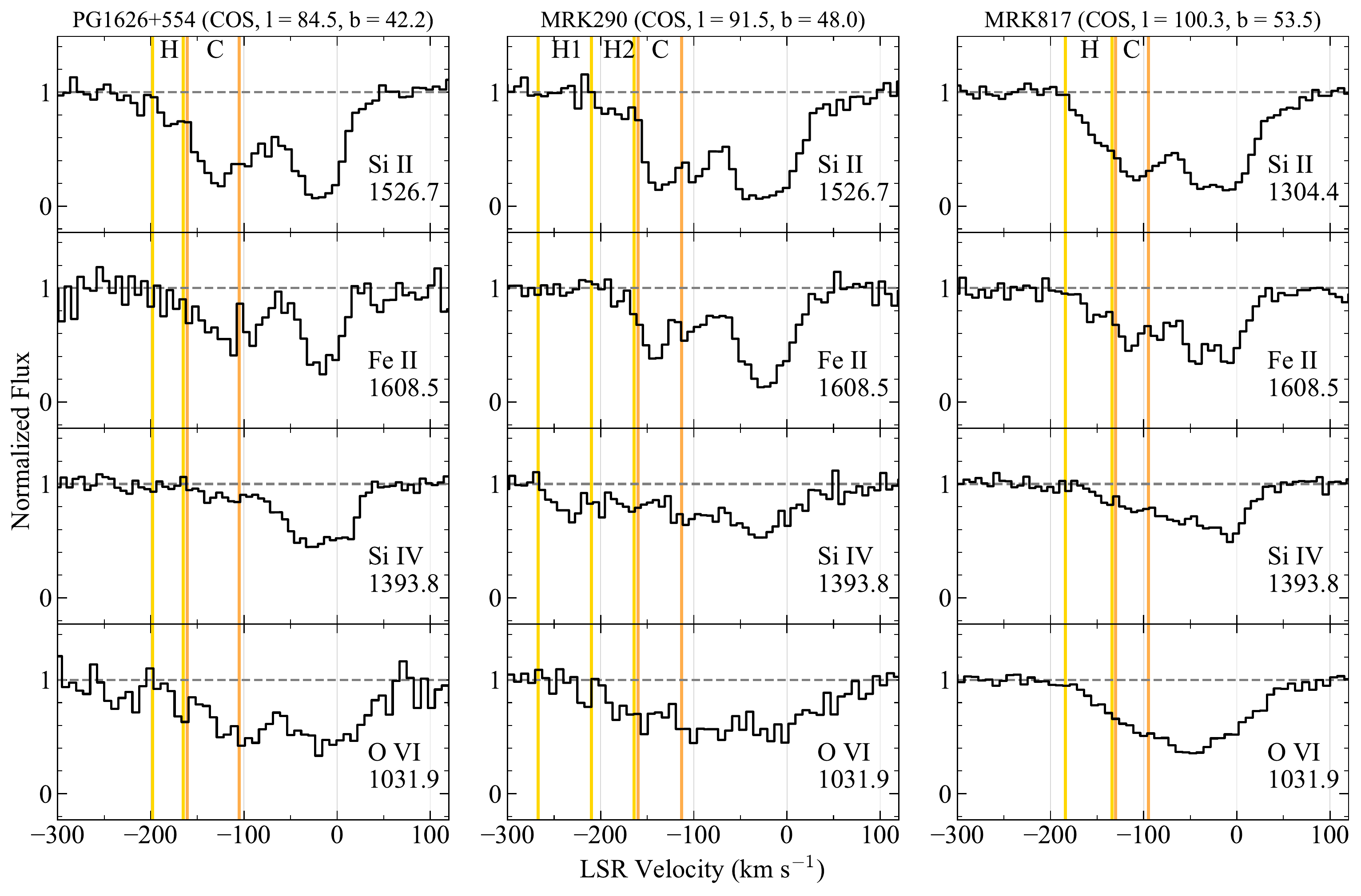}
\caption{Selected continuum-normalized absorption profiles observed toward the ancillary targets (see Tab.~\ref{tab:targets} and Fig.~\ref{fig:outergalmap}) vs. LSR velocity. The sightline target is labeled at the top of each stack of four absorption profiles, and the stacks are presented in order of increasing Galactic latitude from upper left to lower right.  Depending on data availability, for each sightline the \SiII\ and \FeII\ absorption profiles are compared to \SiIV, \CIV, and/or \OVI.  In each stack, the region between orange lines labeled with a `C' at the top is in the velocity range of Complex C, and the region between yellow lines labeled with an `H' is in the velocity range of the high-velocity ridge (see text, S\ref{sec:hvc_summary}).}
\label{fig_suppstacks}
\end{figure*}

% TABLE: Column densities for other sightlines
\ctable[
caption={Column Densities from Ancillary Sightlines},
label={tab:lowion_othertargs},
doinside=\footnotesize
]{llccc}{
\tnote[a]{OA = Outer Arm, C = main Complex C, HVR = Complex C/OA high-velocity ridge. The velocity ranges that are used in this paper to delineate Complex C and the OA are indicated in Fig.~\ref{fig_suppstacks}. If more than one component is present within the velocity range, then the column density in this table is the summed column density from all of the components in the velocity range.}
\tnote[b]{Three-sigma upper limit.}}
{\FL
Sightline          & Cld.\tmark[a] & log $N$(Si~II) & log $N$(S~II) & log $N$(Fe~II) \ML
HS1914+7139 & OA                   &  14.60$\pm$0.03 & 14.50$\pm$0.07 & 14.39$\pm$0.06 \NN
3C 351.0         & C                      & 13.82$\pm$0.05 & $<$13.9\tmark[b] &  13.90$\pm$0.15 \NN
MRK876          & C                      & 14.43$\pm$0.08 & 14.07$\pm$0.10 &  14.47$\pm$0.02 \NN
PG1626+554  & C                      & 14.19$\pm$0.02 & 14.30$\pm$0.06  & 14.16$\pm$0.08 \NN
MRK290         & C                      & 14.65$\pm$0.05 & 14.31$\pm$0.02 & 14.21$\pm$0.04 \NN
MRK817         & C                      & 14.25$\pm$0.05 & 14.42$\pm$0.03 & 13.98$\pm$0.03  \NN \hline
 \                     & \                        & \                          & \                          & \                           \NN \hline
  \                    & \                        & log $N$(Si~IV) & log $N$(C~IV) & log $N$(O~VI) \NN
HS1914+7139 & OA                   & 13.34$\pm$0.01 & 13.52$\pm$0.03 & NA \NN
  \                     & HVR                & 13.31$\pm$0.02 & 13.48$\pm$0.05 & NA \NN
  3C 351.0       & C                   & 12.64$\pm$0.10 & 13.32$\pm$0.08 & NA \NN
  \                    & HVR                 & 12.81$\pm$0.07 & 13.48$\pm$0.05 & NA \NN
  MRK876       & C                      & 13.08$\pm$0.03 & 13.51$\pm$0.06 & 13.88$\pm$0.02 \NN
  \                    & HVR                & 12.53$\pm$0.08 & 13.12$\pm$0.11 & 13.81$\pm$0.02 \NN
  PG1626+554  & C                   & 12.63$\pm$0.06 & 13.37$\pm$0.05 & 14.07$\pm$0.04 \NN
  MRK290       & C                     & 13.04$\pm$0.04 & 13.62$\pm$0.02 & 14.01$\pm$0.03 \NN
    \                  & HVR1              & 12.94$\pm$0.05 & 13.50$\pm$0.03 & $<$13.27 \NN
    \                  & HVR2              & 12.44$\pm$0.08 & 12.87$\pm$0.08 &  13.42$\pm$0.07 \NN
    MRK817     & C                     & 12.66$\pm$0.03 & 13.32$\pm$0.03 & 13.76$\pm$0.02 \NN
     \                 & HVR                & 12.39$\pm$0.07 & 12.89$\pm$0.07 & 13.49$\pm$0.04 \LL        
}

\begin{figure*}
\includegraphics[width=5.8cm]{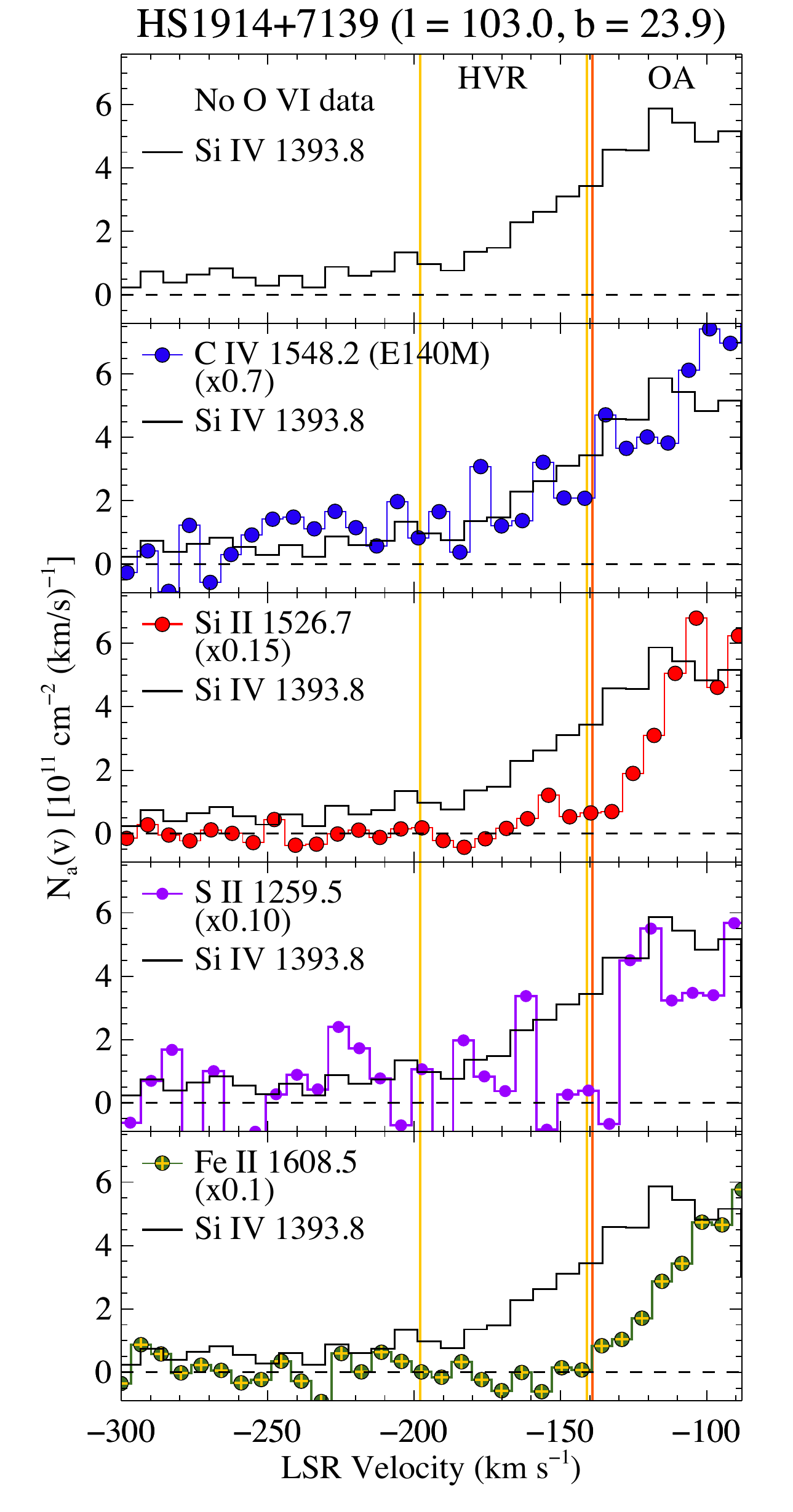}
\includegraphics[width=5.9cm]{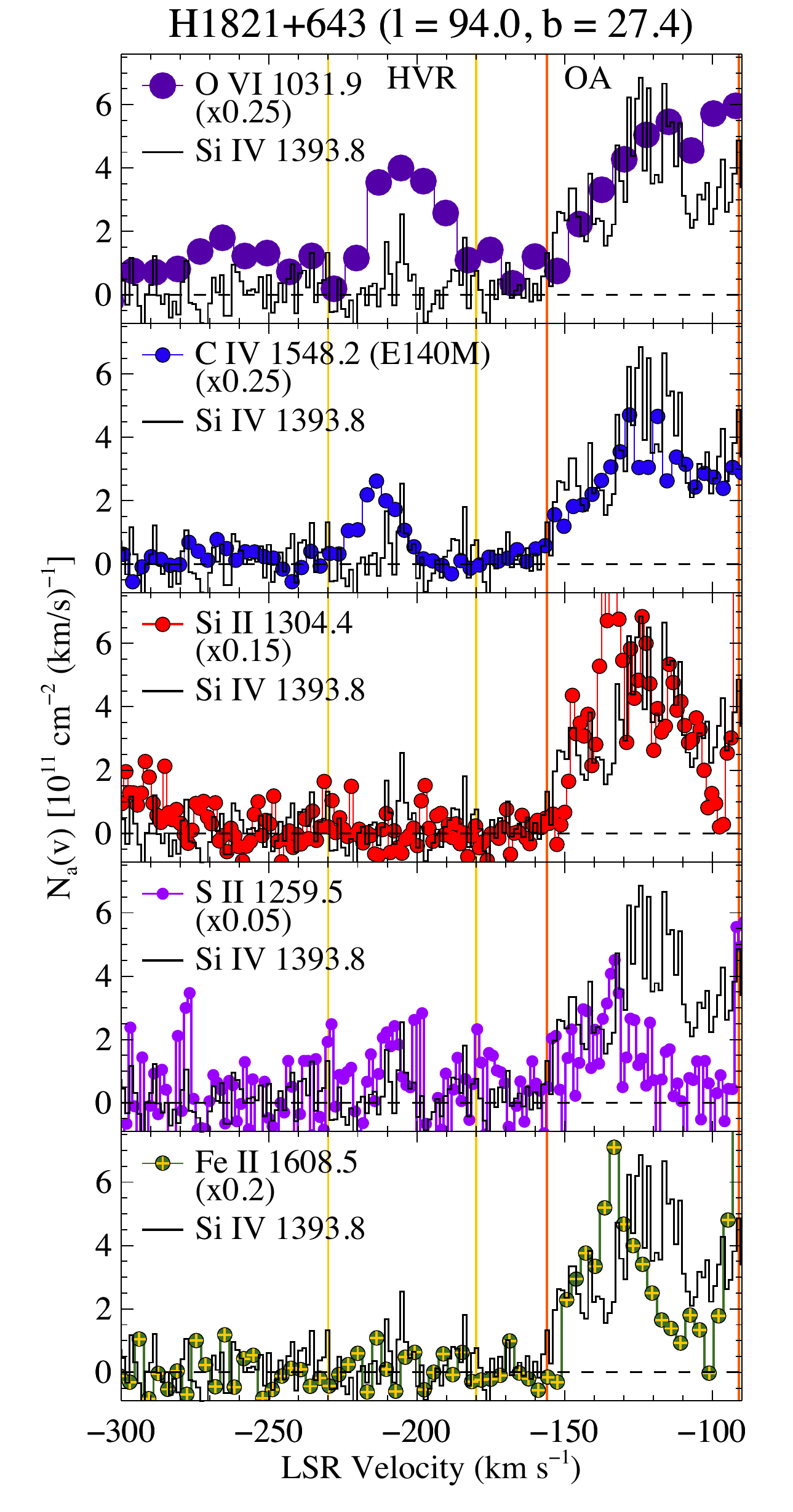}
\includegraphics[width=5.8cm]{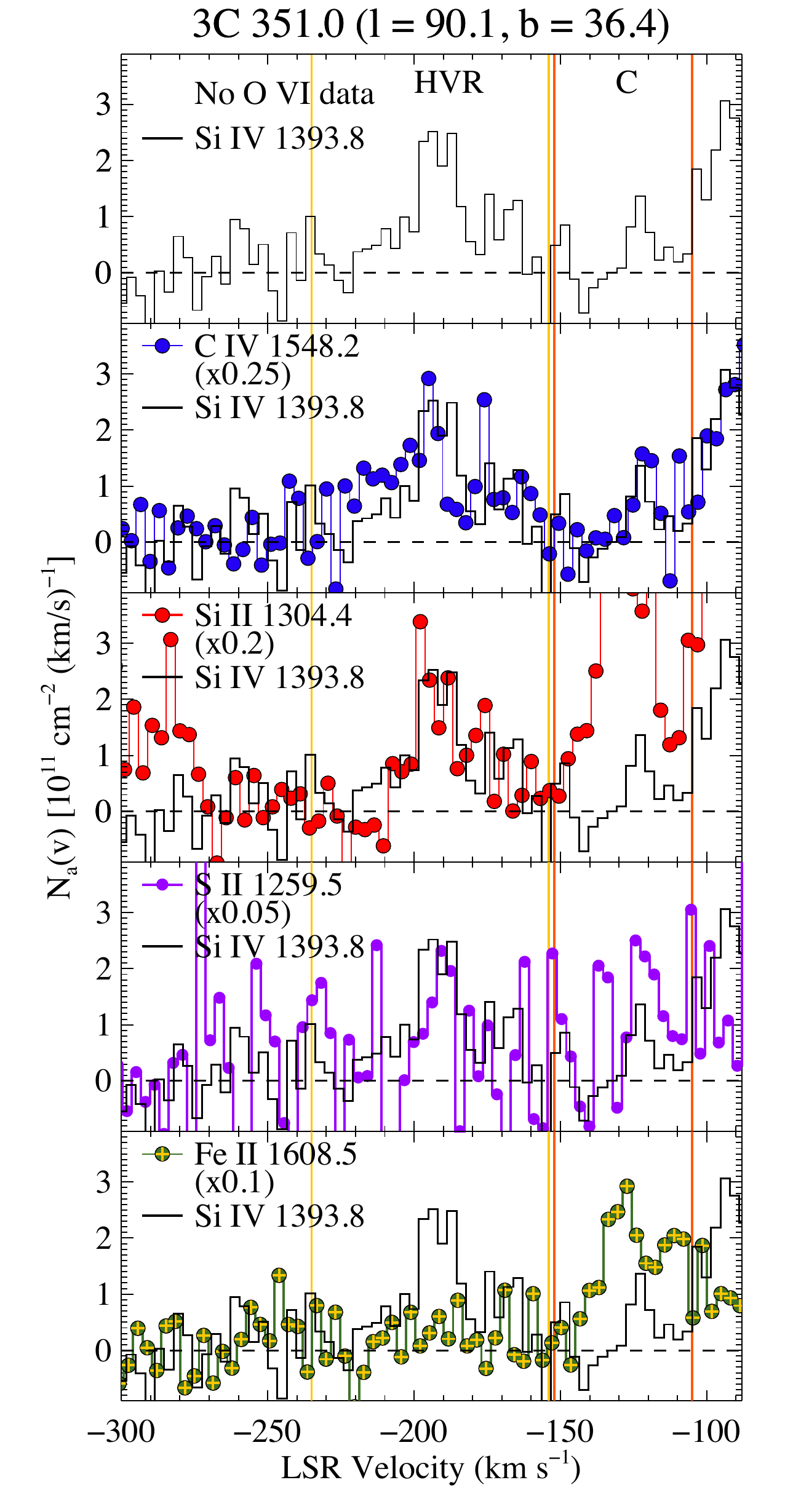}
\caption{Apparent column density profiles (\S\ref{sec:navmethod}), from the ancillary sightline spectra, of low-ionization stages (\SII, \SiII, and \FeII) and high ions (\CIV\ and, when available, \OVI) overplotted on the \SiIV\ $N_{\rm a}(\emph{v})$ profile.  The species plotted, and any scale factors applied to the profiles for purposes of comparison, are shown by the legend in each panel. Vertical lines delineate the regions affiliated with Complex C, the Outer Arm, and the High-Velocity Ridge (labeled `C', `OA', and `HVR'). Each stack shows the profiles from a different sightline, as labeled at the top of the stack, and the stacks are presented in order of increasing latitude. \label{fig:anc_navstacks} }
\end{figure*}

\setcounter{figure}{15}
\begin{figure*}
\includegraphics[width=5.8cm]{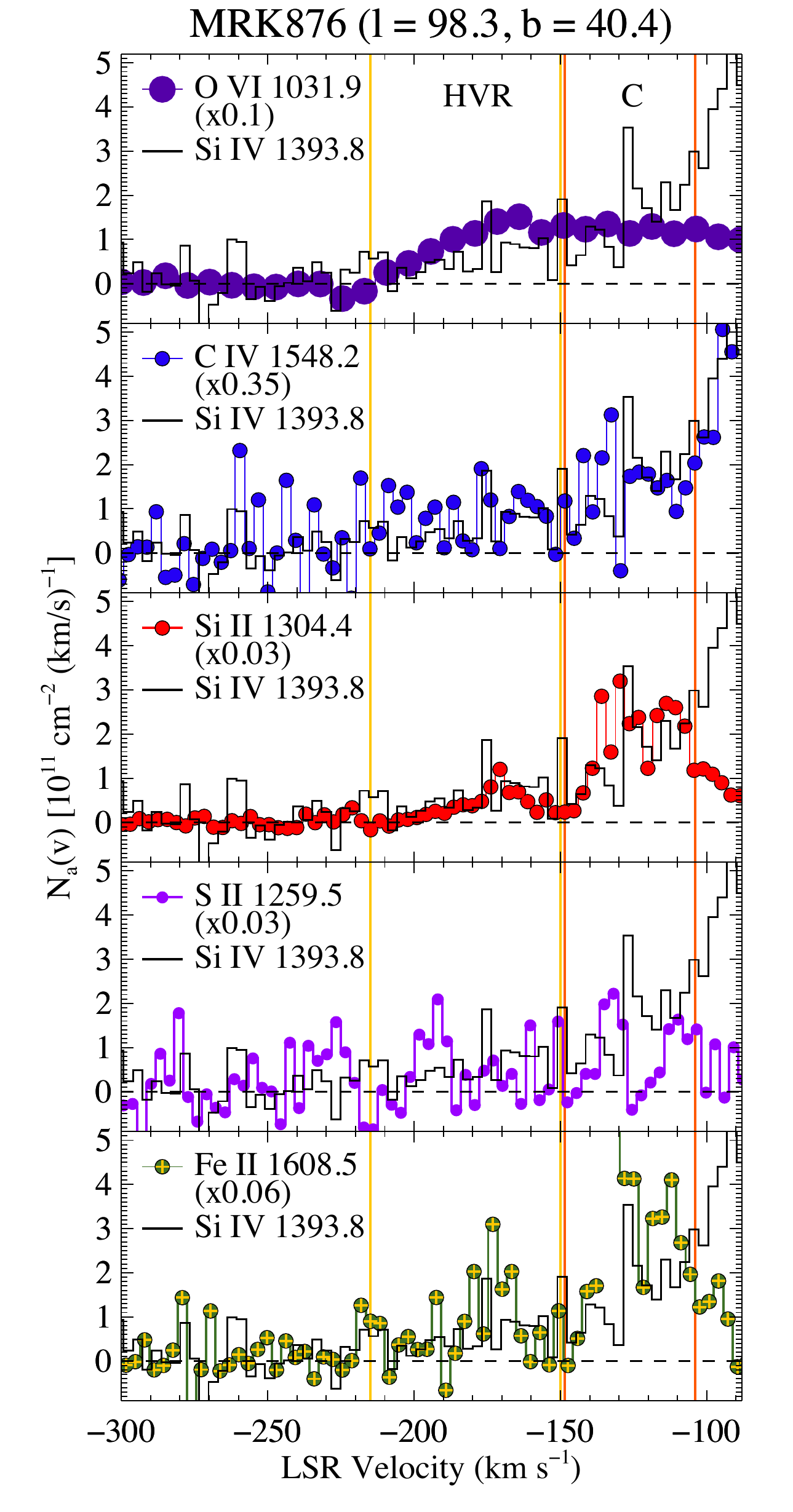}
\includegraphics[width=5.8cm]{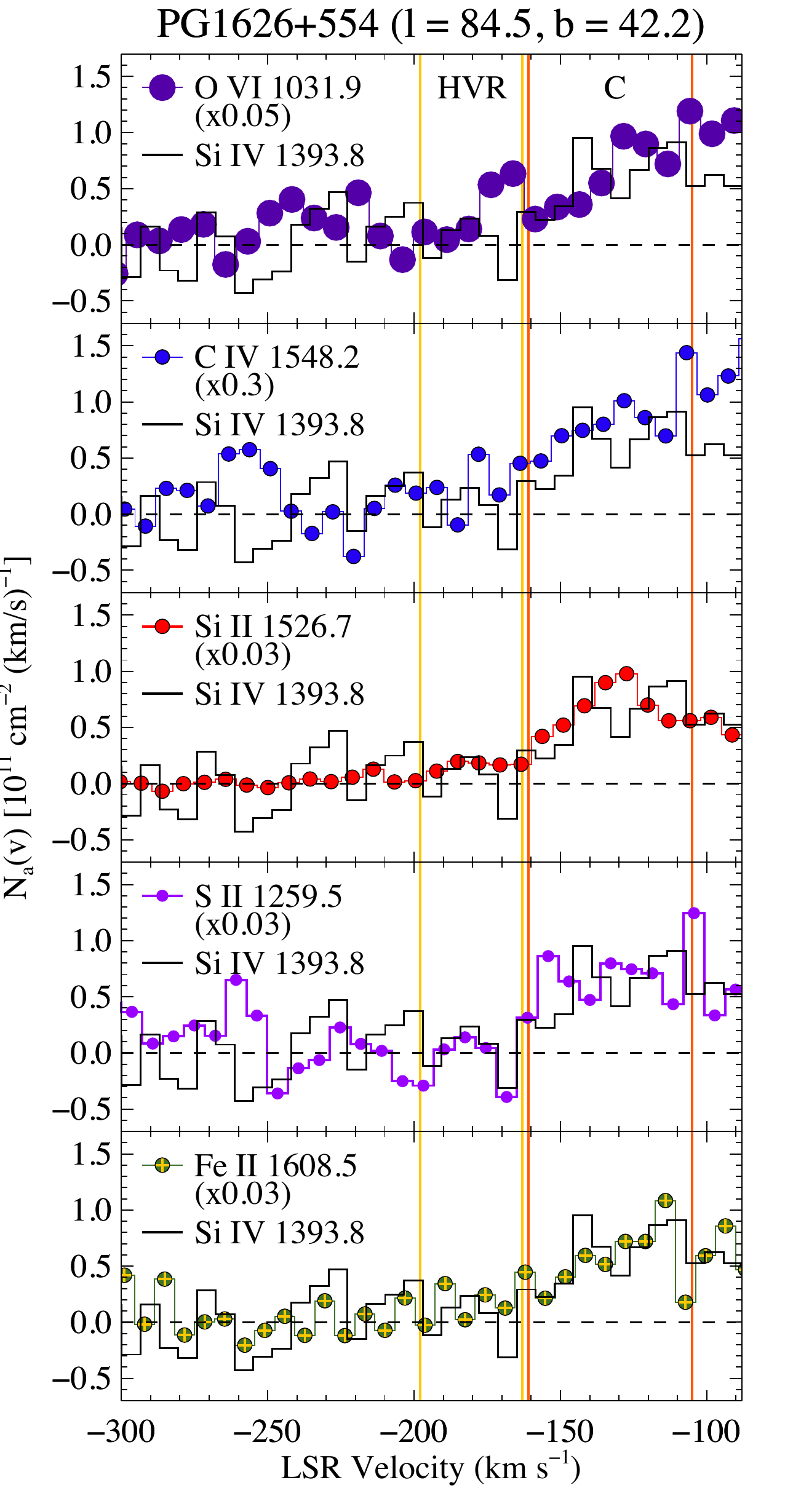}
\includegraphics[width=5.8cm]{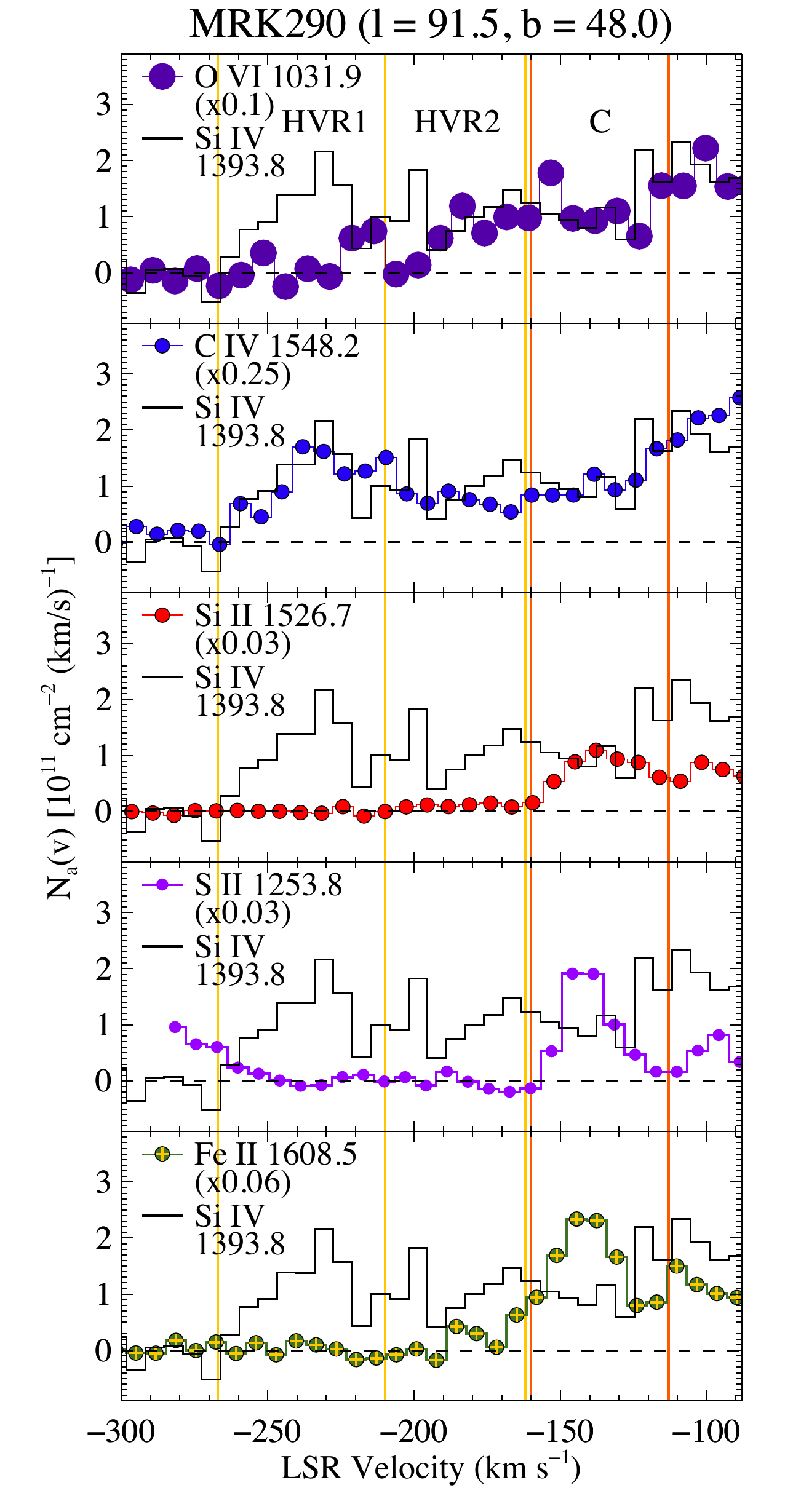}
\caption{(continued)}
\end{figure*}

\setcounter{figure}{15}
\begin{figure}
\includegraphics[width=7.6cm]{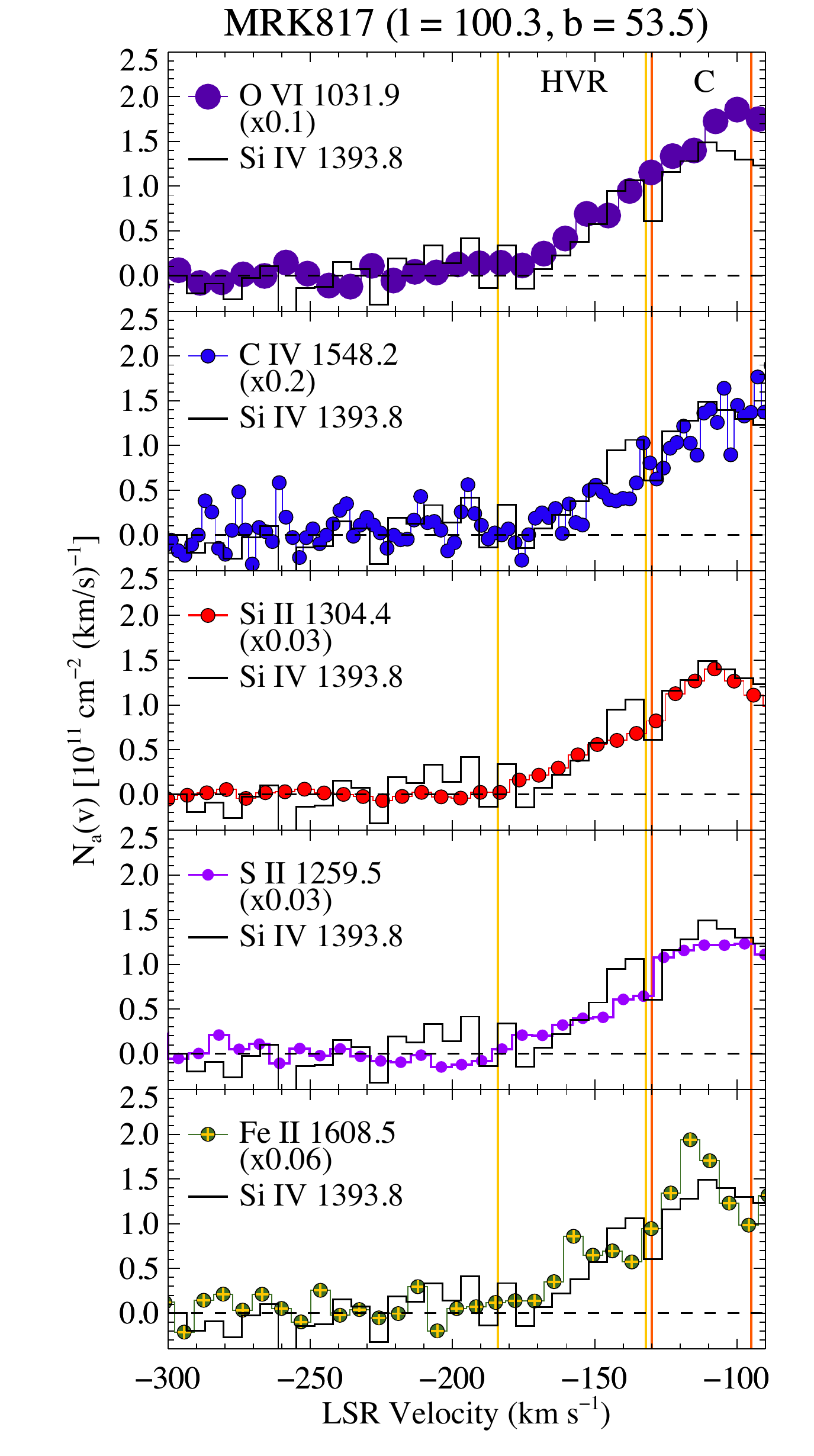}
\caption{(continued)}
\end{figure}

The \SiIV\ line widths (Table~\ref{tab:vp_si2_si4}) indicate temperature upper limits that range from log $[T (K)] < 4.2$ to log $[T (K)] < 5.1$ for the five components. If the \SiIV\ gas is in pressure equilibrium with the \SiII\ phase in the component that is closest in velocity, then the \SiIV\ temperature limits imply density constraints ranging from  $n$(\SiIV\ phase) $> 0.2 n$(\SiII\ phase) to $n$(\SiIV\ phase) $\approx n$(\SiII\ phase).  In turn, using cooling rates in the \SiIV\ phase from \citet{oppenheimer13}, these densities imply that $t_{\rm cool}/t_{\rm ff} \ll$ 1 in the more highly-ionized mixing gas. Thus this reasoning suggests that the mixing material is also rapidly cooling and condensing into the lower-ionization phase. 

The \SiIV\ phase might not be in pressure equilibrium with the \SiII\ phase, but even if the phases are not in pressure equilibrium, the best-fitting models from \citet{oppenheimer13} shown in Figs.\ref{fig:opp13compare} - \ref{fig:bestopp} still indicates that the highly-ionized gas is rapidly cooling and condensing.  Based on the measured \SiIV/\CIV\ and \CIV/\OVI\ ratios, the best \citet{oppenheimer13} models favor log $n_{\rm H} \gtrsim -2$ and log $T \approx $ 4.2.  Using these physical conditions and the cooling rates from the same \citet{oppenheimer13} models, we find that $t_{\rm cool}/t_{\rm ff} \approx$ 0.003 in the mixing highly-ionized gas.  Thus this model also indicates that the mixing and more ionized gas is also rapidly cooling and condensing.

\section{Ancillary Sightlines}
\label{sec:ancmeas}

The previous sections have argued that the OA gas in the direction of H1821+643 is rapidly cooling ``misty'' multiphase plasma.  These measurements can be used to test and constrain CGM physics theories.  In addition, S\ref{sec:hvc_summary} noted many similarities between the OA and HVC Complex C, and these HVCs may have a common origin.  In this section we examine the OA/C high-velocity absorption lines in six additional sightlines near the H1821+643 sightline, including one additional direction through the OA and five sightlines through Complex C (see Table~\ref{tab:clouds} and Figure~\ref{fig:outergalmap}).  The data for these ancillary sightlines have somewhat lower spectral resolution (7 $-$ 20 \kms).  While the lower spectral resolution could mask detailed component structure within the absorption profiles, it is nevertheless adequate to probe whether these sightlines exhibit similar properties as the H1821+643 data.  We will see below that the ancillary data are indeed similar in the following ways: (1) there is a clear kinematical correlation between the low-ionization and high-ionization lines, (2) the low-ionization lines of \SiII\ and \FeII\ usually require some depletion by dust, and (3) the high-ion ratios are similar to the H1821+643 ratios and are best explained by rapidly-cooling non-equilibrium ionization models.

\subsection{Ancillary-Sightline Absorption Spectra}

In order of increasing sightline latitude, Figure~\ref{fig_suppstacks} shows selected absorption lines detected in the ancillary-target spectra.  Each column in this figure presents the absorption lines from a specific sightline with the name of the background AGN and the sightline Galactic coordinates at the top of the column.  Additional plots of the absorption profiles in these sightlines, as well as profiles from other sightlines through Complex C, can be found in the literature \citep[see, e.g.,][]{richter01,richter17,sembach03,wakker03,tripp03,fox04,collins07}. All of the sightlines in Figure~\ref{fig_suppstacks} clearly show absorption in the high-velocity ridge (labeled `H') as well as the main Complex C (labeled `C').  In some directions the HVR and C are fairly cleanly separated, while in other cases the HVR and C are strongly blended and overlapping.  The two ancillary sightlines with higher-resolution STIS E140M data (3C 351.0 and MRK876) reveal multiple components within both the HVR and Complex C.  The rest of the spectra do not show the same amount of intricate multicomponent structure, but the lower resolution of the remaining spectra (recorded with COS and FUSE) is often insufficient to clearly show such profile structure.

Figure~\ref{fig:anc_navstacks} shows the \nav\ profiles of selected species and transitions detected in these spectra.  Each column in this figure presents the \nav\ profiles from a single sightline; in each panel the \SiIV\ 1393.8 \AA\ profile is shown as a fiducial, and the \nav\ profile of another species is scaled and overlaid on the \SiIV\ profile to show similarities or differences and to visually show various column-density ratios (indicated by the scale factor in the panel legends). The absorption assigned to the high-velocity ridge is labeled `HVR' and the main Complex C/OA is labeled `C' or `OA'. The boundary between the HVR and C/OA is placed where the main C/OA has an optical depth minimum before increasing again going into the HVR at more negative velocities. In some cases, the minima are more obvious in Figure~\ref{fig_suppstacks} and may be more clear in particular species.  For example, in the MRK817 spectrum the distinction between the HVR and main C absorption is most obvious in the \FeII\ data. H1821+643 is included in Figure~\ref{fig:anc_navstacks} for purposes of comparison and because this figure shows a broader velocity range than Figure~\ref{fig:high_nav}, which does not show the HVR. All contaminating/unrelated lines (e.g., the H$_{2}$ absorption near \OVI\ 1031.93 \AA ) have been removed from Figures~\ref{fig_suppstacks} and \ref{fig:anc_navstacks} (see S\ref{sec:deblending}).

Inspection of Figures~\ref{fig_suppstacks} and \ref{fig:anc_navstacks} reveals the following features that are similar to the H1821+643 OA data: First, the \SiIV\ and \CIV\ profiles are very similar in both the main Complex C/OA velocity range and in the HVR in all of the sightlines.  No \OVI\ data are available for HS1914+7139 and 3C 351.0, but the remaining four sightlines also exhibit \OVI\ profiles that have a similar shape to the \SiIV\ and \CIV.  Moreover, the high-ion ratios in all of the sightlines are similar, with two exceptions: in the MRK290 spectrum, there is a region (marked as H1 or HVR1 in the figures) that clearly shows \SiIV\ and \CIV\ with little or no corresponding \OVI, and in the HVR toward H1821+643, \CIV\ and \OVI\ are much stronger than \SiIV.  Second, the low-ionization and high-ionization profiles have very similar shapes in the MRK876, PG1626+554, and MRK817 sightlines.  The 3C 351.0 low-ionization and high-ionization data are also kinematically very similar, but 3C 351.0 shows one component that is much stronger in the low-ionization profiles (reminiscent of component 2 in the H1821+643 data).  Toward HS1914+7139 there are small offsets between the low- and high-ionization stages, and the high ions are much stronger in the HVR.  MRK290 seems to have the largest differences between the low- and high-ionization profiles with high-ion absorption spread over $\approx$100 \kms\ in the HVR with no corresponding detections of low ions.   One might wonder if this is related to the location of MRK290 near the upper envelope of Complex C (see Figure~\ref{fig:outergalmap}), but MRK817 is also near the upper edge of Complex C, and that sightline shows an exquisite correspondence between the low- and high-ionization stages (see last column in Fig.~\ref{fig:anc_navstacks}).  There appears to be no correlation of the low-ion to high-ion ratios with longitude or latitude.  Instead, this is likely set by very local conditions.   

Table~\ref{tab:lowion_othertargs} presents the column densities measured in the ancillary directions using Voigt-profile fitting.  Without the highest-resolution STIS E140H data, the component structure model can be inaccurate, but total column densities are reliable, so when multiple components are present in the C/OA and/or HVR velocity range, the component columns are summed in Table~\ref{tab:lowion_othertargs}. The high-ionization profiles are often smooth and broad, so for these species, the column densites were obtained by direct integration of the \nav\ profiles over the velocity ranges indicated in Figure~\ref{fig:anc_navstacks}. The upper half of Table~\ref{tab:lowion_othertargs} reports the columns of \SiII\, \SII, and \FeII, and the lower half contains the measurements of \SiIV, \CIV, and \OVI. In the ancillary sightlines, \OI\ is difficult to measure for a variety of reasons and thus is not listed in this table.  Also, the HVR is generally more highly ionized than the main C/OA components, and measurement of the low-ionization stages is more challenging in the HVR, so analysis of the low-ionization phase will focus on the main C/OA absorption.

\subsection{Ancillary Sightline Low-Ionization Phases}
\label{sec:ancionization}

Comparing the $N$(\SiII), $N$(\SII), and $N$(\FeII)  from Table~\ref{tab:lowion_othertargs} to the 21cm $N$(\HI) measurements in Table~\ref{tab:targets}, we see that in a sightline-to-sightline comparison, the column densities of low-ionization metals do not increase in direct proportion to $N$(\HI).  For example, we see that $N$(\HI) toward MRK817 is an order of magitude higher than $N$(\HI) toward H1821+643, and yet the total low-ionization metal column densities are almost identical (the MRK817 low-ion columns are only $0.1 - 0.2$ dex higher than the H1821+643 columns).  The same result emerges from comparison of the MRK876, PG1626+554, and MRK290 measurements to those observed toward H1821+643.  This is another indication that these HVCs are composed of many clouds that are smaller than the $\approx 10'$ beam of the single-dish observations used to obtain the 21cm measurements in Table~\ref{tab:targets} (a $\approx 10'$ beam corresponds to a physical scale of $\approx$30 pc if the HVCs are $\approx$ 10 kpc away). If there are numerous smaller structures inside the single-dish beam pointed toward each of the AGN, then the radio $N$(\HI) could be significantly higher than the \HI\ column density on the pencil beam probed by the ultraviolet spectroscopy, which could lead to a poor correlation between the 21cm \HI\ column and the low-ion column densities, as observed.  Recently, \citet{marchal21} have obtained 21cm emission maps of a portion of Complex C at $1.1'$ angular resolution, and these maps confirm the presence of clustered 21cm-emitting structures that are unresolved at $10'$ resolution. 

The ratios of low-ionization metal column densities are also similar in all of the sightlines, with a scatter of $\approx$0.2 dex.  While the detailed components are not as well constrained in the lower-resolution data, the similarity of the ratios from sightline to sightline across the full velocity range of the OA/C structures indicates that the physical conditions in the ancillary directions are close to those derived from the H1821+643 data.   The additional sightlines also require some dust depletion.  Using the method described in S\ref{sec:dustdepletion}, upper and lower bounds on the amount of depletion required in each direction are summarized in Table~\ref{tab:ancillary_dustdepl}. Since reliable $N$(\OI) measurements were not obtained for the ancillary sightlines, some other means to place an upper limit on the ionization parameter was needed to derive upper and lower bounds on the amount of Si and Fe depletion in each direction.  The results in Table~\ref{tab:ancillary_dustdepl} were obtained by using \SiIV\ to place an upper limit on the ionization parameter (if $U$ is too high, the photoionized gas will produce more \SiIV\ than observed).  This is a conservative upper limit on $U$ because some of the \SiIV\ must come from the more highly ionized phase.  This translates to a conservative lower limit on the Fe depletion because as $U$ decreases, more Fe depletion is needed (Figure~\ref{fig:photomodrat}).  The Si depletion is less sensitive to $U$.  The range of depletions indicated in this table considers both the low- and high-metallicity models (the higher metallicity metallicity models require somewhat greater depletion).  

There are no obvious spatial trends in the depletion patterns in Table~\ref{tab:ancillary_dustdepl}.  The two sightlines with the greatest depletion (PG1626+554 and MRK817) are near the higher-latitude boundary of Complex C (see Fig.~\ref{fig:outergalmap}), but a third sightline near the upper boundary (MRK290) indicates only light depletion, and the H1821+643 sightline, which is among the lowest-latitudes probed, shows depletion amounts similar to PG1626+554 and MRK817.  There is also no difference in depletion patterns between the OA and Complex C.  The dust evidently has a patchy distribution in these high-velocity structures.

\ctable[
caption={Dust Depletion Levels in Ancillary Sightlines},
label={tab:ancillary_dustdepl},
doinside=\footnotesize
]{lcccc}{
\tnote[a]{Logarithmic depletion by dust required to achieve agreement with the ionization models and the $\pm 1\sigma$ measurement uncertainties.} 
\tnote[b]{The 3C 351.0 sightline does not provide sufficient information to constrain the depletion levels in this direction and thus is not included in this table.} 
}
{\FL \ & \multicolumn{2}{c}{\underline{Si Depletion (dex)}\tmark[a]}  & \multicolumn{2}{c}{\underline{Fe Depletion (dex)}\tmark[a]} \NN
Sightline\tmark[b]            & Min.   & Max.  & Min. & Max. \ML
HS1914+7139  & $-0.2$ & $-0.8$ & $-0.1$ & $-0.7$ \NN 
MRK876           & \ $0.0$  & $-0.6$ & 0.0 & $-0.2$ \NN 
PG1626+554    & $-0.5$ & $-0.9$  & $-0.3$ & $-0.7$ \NN
MRK290           & \ $0.0$  & $-0.5$ & $-0.2$ & $-0.6$ \NN
MRK817           & $-0.5$ & $-1.0$ & $-0.6$ & $-0.9$ \LL
}

\subsection{Ancillary Sightline High-Ionization Phases}

\begin{figure}
\includegraphics[width=8.0cm]{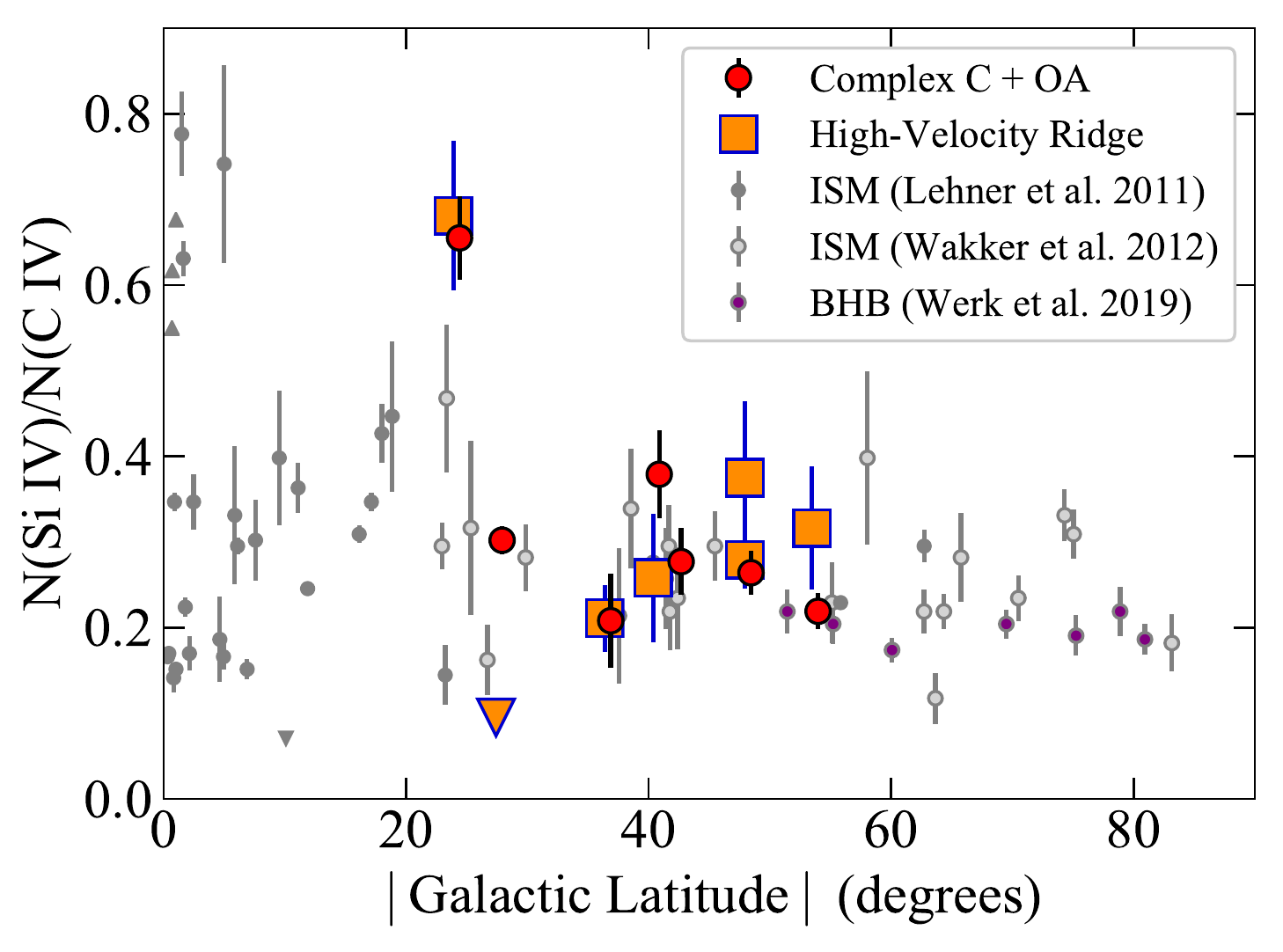}
\includegraphics[width=8.0cm]{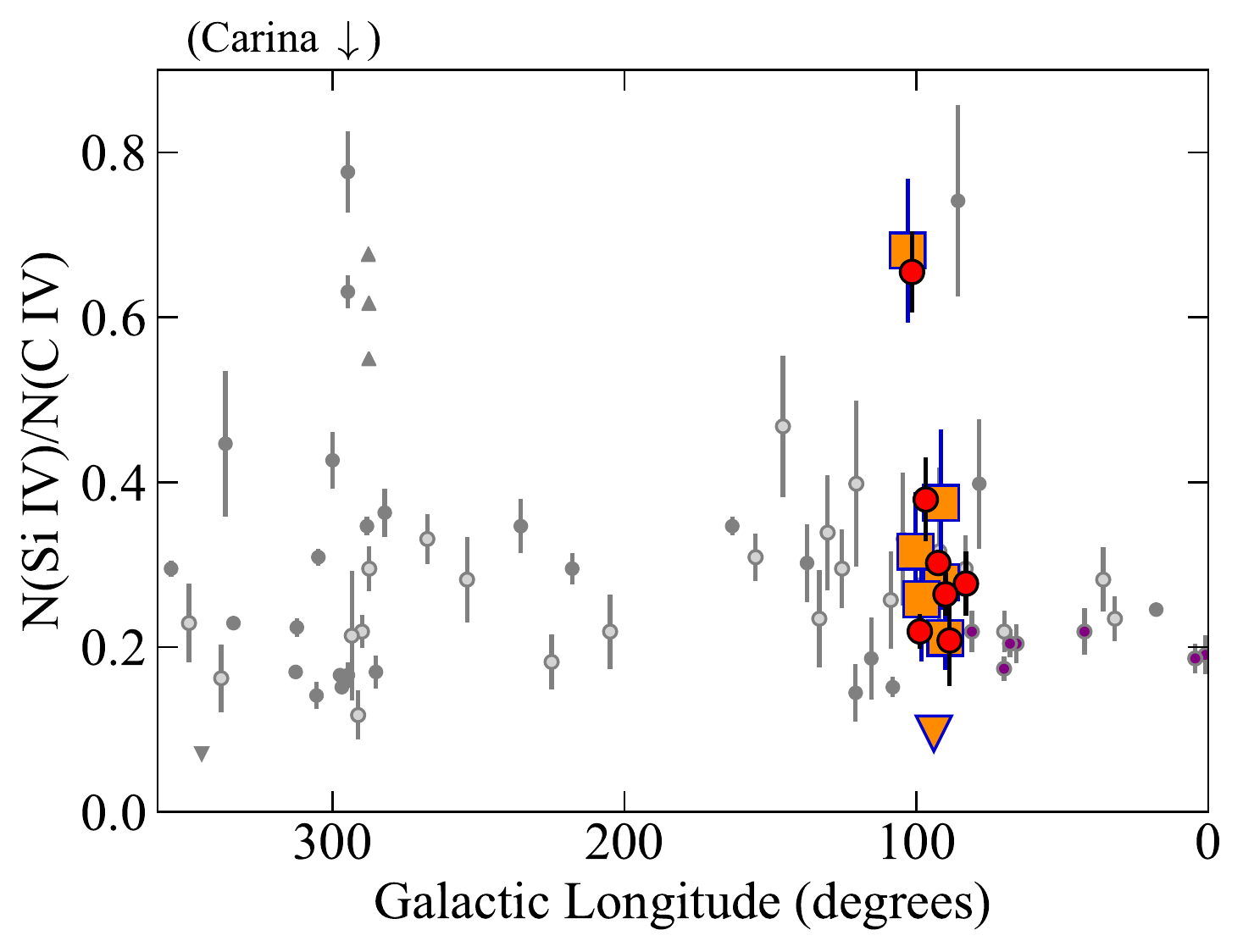}
\caption{The \SiIV/\CIV\ column-density ratios measured in the sightlines of this paper, with $1\sigma$ error bars, in the velocity range of Complex C and the Outer Arm (red circles) and the High-Velocity Ridge (orange squares), vs. the \SiIV/\CIV\ ratios measured in the ISM by \citet{lehner11}, \citet{wakker12}, and \citet{werk19} (see legend in the top panel).  Upward- and downward-pointing arrows indicate lower and upper limits, respectively. The top panel shows this ratio as a function of the absolute value of Galactic latitude, and the bottom panel plots the ratio vs. Galactic longitude.  In the ISM, this ratio is anomalously high in sightlines through the Carina Nebula at longitude $\approx 300^{\circ}$. \label{fig:si4c4map}}
\end{figure}

\begin{figure}
\includegraphics[width=8.0cm]{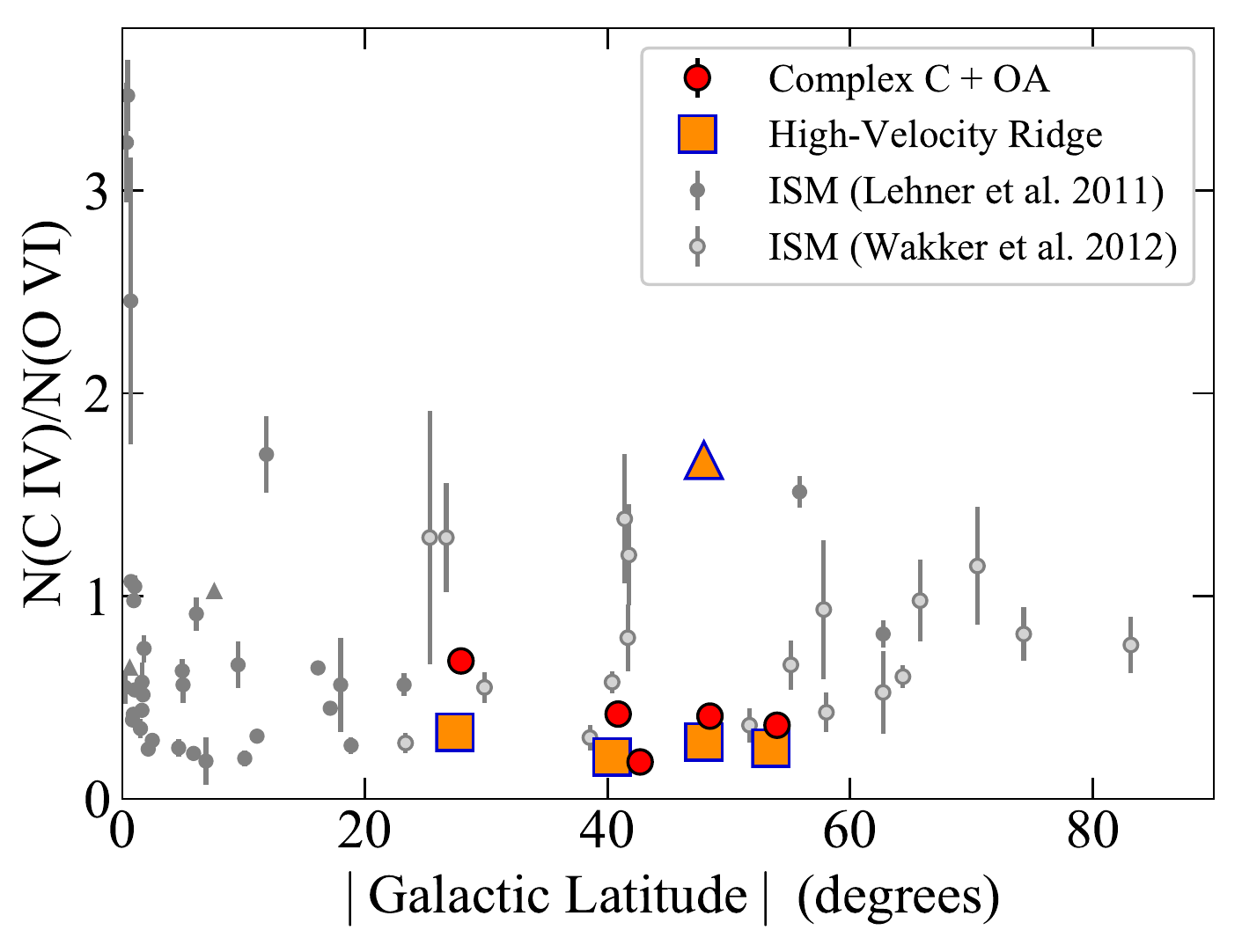}
\includegraphics[width=8.0cm]{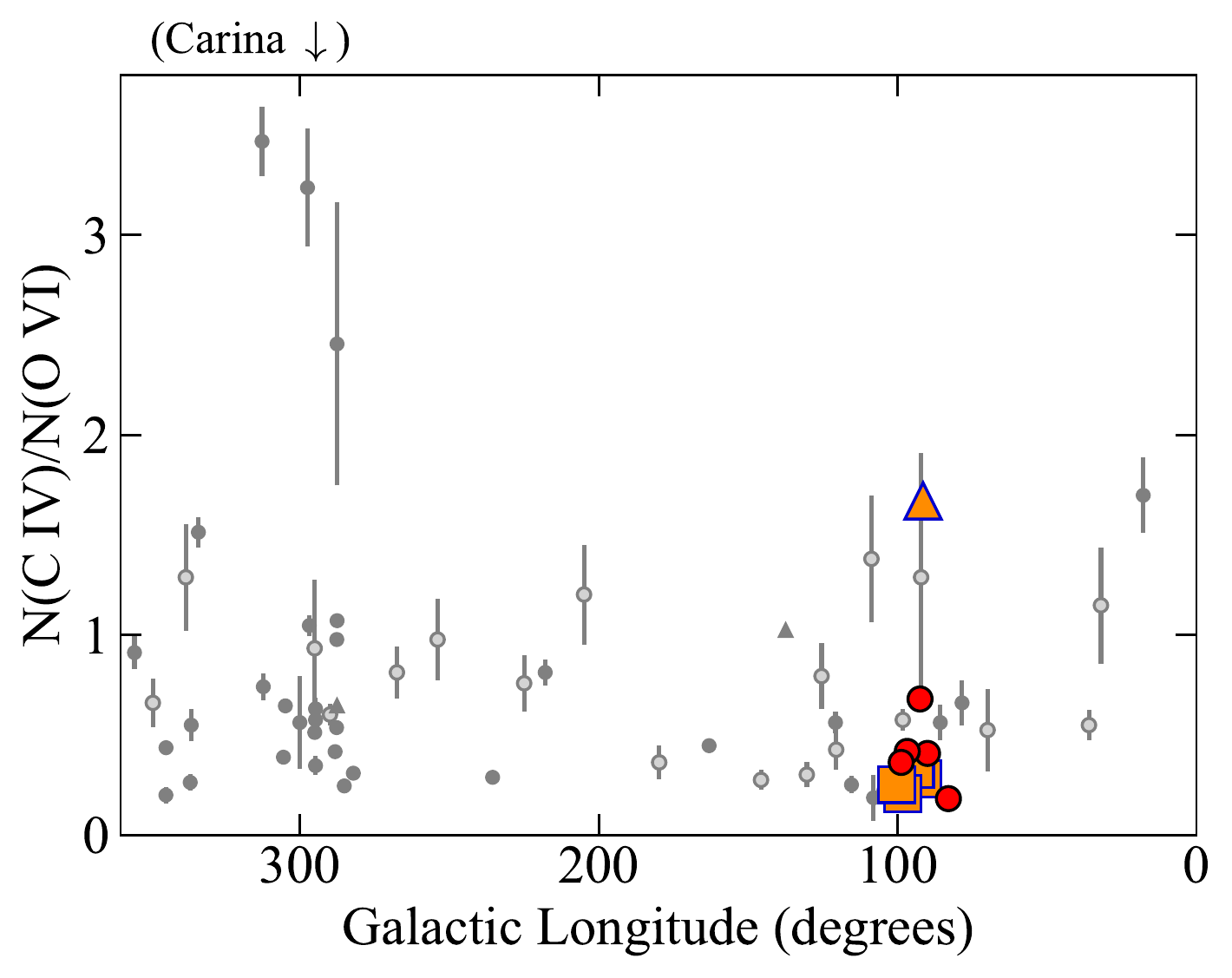}
\caption{Comparison of the \CIV/\OVI\ ratio observed in the sightlines studied in this paper vs. ISM measurements (as in Fig.~\ref{fig:si4c4map}).\label{fig:c4o6map}}
\end{figure}

\begin{figure}
\includegraphics[width=8.3cm]{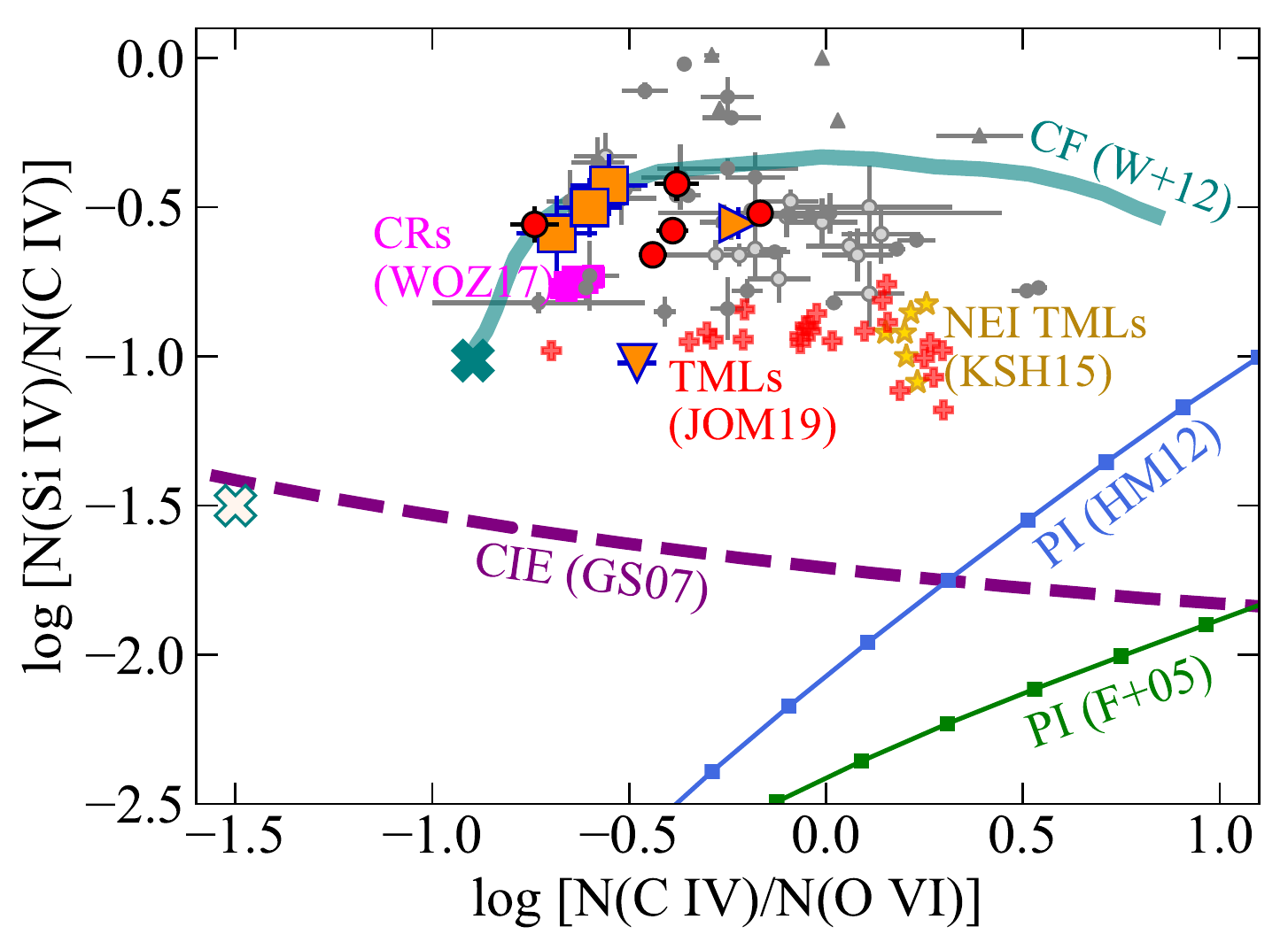}
\caption{The  \SiIV/\CIV\ vs. \CIV/\OVI\ ratios observed in the HVCs and ISM (as in Figure~\ref{fig:si4c4map}) compared to the ratios predicted by several theoretical models, including: single-phase collisionally ionization equilibrium \citep[CIE,][GS07, purple-dashed line]{gnat07}, single-phase photoionization (PI) using the ionizing flux field of \citet[][F+05, solid green line with squares]{fox05} or \citet[][HM12, solid blue line with squares]{haardt12}, turbulent mixing layers (TMLs) from \citet[][JOM19, red plus symbols]{ji19} and non-equilibrium TMLs calculated by \citet[][KSH15, gold stars]{kwak15}, ionization models with cosmic ray (CR) heating \citep[][WOZ17, pink squares]{wiener17}, and the isochoric cooling flow (CF) model from the appendix of \citet[][W+12, thick teal line]{wakker12} including photoionization. Neglecting photoionization causes the W+12 model to collapse, in this ratio-ratio plot, to a single point indicated by the teal X. The isobaric CF model from W+12 predicts a single point in this plot indicated with a white X (regardless of whether photoionization is included or not).  The two OA/HVR measurements with anomalously high \SiIV/\CIV\ ratios are not shown here because \OVI\ data have not been obtained for that sightline. The cluster of ISM measurements with log $N$(\SiIV/\CIV) $> \ -0.3$ are mostly from the Carina Nebula.\label{fig:ratrat1}}
\end{figure}

\begin{figure}
\includegraphics[width=8.7cm]{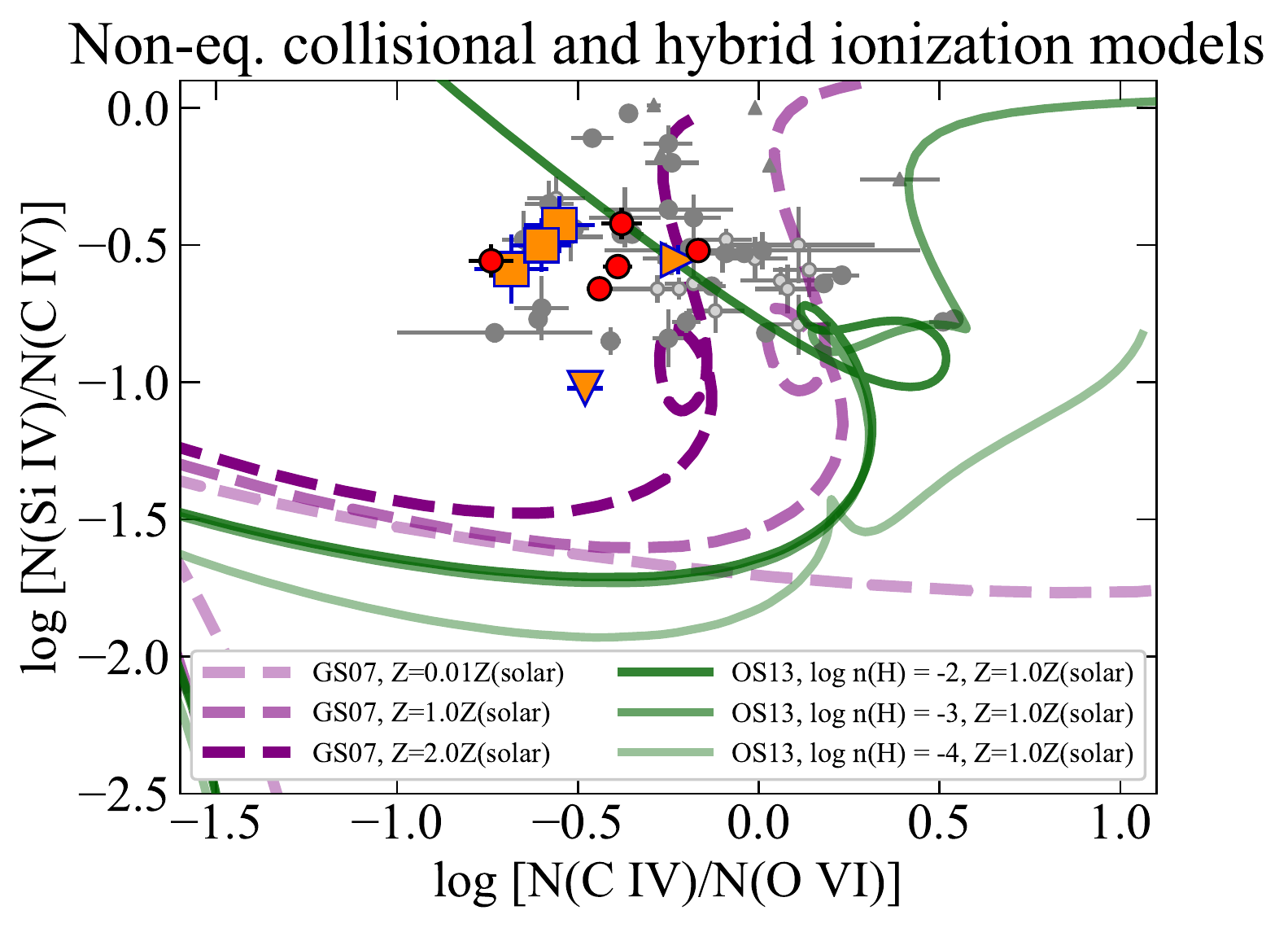}
\caption{Observed \SiIV/\CIV\ vs. \CIV/\OVI\ ratios, as in Figure~\ref{fig:ratrat1}, compared to the isochoric non-equlibrium models of rapidly cooling gas from \citet{gnat07} and \citet{oppenheimer13}. For the \citet{gnat07} models, which only include collisional ionization, the light-, medium-, and dark-purple dashed lines show the ratios of cooling gas with metallicity $Z =$ 0.1, 1.0, and 2.0 $Z_{\odot}$, respectively. The lowest-metallicity model is very similar to collisional ionization equilibrium (compare the light-purple line to the CIE model in Fig.~\ref{fig:ratrat1}). The \citet{oppenheimer13} calculations show the ratios in rapidly cooling collisionally ionized gas with additional photoionization by the \citet{haardt12} UV background with $n_{\rm H} = 10^{-4}, 10^{-3},$ and $10^{-2}$ cm$^{-3}$ (light-, medium-, and dark-green solid curves, respectively) with $Z = 1.0 \ Z_{\odot}$. As was found for the H1821+643 OA data in S\ref{sec:1821highionmodels}, the high-$Z$ non-equilibrium models are better able to simultaneously match the \SiIV, \CIV, and \OVI\ column densities than most of the models shown in Figure~\ref{fig:ratrat1}. The \textit{isobaric} cooling models (not shown) from \citet{gnat07} and \citet{oppenheimer13} fit the observed column densities more poorly.\label{fig:ratrat2}}
\end{figure}

As we can see from Figures~\ref{fig_suppstacks} and \ref{fig:anc_navstacks}, the Complex C/OA and HVR absorption features are detected in very similar velocity ranges in all seven sightlines studied in this paper, and there appears to be no dependence of the metal or \HI\ column densities on Galactic latitude.  The kinematic similarities of the absorption over a 30-degree range in latitude, combined with distance constraints that place these HVCs in the same region of the Galaxy (S\ref{sec:hvc_summary}), suggest that the absorption lines in these 7 directions may have a related physical origin.  To further compare these sightlines to each other and to the ISM in the Milky Way disk, Figures~\ref{fig:si4c4map} and \ref{fig:c4o6map} plot the linear \SiIV/\CIV\ and \CIV/\OVI\ column density ratios observed toward all of the targets as a function of Galactic longitude and latitude for the main Complex C/OA absorption (filled red circles) and the HVR (filled orange squares).  In a few cases in the HVR these ratios are limits; these are indicated with filled orange triangles pointing in the direction of the limit.  To compare these ratios to those in the underlying Galactic disk, Figures~\ref{fig:si4c4map} and \ref{fig:c4o6map} also show the ratios measured by \citet{lehner11} and \citet{wakker12}\footnote{\citet{wakker12} also report measurements in HVCs detected in their sample.  To provide a comparison to the lower-velocity ISM only, the \citet{wakker12} HVC measurements are excluded from Figs. ~\ref{fig:si4c4map} and \ref{fig:c4o6map}.} in the ISM at low velocities.  Finally, the \SiIV/\CIV\ ratios in the disk-halo interface observed toward high-latitude blue-horizontal branch (BHB) stars by \citet{werk19} are also shown (\OVI\ measurements are not available toward the BHB stars).

With four exceptions, the high-ion ratios in the HVCs are very similar to each other. Starting with \SiIV/\CIV, 11/14 measured \SiIV/\CIV\ ratios in the HVCs are between 0.2 and 0.4, with mean = 0.32 and standard deviation $\sigma$ = 0.18.  The two measurements with anomalously high \SiIV/\CIV\ ratios are from the OA and HVR components toward the star HS1914+7139, and it is possible that these two cases are affected by the star and its circumstellar environment.  Also, the continuum placement is more ambiguous in this stellar spectrum than in the AGN spectra, and it is possible that continuum placement introduced systematic error in the HS1914+7139 measurements, although the \nav\ profiles of the two \SiIV\ lines are in good agreement with the adopted continuum placement.  Such high \SiIV/\CIV\ ratios are rare in the ISM \citep{sembach97,lehner11,wakker12}, although similarly high \SiIV/\CIV\ ratios are observed on several sightlines toward the Carina Nebula (see Fig.~\ref{fig:si4c4map}), an X-ray bright giant \textsc{H~ii} region comprising $>10$ star clusters. Such an environment is unlikely at the location of HS1914+7139. If the HS1914+7139 measurements are excluded, the mean \SiIV/\CIV\ reduces to 0.26 with $\sigma$ = 0.08.  There is more scatter in the \CIV/\OVI\ ratios with mean = 0.48 and $\sigma$ = 0.44. Here the oddity is the HVR1 component toward MRK290, which has a normal \SiIV/\CIV\ ratio with little corresponding \OVI.  If this HVR1 region is excluded, the \CIV/\OVI\ mean reduces to 0.35 with $\sigma$ = 0.15. 

It is interesting that the HVCs ratios are generally close to the ratios typically measured in low-velocity ISM gas.  This might occur naturally if the high ions are generated by a similar physical mechanism (e.g., turbulent mixing layers or non-equilibrium rapidly cooling gas) in the HVCs and the ISM.    Figure~\ref{fig:ratrat1} compares the HVC and ISM \SiIV/\CIV\ vs. \CIV/\OVI\ ratios to several theoretical production mechanisms.  For comparison to the results from S\ref{sec:1821highionmodels}, this figure shows the ratios predicted by simple single-phase photoionization models, using either the \citet{fox05} or \citet{haardt12} ionizing radiation fields, and by a single-phase CIE model.  As found for the H1821+643 OA data, these single-phase models are strongly ruled out.   The literature often suggests that \SiIV, \CIV, and \OVI\ are produced by photoionization in circumgalactic halos but are collisionally ionized in galactic disks.  Figure~\ref{fig:ratrat1} shows the utility of ratio-ratio diagrams for testing ionization mechanisms -- photoionization and collisional ionization typically lead to very different ratios.  This also underscores the implication of the HVC and ISM similarity, i.e., that the high ions may be generated by the same physical process in these two contexts.

Figure~\ref{fig:ratrat1} also shows several more complex models that are significantly closer to matching the observed ratios, including turbulent mixing layers (TMLs) from \citet{ji19} and \citet{kwak15}, ionization models including cosmic-ray heating \citep{wiener17}, and the \citet{wakker12} cooling-flow models discussed in S\ref{sec:wakkerflows}.\footnote{Figure 11 in \citet{wakker12} also shows the predictions of shock-heating and conductive-interface models; these models are not successful at fitting these ratios and are not shown here for the sake of clarity.}  \citet{ji19} explored TMLs models with a wide variety of simulation parameters including equlibrium and non-equilibrium ionization, magnetic fields, and photoionization as well as differing simulation box sizes, metallicities, cooling rates, density contrasts, shear velocities, and ambient pressures.  Their model results are time-dependent but typically converge to roughly constant column densities and column-density ratios after $\approx$20 Myr.  Each red plus symbol in Figure~\ref{fig:ratrat1} shows the ratios from one of the \citet{ji19} runs after 20 Myr.  \citet{kwak15} have calculated TML column densties in the HVC context; their time-averaged results (gold stars) are similar to those of \citet{ji19}, although some of the \citet{ji19} runs attain lower \CIV/\OVI\ ratios that are closer to the observed values.  All TML variations in both of these studies fall short of the observed \SiIV/\CIV\ ratios by factors of $\approx 2-3$.  The cosmic ray models (pink squares) have even lower \CIV/\OVI\ ratios by still fall short of the observed  \SiIV/\CIV\ by $\approx 0.1 - 0.3$ dex.  The \citet{wakker12} isochoric cooling-flow, including self-photoionization, may be the most successful model shown in Figure~\ref{fig:ratrat1}: this model can explain the full range of \CIV/\OVI\, and some of the HVCs are well matched by this model.  However, this model also predicts \SiIV/\CIV\ ratios that are too high compared to some HVC and many ISM measurements, possibly due to the absence of photoionization by external UV radiation.  Indeed, it is possible that none of the models properly incorporate both photoionization by external flux and self-photoionization from nearby hot gas, and this could affect the ratios at the levels of the observed discrepancies.   The scatter in the observations is larger than in any of the models, and there may also be stochastic factors that need to be considered.

In Figure~\ref{fig:ratrat2}, the ancillary measurements are compared to the non-equilibrium and hybrid ionization models that were able to successfully explain the high-ionization lines observed in the OA toward H1821+643, including the \citet{gnat07} rapidly cooling models (purple dashed lines) and the \citet{oppenheimer13} rapidly-cooling hybrid-ionization models (green solid lines).  As was found for the H1821+643 data, along as the metallicity and gas density are high enough, both of these models can fit the ancillary HVC and ISM ratios.  Moreover, it is plausible that reasonble variations in the parameters of these models could explain the scatter in the observations.  Overall, the non-equilibrium models in Figure~\ref{fig:ratrat2} appear to provide the best fit to the high-ionization lines in the OA/C and HVR structures.

Figures~\ref{fig:ratrat1} - \ref{fig:ratrat2} show only ratios, and it is worth noting that many of these models predict ion column densities that are lower than the observed columns by large factors.  So, similarly to the discussion in S\ref{sec:opp13model}, some of these models require hundreds of individual interfaces to match the observed column densities.  Again, this may be plausible if the absorption components comprise hundreds of low-ionization cloudlets, as in the \citet{mccourt18} shattering theory, with high-ionization interface layers on their surfaces.

\section{Discussion}
\label{sec:discussion}

This paper has presented data and models that suggest that the OA and Complex C HVCs are misty structures that are rapidly cooling.  This condensing gas has higher densities than the ambient gaseous halo, and this material could ``precipitate'' onto the underlying disk thereby providing fuel for new star formation.  Describing these HVCs as misty is a robust statement, but the smallest size of the droplets in the mist is an open question.  At a minimum, the overall angular extent of the coherent 21cm emission from the OA and C indicates that these structures are $\approx$20 kpc across, but the low-ionization metal lines and photoionization models require individual-component cloud sizes of $2 - 40$ pc.  These numbers imply roughly 500 $-$ 10000 ``mist droplets'' in the HVCs.  However, it remains possible that each component detected in metal absorption is itself composed of hundreds of cloudlets so that these numbers would go up to roughly 50000 $- \ 10^{6}$ droplets.  Indeed, some of the models that provide reasonable reproduction of the high-ionization line properties favor this scenario of tens to hundreds of cloudlets per component.

Several results in the literature corroborate this conclusion that the HVCs are misty with internal clouds with sizes $\approx 10$ pc.  First, \citet{hsu11} and \citet{marchal21} have obtained 21cm-emission maps of portions of Complex C, with good angular resolution ($1.1' - 3.5'$), that reveal a complex web of small \HI\ filaments and blobs, and the filaments themselves often resolve into strings of adjacent blobs in high-resolution observations.  Similar 21cm filament/blob morphologies are evident in the Smith Cloud HVC \citep{lockman08}, HVC Complex A \citep{barger20}, and the disk-halo interface \citep{lockman02,blagrave17}.   \citet{marchal21} decompose the 21cm features into cold-, warm, and ``lukewarm''-neutral phases, and the physical properties they infer for the warm-neutral phases are encouragingly similar to the results obtained here from the H1821+643 data: in their `F' (`A') regions they find mean $T = 9.5 \times 10^{3}$ K ($1.1 \times 10^{4}$ K) and mean $n = $ 0.4 cm$^{-3}$ (0.4 cm$^{-3}$) with a mean cloud size of 28 pc (26 pc).  Comparing to Table~\ref{tab:h1821_physcon}, we see that these temperatures and densities are very close to the values derived from the H1821+643 spectra.  The 21cm blob sizes are also similar to the H1821+643 sizes in the low-metallicity model, but they are a factor of $3-14$ larger than the sizes in the high-$Z$ model.

Information from other studies on cloud spatial sizes points to similar conclusions. Using sightline pairs derived from 23 hot stars in the Magellanic Clouds, \citet{howk02} find significant variations in \OVI\ column densities on angular scales ranging from $0.05^{\circ} - 5.0^{\circ}$, which corresponds to scales of $<6 - 800$ pc for an assumed location within 5 kpc from the Sun.  In a follow-up study of horizontal branch stars in a globular cluster, \citet{lehner04} find no evidence of $N$(\OVI) variations exceeding $\gtrsim 25$\% toward four stars, indicating that the \OVI-bearing gas has a size $\gtrsim$ 10 pc in that direction.  Comparing the spectra toward two AGN probing Complex A, \citet{richter17} show highly significant and consistent variations in the \textsc{C~ii} and \SiII\ absorption profiles, which indicates that the low-ionization gas changes on scales $\lesssim 90$ pc.  In their analysis of Ca~\textsc{ii}, \FeII, \SiIV, and \CIV\ toward seven BHB stars, \citet{werk19} conclude that the \FeII + Ca~\textsc{ii} phase arises in a small region with size $\approx$ 10 pc, but they favor a significantly larger size for the \SiIV + \CIV\ phase; toward these stars, the \SiIV\ + \CIV\ may be revealing the larger coherent structure in which the small clouds are embedded.  Most relevant to this study, \citet{dixon06} detected $3.6\sigma$ \OVI\ emission from HVC Complex C in the direction \textsl{l,b} = 107.0,48.8$^{\circ}$; by analyzing the emission intensity combined with the \OVI\ column density from absorption, \citet{dixon06} estimate that the path length of the \OVI\ emission region in this part of Complex C is $\approx 1 - 14$ pc.    Similarly, in the high-velocity Magellanic Stream, \citet{dixon06} find an \OVI\ pathlength of 16 $-$ 200 pc based on a $3.3\sigma$ emission detection.  These cloud sizes are far too small to produce \OVI\ by photoionization.  Current size constraints could still allow photoionized \SiIV\ and/or \CIV, but ionization mechanisms that can explain the \OVI\ are likely to affect \SiIV\ and \CIV\ as well.

The finding of small clouds embedded within larger coherent structures suggests that gas structures may frequently be misty in various contexts, but important questions remain open.   For example, this paper has presented consistent evidence of Si and Fe depletion in several sightlines through the OA and C, and this has potentially interesting implications.  The survival of dust could provide insight regarding the physics of the HVCs, and insights on dust survival in high-velocity objects could help to explain molecular gas in galactic outflows \citep{farber21}.  How ubiquitous is the dust?  How does it affect the heating/cooling and general gas physics?  Does the presence of dust provide a clue about the origin of the high-velocity gas?  Observationally, new ultraviolet observations of HVCs with the highest spectral resolution and good S/N would provide better statistics on the prevalence of dust.

The metallicity of the HVCs is also a pesky  open question -- this may be more complicated than indicated by past literature.   If the highly-ionized gas has a high enough metallicity so that it has cooled rapidly into a non-equilibrium overionized state, then there may be nearly as much \HI\ in the highly-ionized gas as in the low-ionization gas, and this in turn might lead to the conclusion that the low-ionization gas has a significantly higher metallicity than previously thought (S\ref{sec:opp13model}).   Recently, \citet{heitsch21} have discussed metallicity measurement issues from a theoretical/simulation perspective, and they find that mixing with ambient material has a profound effect on HVC metallicities. Given the location of the OA and Complex C, high metallicities are plausible.  As discussed above, this material could be material elevated from the disk by the same process that produced well-known stellar streams in the same region, or it could be Galactic-fountain ejecta \citep[e.g.,][]{fraternali15}; in these scenarios, high-$Z$ gas could be present.  However, in other locations in the CGM at much greater distances from the disk, one might wonder if high metallicities would be unlikely, although it is notable that studies of the CGM at relatively large impact parameters have found high metallicities \citep[e.g.,][]{meiring13,prochaska17}.  It would be valuable to obtain new observations with the STIS E140H mode (like the H1821+643 spectra presented here) to investigate clouds in other locations.

\section{Summary}
\label{sec:summary}

To observationally study the physics of circumgalactic gas in the inner CGM and disk-halo interface, this paper has presented a combination of new high spectral resolution (2.6 \kms ) STIS observations of H1821+643 combined with archival STIS, FUSE, and COS observations of H1821+643 and six other nearby targets (5 AGN and 1 star).  These targets pierce the Outer Arm and Complex C HVCs.  The similarity of published constraints on the distances, kinematics, and metallicities of the HVCs suggest that these objects could be related.  In addition, there are a number of stellar streams located in the same general outer-Galaxy (three-dimensional) location.  Some of these stellar streams have metallicities and/or kinematics that are similar to those of the OA and Complex C, while others have significantly different kinematics and/or metallicities.  Even with different metallicities and kinematics, these streams could have a causal connection to the HVCs because they could roil the gas in this region, which in turn could make thermal instability and precipitation more likely to occur \citep[e.g.,][and references therein]{voit21}. Based on the absorption spectra of these targets, the following findings are reported:

(1)  With the benefit of improved resolution, the \SiII\ absorption in the OA toward H1821+643 transforms from what initially appears to be a single component (at 7 \kms\ resolution) into five discrete components.  Closer inspection of the older 7 \kms\ spectra reveals significant inflections in the profile, and \citet{tripp12} fitted the older \SiII\ data with three components, but it is not possible to recognize all five components in the 7 \kms\ data.  The total column densities (summed over all components) are reasonably well constrained in the 7 \kms\ data, but there are significant systematic errors in the individual-component properties in the lower-resolution measurements (see Fig.~\ref{fig:si2si4fitparams}).  

(2) Voigt-profile fitting to species with a range of masses somewhat favors non-thermal broadening (Fig.~\ref{fig:vp_therm_nontherm}), which could indicate that the line broadening is dominated by turbulence.  

(3) Similarly to the \SiII\ data, the 2.6 \kms\ resolution \SiIV\ absorption profiles in the OA toward H1821+643 show clear evidence of multiple narrow components, and this component structure is corroborated (but less well-constrained) by the 7 \kms\ spectra.  The \SiIV\ line widths imply temperatures lower than the temperature where \SiIV\ peaks in CIE, although the line widths are only $1 - 2 \sigma$ below the widths expected in CIE.  The component centroids and lines in the \SiIV\ profiles are similar, though clearly not identical, to the \SiII\ component structure (Figs.~\ref{fig:vpstack}, \ref{fig:high_nav} - \ref{fig:si2si4fitparams}).  The \CIV\ E140H spectra are noisier than the \SiIV\ data, and the \OVI\ profile has lower spectral resolution, but despite these limitations, it is clear that \SiIV, \CIV, and \OVI\ have similar overall absorption profiles, and the high-ion profiles are also similar to the low-ion profiles.  This similarity suggests that there is a physical relationship between the low- and high-ionization species.  For example, the highly ionized gas might originate in a mixing layer on the surface of the low-ionization clouds. 

(4) Photoionization modeling of the low-ionization H1821+643 lines indicates that some dust depletion of Si and Fe by $-0.2$ to $-0.9$ dex is required to reconcile these models with the observations (Table~\ref{tab:h1821_dustdepl}).  

(5) If the majority of the \HI\ is located within the phase that gives rise to the H1821+643 low-ionization lines, then photoionization models can fit the data using metallicity [M/H] $\approx -0.7$ (as found previously) with $N$(\HI) = $10^{18.0}$ to $10^{18.4}$ cm$^{-2}$ and $N_{\rm total}$(H) = $10^{19.0}$ - $10^{19.3}$  cm$^{-2}$ in the three best-measured low-ionization components.  However, it is possible that a substantial portion of the \HI\ originates in the high-ionization phase (see below), in which case the low-ionization lines can be fitted with solar-metallicity gas with $N$(\HI) = $10^{17.3}$ to $10^{17.7}$ cm$^{-2}$ and $N_{\rm total}$(H) = $10^{18.5}$ - $10^{18.8}$  cm$^{-2}$.  

(6) In either the low-$Z$ or solar$-Z$ case, the H1821+643 photoionization models imply $n_{\rm H} = 10^{-0.8} - 10^{-0.3}$ cm$^{-3}$ and small cloud sizes ranging from 2 to 37 parsecs (Table~\ref{tab:h1821_physcon}). There are systematic uncertainties in these models, e.g., due to the uncertain ionizing flux impinging on the clouds, but nevertheless we can robustly conclude that the cooling times are short and $t_{\rm cool}/t_{\rm ff}$ and $t_{\rm cool}/t_{\rm sc} \ll 1$.  

(7) Models that produce the highly-ionized gas in the H1821+643 spectrum by photoionization or equilibrium collisional ionization suffer from several problems, but non-equilibrium collisional ionization or hybrid (non-equilibrium collisional ionization + photoionization) models can successfully explain the high-ion ratios and overall column densities.  The static non-equilibrium-collisional ionization models require relatively high metallicities ($Z \gtrsim 1\ Z_{\odot}$) and gas that is much cooler (log $T \approx$ 4.2) than expected in CIE.  High metallicities are plausible given the constrained location of the gas above the disk in the outer Milky Way, and the cool temperatures are consistent with the narrowness of the \SiIV\ components.  To match the overall column densities, these models predict that a substantial portion of the \HI\ actually arises in the high-ionization gas.   This is allowed by the data if the low-ionization phase likewise has a high metallicity; in this case the \HI\ is roughly evenly split between the low- and high-ionization phases.  Cooling-flow non-equilibrium models can produce the high ions with lower metallicity as long as self-photoionization is included, but available models do not include photoionization by the external UV background.   

(8) The total hydrogen column densities in the H1821+643 OA components are much larger than expected based on cloud-shattering theory.  This suggests that the components may themselves be composed of tens to hundreds of cloudlets.  Many models for the production of the high ions also require tens to hundreds of high-ion layers to match the observed high-ion column densities.  These two results fit together -- if there are tens to hundreds of cloudlets in the gas, then there could also be tens to hundreds of highly-ionized interface layers on the cloudlet surfaces.  The absorber sizes indicated by the ionization models (few to tens of parsecs) robustly establish that the outer-Galaxy HVCs are misty -- even parsec-scale clouds are much smaller than the region over which coherent HVCs can be recognized in 21cm emission maps. 

(9) Investigation of archival UV spectra of six other sightlines near H1821+643, including the low-ion ratios, high-ion ratios, and the kinematical correlation of the low- and high-ionization absorption profiles, indicates that these other nearby absorbers have similar origins and a similar physical nature.   Overall, the full dataset suggests that these HVCs comprise somewhat dusty, misty clouds that are condensing out of the halo and will subsequently rain down on the disk.  As such, these appear to be a local example of circumgalactic precipitation.

\section*{Acknowledgements} This research benefitted from stimulating discussions with Chris Howk, Mark Voit, Martin Weinberg, Bob Benjamin, and Neal Katz.  This work is based on observations made with the NASA/ESA Hubble Space Telescope and archival data from the Mikulski Archive for Space Telescopes at the Space Telescope Science Institute (STScI), operated by the Association of Universities for Research in Astronomy, Inc., under NASA contract NAS5-26555, with support from grant HST-GO-15321.001-A from STScI.  Portions of this paper were informed by lively presentations and discussions at the 2021 \textsl{Fundamentals of Gaseous Halos} program at the Kavli Institute for Theoretical Physics, which was supported in part by the National Science Foundation under Grant No. NSF PHY-1748958.  This research made use of several \textsc{python} packages: \textsc{numpy} \citep{harris20}, \textsc{matplotlib} \citep{hunter07}, and \textsc{astropy} \citep{astropy18}.

\section*{Data Availability} The data used in this paper are available from the Mikulski Archive for Space Telescopes (https://archive.stsci.edu).

\appendix

\section{Behavior and Mitigation of Hot Pixels in STIS MAMA E140H Spectra of H1821+643}
\label{sec:hotpix}

\subsection{Variability of Dark Counts and Hot Pixels}

It is known that the STIS far-UV MAMA detector suffers from a complex background due to dark counts \citep{brown00}. Part of the detector has a constant dark background, but a large portion of the detector is affected by an irregular ``blob'' of dark counts \citep[Fig.7.21 in][]{branton21} with a variable level that depends on detector temperature and the duration of time that the detector high voltage (HV) has been on \citep[Fig.7.22 in][]{branton21}.  The dark backgrounds have also increased over the years since STIS was deployed, and they also depend on telescope pointing: the dark backgrounds become more elevated during continuous-viewing zone observations because the detector becomes warmer when HST is pointed in a CVZ direction.  The ``continuum'' of dark counts is relatively low, however, and this background is mainly a problem for faint targets and/or STIS configurations that have low count rates from the source such as the E140H observations of H1821+643 presented here.

If the dark counts were broadly distributed in a more-or-less smoothy changing blob (as depicted in most literature on this issue), they would mainly reduce the S/N of the final STIS spectra.  Unfortunately, the dark counts create a more insidious problem for STIS spectroscopy in a way that is generally not discussed in available literature: \textsl{the MAMA detectors also have discrete warm and hot pixels that are temporally variable and correlated with the ``continuum'' dark counts.}  

\begin{figure*}
\includegraphics[width=16.0cm]{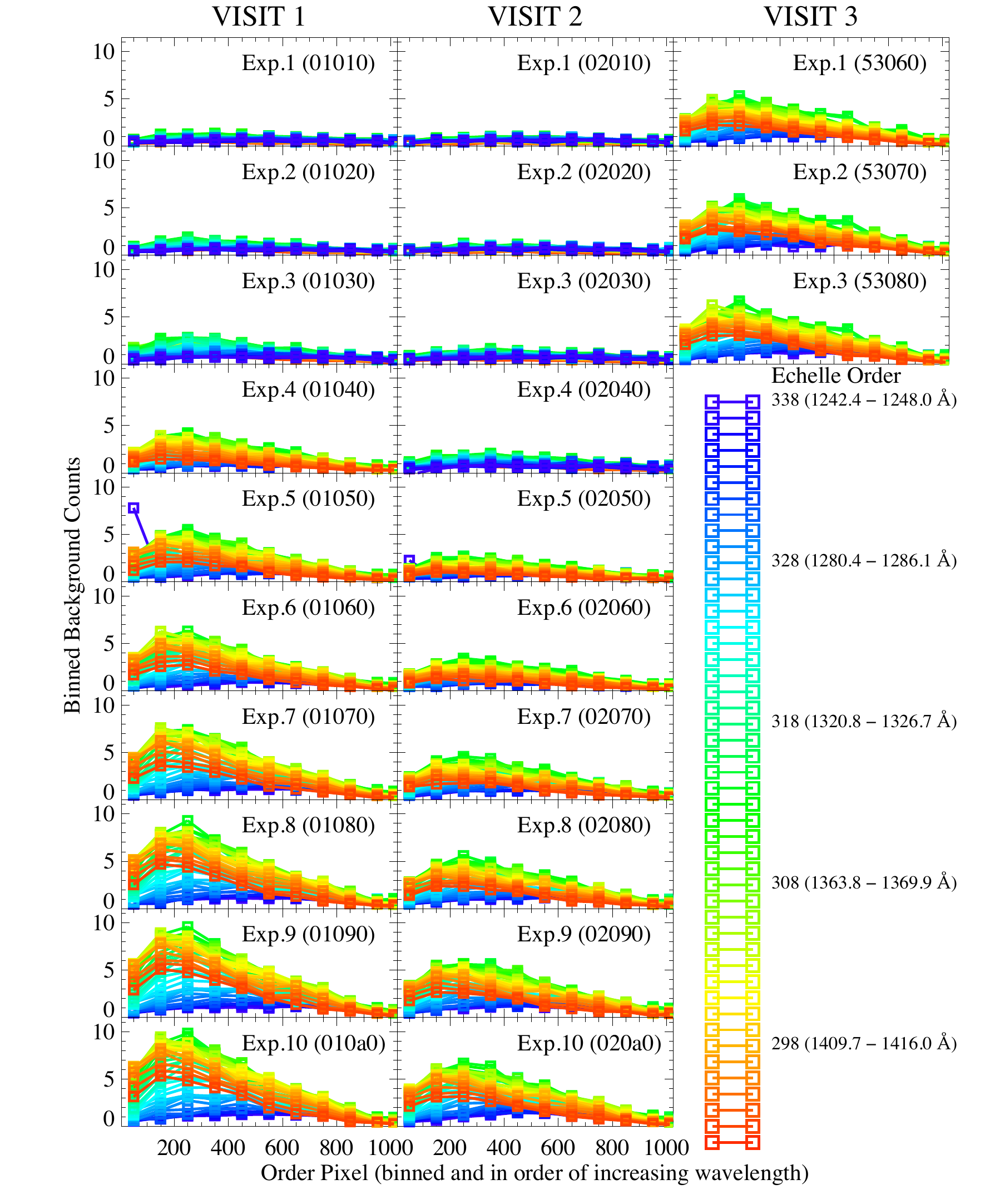}
\caption{Overall trends in background dark counts recorded in the STIS E140H observations of H1821+643 at the grating tilt with $\lambda_{\rm cen}$ = 1343 \AA. The left, center, and right columns show the exposures in visits 1, 2, and 3, respectively; the three visits occurred on 2018 May 10, 2018 July 4, and 2019 July 19 (see Table~\ref{tab:stis_log}). Each exposure is shown in an individual panel starting with the first exposure of the visit at the top of the column and progressing chronologically to the last exposure at the bottom. The last five digits of the MAST ID numbers are shown in parentheses in each panel. The echelle orders are color coded (see legend at lower right), and all of the observed orders are plotted for each exposure. To show the broader trends in the dark backgrounds, the background counts are averaged over 100-pixel bins. The dark backgrounds increase as the time into the visit increases and the detector temperature goes up. The backgrounds are lowest at short wavelengths, they increase to a maximum at $\lambda \approx$ 1340 \AA , and then decrease somewhat at $\lambda >$ 1340 \AA . \label{fig:broad_dark_bkg}}
\end{figure*}

\begin{figure*}
\includegraphics[width=16.0cm]{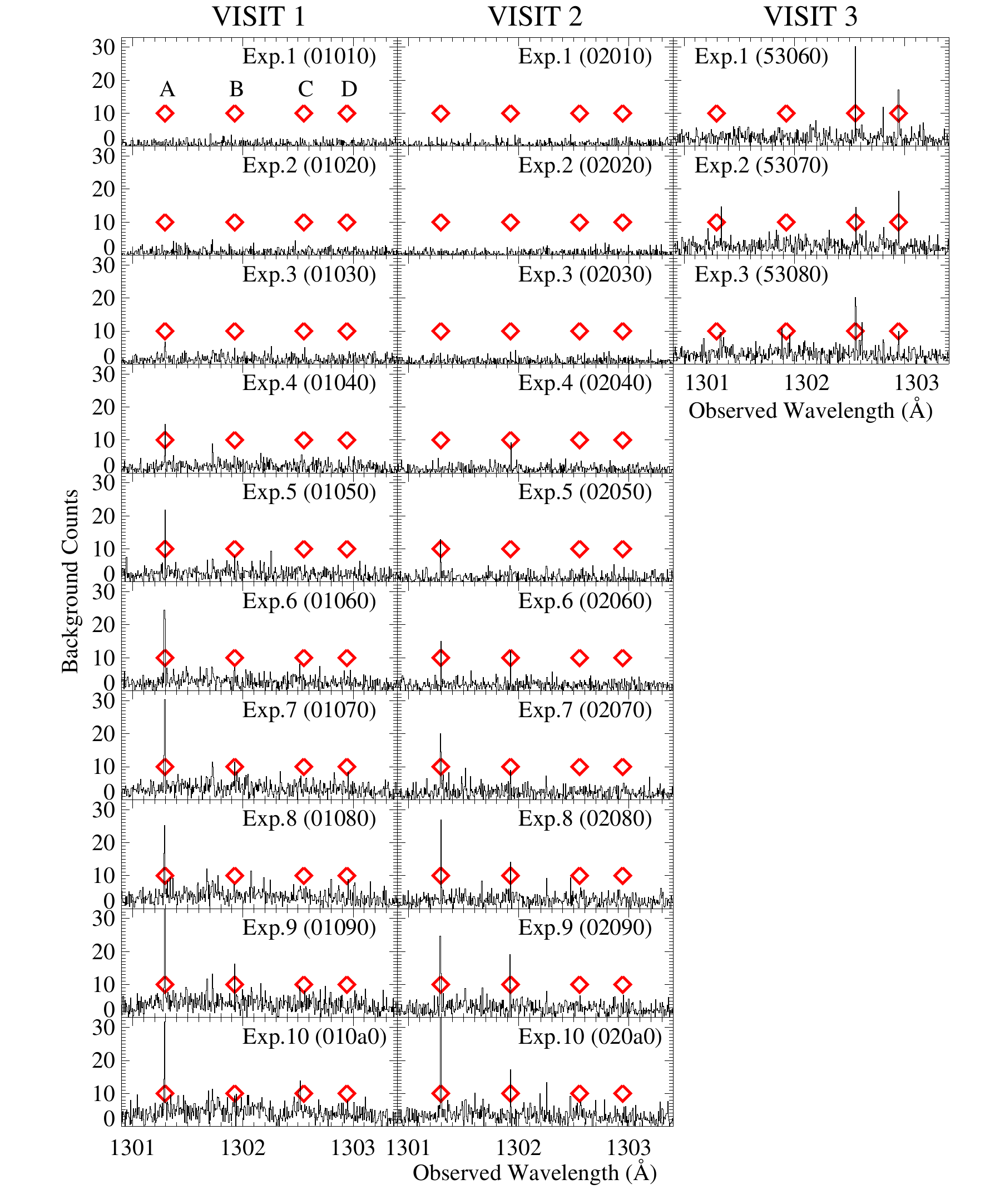}
\caption{Background counts recorded between 1301 and 1303.5 \AA\ in STIS E140H echelle order 323 in each exposure obtained during the three visits carried out for this program. The arrangement of the visits and exposures in the columns and panels is the same as in Fig.~\ref{fig:broad_dark_bkg}, so these figures can be directly compared panel-to-panel.  In this figure the counts are not binned. At the beginning of Visits 1 and 2, the detector was cold and the backgrounds are low.  As these visits progressed in time, the dark background continuum increased, and \textsl{discrete hot pixels also appeared and (usually) grew stronger} as well.  For purposes of discussion, the red diamonds labeled, A, B, C, and D mark the locations of four prominent hot pixels apparent in the three visits. \label{fig:all_hotpix_example} }
\end{figure*}

This is demonstrated in Figures~\ref{fig:broad_dark_bkg} $-$ \ref{fig:all_hotpix_example} using the new H1821+643 data obtained with the $\lambda_{\rm cen}$ = 1343 \AA\ grating tilt.  H1821+643 was observed with this grating tilt in three visits (Table~\ref{tab:stis_log}).  From left to right, the three columns in Figures~\ref{fig:broad_dark_bkg} $-$ \ref{fig:all_hotpix_example} show Visits 1, 2, and 3, and from top to bottom, each panel shows the background counts in each individual exposure starting from the first exposure.

Figure~\ref{fig:broad_dark_bkg} shows the broad behavior of the dark counts: in this figure, the pipeline background counts extracted by CALSTIS in each order in each exposure are plotted, using the color code shown in the legend at right, versus the pixel in the order (numbered in order of increasing wavelength).  In this figure only, the background counts are binned by 100 pixels to show broader trends.  This figure shows several of the known trends:  The ``continuum'' dark counts depend on detector region -- in echelle orders covering the shortest wavelengths (dark-blue curves), the dark counts are relatively low in all exposures, but the dark counts become increasingly elevated in orders covering increasing wavelengths until they peak in orders near $\approx$1350 \AA\ (green/yellow curves) and then decrease somewhat in orders at longer wavelengths (red curves).  The dark-count time dependence is also clear. At the beginning of Visits 1 and 2, the MAMA detector was relatively cool and the dark counts are low in all orders, but as the visits progressed and the HV-on time and detector temperature increased, the dark counts increased in the orders that are located in the dark-blob region.   The dark counts are elevated in the first Visit-3 exposure because the visit started with the $\lambda_{\rm cen}$ = 1453 \AA\ grating tilt and then switched to the $\lambda_{\rm cen}$ = 1343 \AA\  tilt (Table~\ref{tab:stis_log}). The dark counts are also lower on the long-wavelength side of each order.

Figure~\ref{fig:all_hotpix_example} illustrates how the discrete hot-pixel brightnesses are correlated with the overall dark background, but sometimes in a complicated way. This figure shows the background counts in a single order (echelle order 323), with no binning applied, in the region between 1301.0 and 1301.5 \AA .  The sequence of the panels and columns is that same as in Figure~\ref{fig:broad_dark_bkg}.  To guide this discussion, the location of four hot pixels are marked with red diamonds and labeled \textsl{`A'}, \textsl{`B'}, \textsl{`C'}, and \textsl{`D'} (see upper-left panel).  First consider hot pixels \textsl{A} and \textsl{B} and Visits 1 and 2.  In the first two or three exposures of Visits 1 and 2, when the detector was cold, hot pixel \textsl{A} was not evident.  As the detector warmed up, hot pixel \textsl{A} appeared and increased in brightness, positively correlated with the overall increase seen in the dark counts evident in Figure~\ref{fig:broad_dark_bkg}.  Its increases in brightness were not monotonic however: In Visit 1, this hot pixel brightness decreased a small amount (compared to the previous exposure) in Exposures 8 and 10.   Hot pixel \textsl{B} shows the same trend, mostly increasing with the continuum dark counts, but with undulations in brightness in some contiguous exposures.  For example, in Visit 2 hot pixel \textsl{B} appeared in Exposure 4 but then fluctuated up and down in Exposures 5, 6, and 7 until it gained some persistence in the last three exposures.    Even though Visit 1 started hotter (more dark counts) than Visit 2, hot pixel \textsl{B} was mostly fainter in Visit 1 than Visit 2.  Evidently, the behavior of the hot pixels is not a simple function of detector temperature or HV-on time.  This conclusion is also suggested by comparing the locations of the hot pixels in Visits 1 and 2 to their locations in Visit 3 carried out a year later.  In Visit 3, hot pixels \textsl{A} and \textsl{B} disappeared, but hot pixels appeared at different locations such as the locations marked \textsl{C} and \textsl{D} in Figure~\ref{fig:all_hotpix_example}. 

\begin{figure*}
\includegraphics[width=17.0cm]{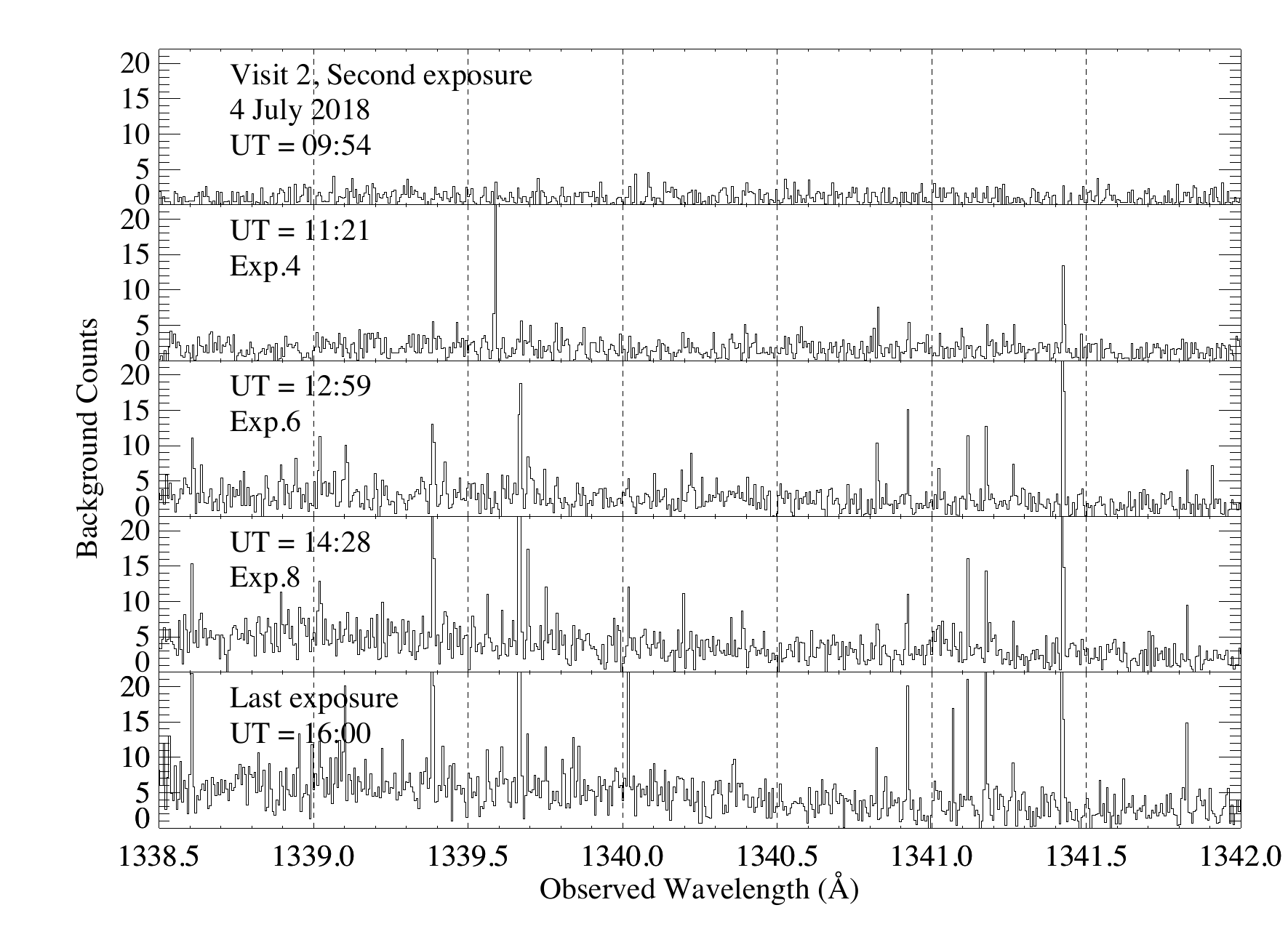}
\caption{Background counts in a portion of echelle order 314, an order that was badly affected by hot pixels in Visit 2.  Each panel shows the unbinned background counts for every other exposure starting with the second exposure (top) and ending with the final exposure in the visit (bottom).  Similarly to the backgrounds shown for a different order in Figure~\ref{fig:all_hotpix_example}, as the visit progressed and the detector warmed up, the hot-pixel problem became increasingly severe, and the final exposure is riddled with hot pixels.  Some of the hot pixels show elevated counts that monotonically increase with time, but other hot pixels can be seen to have counts that vary up and down with time. This order is close to the region where the MAMA dark counts are brightest (see Fig.~\ref{fig:broad_dark_bkg}) and thus represents one of the orders that is most badly affected by this problem. \label{fig:bad_hotpix_example}}
\end{figure*}

Figure~\ref{fig:bad_hotpix_example} shows the hot pixels in Visit 2 in echelle order 314, which is close to the region of the detector most severely impacted by dark counts.  We see that by the end of Visit 2, this order was riddled with hot pixels.  Not only are there hot pixels in many locations, there are also ``warm'' pixels that are not as obvious but are persistently higher than adjacent pixels in multiple contiguous exposures. Also, the hot spikes are not isolated to single pixels; in many cases two or even three adjacent pixels are elevated. 

\subsection{Hot-Pixel Mitigation Procedure}

Figures~\ref{fig:all_hotpix_example} - \ref{fig:bad_hotpix_example} only show the extracted backgrounds to make the warm/hot pixels easier to inspect (without being mixed with counts from the astronomical target) and to show the dark-count continuum as well as the discrete hot pixels.  Plots of the gross (source+background) counts similarly reveal many warm/hot pixels.  These pixels are often located in regions where there are no lines of interest, but of course they also frequently impact the absorption lines targeted for the program science.  This problem is not easily noticed in many STIS echelle programs because the targets are often much brighter (or use a lower-resolution grating) so that the count rates from the source are much higher than the dark count rates, and the hot pixels and dark counts have a small or negligible effect. The hot pixels are a serious problem in this paper; if some strategy is not used to mitigate the hot-pixel effects, the science data will be badly corrupted in many absorption profiles.

\begin{figure}
\includegraphics[width=8.0cm]{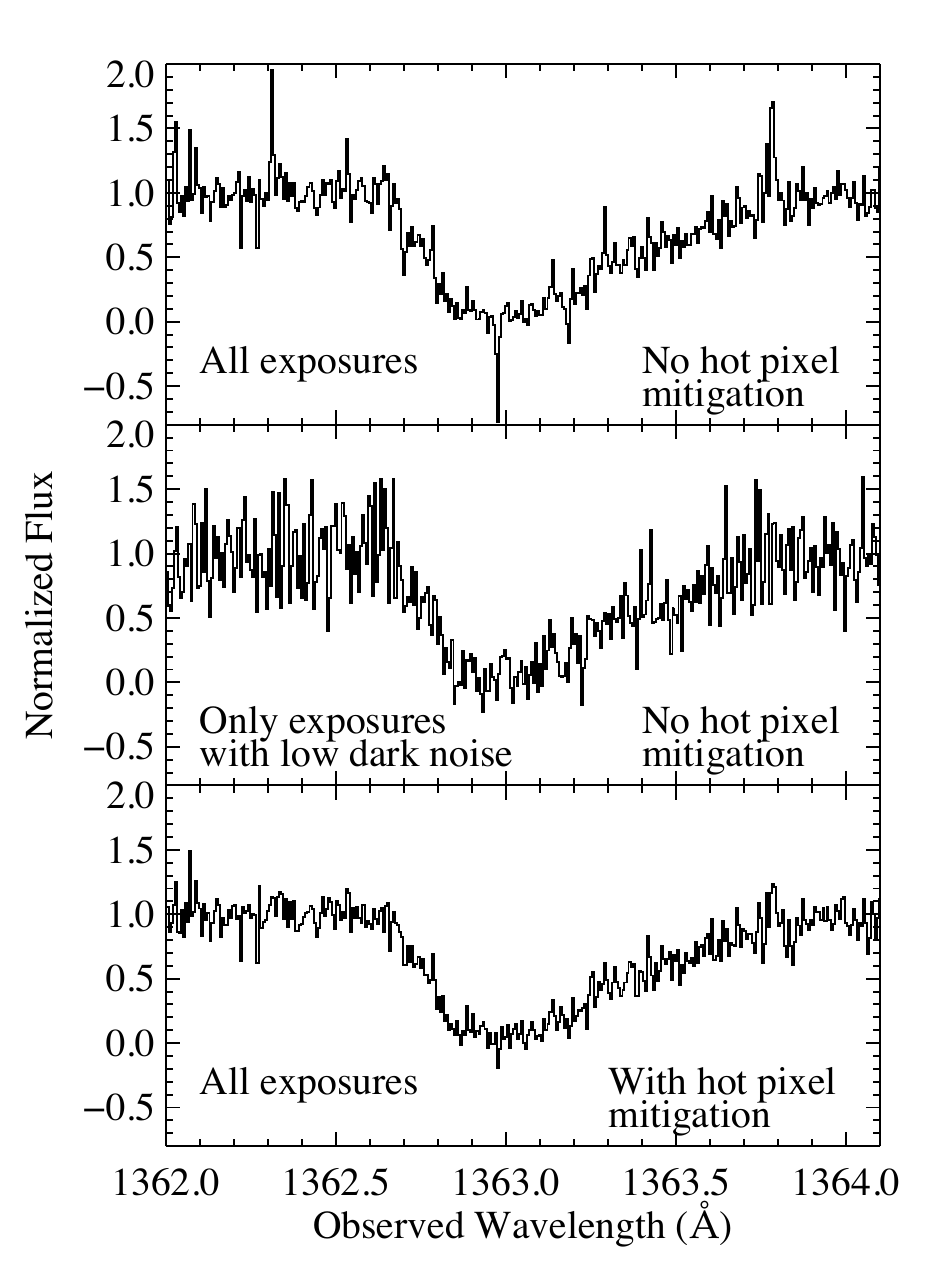}
\caption{Examples of the STIS E140H spectrum of H1821+643 near the extragalactic \HI\ Ly$\alpha$ line at $z_{\rm abs}$ = 0.1212 \citep{tripp01}, extracted with three methods.  The top panel shows the spectrum resulting from coaddition of the pipeline-extracted spectrum using the method of \citet{tripp01}. This includes the pipeline flagging of hot pixels but has no additional hot-pixel mitigation applied.  The middle panel plots a spectrum extracted with the same method but using only the exposures with the lowest dark noise.  The bottom panel presents the spectrum obtained by combining all of the exposures with the hot-pixel mitigation technique described in this appendix. \label{fig:three_hotpix}}
\end{figure}

The standard CALSTIS pipeline does include an algorithm to suppress hot pixels.  The pipeline flags pixels that have a dark-count rate $> 5\times$ the median dark rate recorded by the entire detector, and this flag can be used to mask out those pixels when the data are coadded.  The top panel of Figure~\ref{fig:three_hotpix} shows a spectrum produced by coadding all exposures from all of the visits with the pipeline hot-pixel flagging.  While this approach does reduce the impact of hot pixels, the top panel shows both hot-pixel positive spikes and negative (spurious) features that look like absorption lines but are actually artifacts due to subtraction hot pixels in the extracted background.  Another option is to only use exposures with low dark-noise levels.  The middle panel in Figure~\ref{fig:three_hotpix} presents a spectrum of the same region produced using only exposures with low dark noise.  Here we see that spurious hot-pixel effects have been suppressed, but at great cost in terms of reduced S/N.  Moreover, inspection of even the lowest-noise exposures reveals that they are not entirely free of warm/hot pixels.

Fortunately, the temporal and visit-to-visit variability of the warm/hot pixels enables a third option that is used in this paper.  Since the warm/hot pixels are not present in every exposure, it is possible to identify the hot pixels with a sigma-clipping algorithm and mask them out in only the exposures where they are present at a deleterious level.  The sigma clipping is applied to the extracted gross and background spectra from each individual exposure.  In every pixel, the mean counts in two 10-pixel regions, offset to lower and higher wavelengths by 10 pixels, is calculated along with the standard deviation $\sigma$.  Then, if the counts in that pixel exceed the mean in the adjacent regions by $>3\sigma$, it is flagged as a hot pixel in that exposure and masked out of the final coadded spectrum.  Since the final spectrum is produced by accumulating gross and background counts, it is also necessary to keep track of the effective exposure time in each pixel so that when the final count rate is calculated (= counts/effective exposure time), the fact that some pixels were masked out in some exposures will be accounted for.  Overall, this simple procedure effectively identified hot pixels, but some problems were observed.  First, when there is a real peak in an absorption profile that is adjacent to or in between regions of low flux (where the absorption is strong and the flux level is low), pixels in the peak were sometimes incorrectly flagged as hot pixels.  Second, in some cases especially the \textsl{warm} pixels were not flagged because they are not $> 3\sigma$ above the levels in the adjacent regions.  These warm pixels could be clearly identified as such by inspecting plots analogous to Figure~\ref{fig:all_hotpix_example}, and if igored, they affected the final spectrum.  In principle this could be rectified by choosing a more aggressive sigma level (e.g., 2$\sigma$) for the flagging algorithm, but this would lead to more good pixels being excluded.  Third, the hot pixels often afffect multiple adjacent pixels, and this algorithm sometimes flagged only the highest pixel in the multi-pixel spike (again, the multi-pixel nature can be seen from plots like Figure~\ref{fig:all_hotpix_example}).  Finally, occasionally the pipeline-extracted  pixels had negative counts.  These were not flagged by the algorithm, which only searched for positive hot pixels.  To deal with these occasional problems, plots of the gross and background counts, with hot pixels marked, were interactively inspected.  When one of these issues occurred, the pixel could be marked as a hot pixel or, in the case of an incorrectly marked hot pixel, the flag could be removed.  The algorithm by itself was deemed to work well for the vast majority of the pixels, but this interactive option, while labor intensive, did occasionally provide some improvement of the final spectrum.

\begin{figure}
\includegraphics[width=9.2cm]{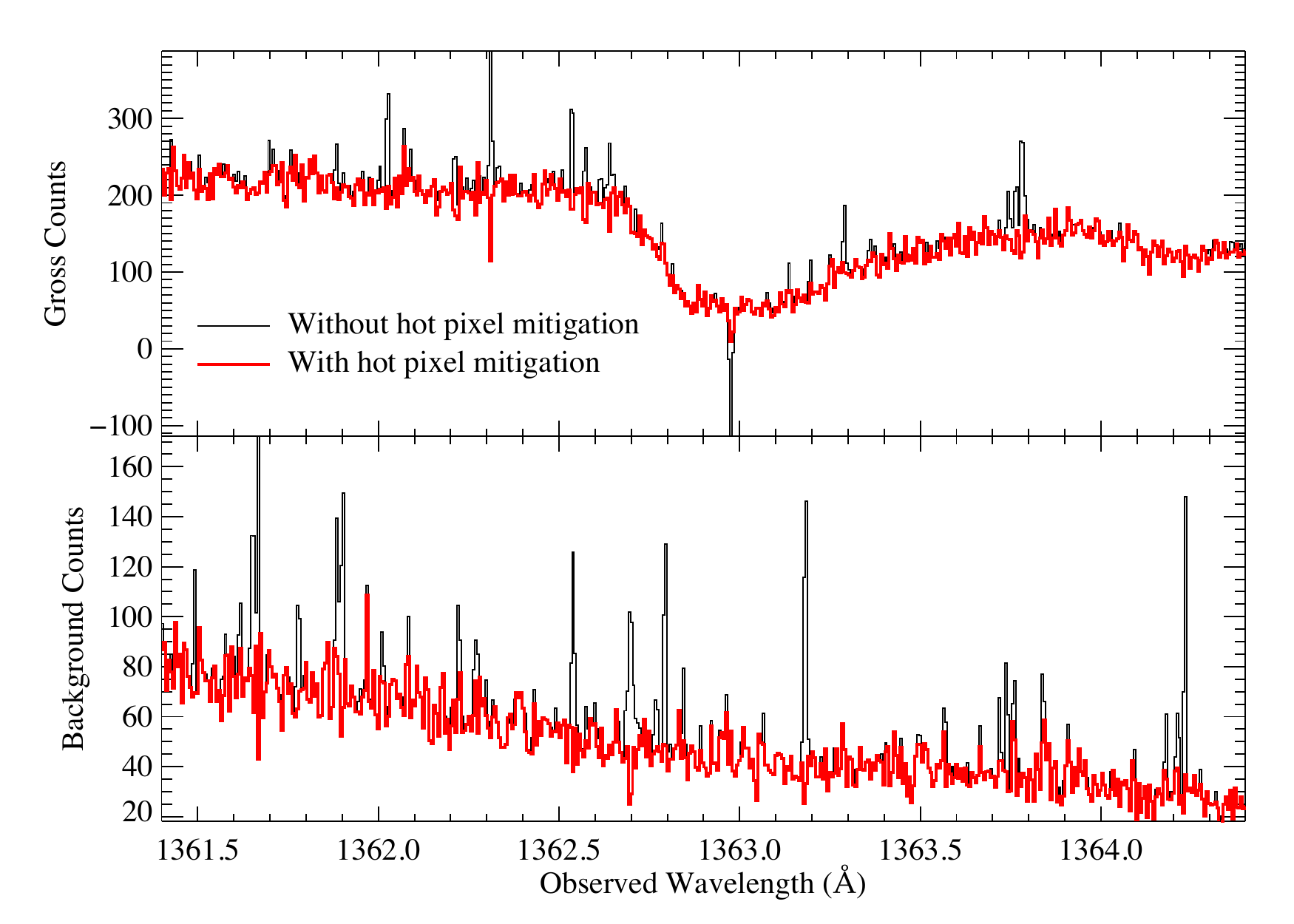}
\caption{Demonstration of the hot-pixel mitigation, applied to order 309.  The top panel shows the gross counts (target + background) extracted without the hot-pixel mitigation (black histogram) versus the gross counts extracted with the hot-pixel mitigation applied (red histogram).  Likewise, the bottom panel shows the background counts in this same order and wavelength range with (red) and without (black) the hot-pixel mitigation. \label{fig:gross_background_demo}}
\end{figure}

Figure~\ref{fig:gross_background_demo} demonstrates the total gross and background counts, accumulated from all exposures, with and without the hot-pixel mitigation, in the same region shown in Figure~\ref{fig:three_hotpix}.  Without hot-pixel mitigation, many warm/hot pixels are easily seen in both the gross and background spectra.  Near 1363 \AA , in the core of this saturated absorption line, an example of a pixel with negative counts from the pipeline can be seen (note: since the background has not yet been subtracted from the gross in this figure, the counts in the line core are offset from zero).  When the hot-pixel mitigation is employed, most of the problematic pixels are overcome.  The lower panel of Figure~\ref{fig:three_hotpix} shows the final spectrum obtained with the hot-pixel mitigation, after background subtraction and continuum normalization, for comparison to the pipeline algorithm and the spectrum extracted using only low-noise exposures.  This hot-pixel suppression procedure greatly reduces the hot-pixel problems while still attaining good S/N.  It is probably impossible to mask out the hot pixels entirely in this dataset because in some cases the hot pixels were already activated even in the first visit exposures when the detector was still cool.  In the future, it could be helpful to employ special observation scheduling to minimize the HV-on time when STIS MAMA exposures are recorded.

% Don't change these lines
\bsp	% typesetting comment
\label{lastpage}
\end{document}